\documentclass[11pt]{article}
\pdfoutput=1 
\usepackage{./jheppub}

\usepackage[T1]{fontenc}

\usepackage{graphicx}
\usepackage{amsfonts,amsmath,amssymb}
\usepackage{verbatim}
\usepackage{array}
\usepackage{mathrsfs}
\usepackage{mathrsfs}
\usepackage{yfonts}
\usepackage{dsfont}
\usepackage{bbm}
\usepackage{colonequals}
\usepackage{amscd}
\usepackage{xcolor}

\usepackage{url}

\usepackage{bm}

\usepackage{csquotes}

\usepackage{subcaption}

\usepackage{wrapfig}

\usepackage{suffix,mathtools,cancel,bbm}

\usepackage{multirow}

\usepackage{enumerate}

	\usepackage{tikz, tikz-3dplot, pgfplots}
	\usepackage{tkz-graph}
	\usetikzlibrary[positioning,patterns] 
	\usepackage{genyoungtabtikz}

\newcommand{\bn}{\begin{enumerate}}
\newcommand{\en}{\end{enumerate}}

\def\det{{\rm det}}

\def\deg{{\rm deg}}

\newcommand{\beq}{\begin{equation}}
\newcommand{\eeq}{\end{equation}}

\newcommand\nn{\nonumber}

\newcommand{\cA}{\mathcal{A}}
\newcommand{\cB}{\mathcal{B}}
\newcommand{\cC}{\mathcal{C}}

\newcommand{\cF}{\mathcal{F}}

\newcommand{\cI}{\mathcal{I}}

\newcommand{\cK}{\mathcal{K}}
\newcommand{\cL}{\mathcal{L}}
\newcommand{\cM}{\mathcal{M}}
\newcommand{\cN}{\mathcal{N}}
\newcommand{\cO}{\mathcal{O}}

\newcommand{\cS}{\mathcal{S}}

\newcommand{\cV}{\mathcal{V}}
\newcommand{\cW}{\mathcal{W}}

\newcommand{\cY}{\mathcal{Y}}

\numberwithin{equation}{section}

\def\bea{\begin{eqnarray}}
\def\eea{\end{eqnarray}}

\DeclarePairedDelimiterX\MeijerM[3]{\lparen}{\rparen}%
{\begin{smallmatrix}#1 \\ #2\end{smallmatrix}\delimsize\vert\,#3}

\newcommand\MeijerG[8][]{%
  G^{\,#2,#3}_{#4,#5}\MeijerM[#1]{#6}{#7}{#8}}

\WithSuffix\newcommand\MeijerG*[7]{%
  G^{\,#1,#2}_{#3,#4}\MeijerM*{#5}{#6}{#7}}

\def\cN{\mathcal{N}}

\def \beg#1{\begin{#1}} 

\def \bea{\beg{eqnarray}}
\def \eea{\end{eqnarray}}
\def \ee{\end{equation}}

\def \restr#1#2{{\left.\kern-\nulldelimiterspace#1\vphantom{\big|}\right|_{#2}}}

\def \nn{\nonumber}

\usepackage{colortbl}
\definecolor{mygray}{gray}{0.93}

\title{\boldmath Discrete and higher-form symmetries
in SCFTs from wrapped M5-branes }

\author[a]{Ibrahima Bah,}
\author[a]{Federico Bonetti,}
\author[b]{and Ruben Minasian}
\affiliation[a]{Department of Physics and Astronomy, Johns Hopkins University, 3400 North Charles Street, Baltimore, MD 21218, USA}
\affiliation[b]{Institut de Physique Th\'{e}orique, Universit\'{e} Paris Saclay, CNRS, CEA, F-91191, Gif-sur-Yvette, France}

\emailAdd{iboubah@jhu.edu, fbonett3@jhu.edu, ruben.minasian@ipht.fr}

\abstract
{
We analyze topological mass terms of BF type 
arising in supersymmetric M-theory compactifications to $AdS_5$.
These describe spontaneously broken higher-form gauge symmetries
in the bulk. Different choices of boundary conditions for the BF terms
yield  dual field theories with distinct
global discrete symmetries.
We discuss in detail
these symmetries and their 't Hooft anomalies for
4d $\mathcal N = 1$ SCFTs arising
from   M5-branes wrapped on a   Riemann surface
without punctures, including theories
from M5-branes at a $\mathbb Z_2$ orbifold
singularity.
The anomaly polynomial is computed via inflow and contains
background fields for discrete global 0-, 1-, and 2-form  symmetries
and 
continuous 0-form symmetries,
as well as  axionic background fields. The latter are properly interpreted  in the context
of anomalies in the space of coupling constants.

}

\usepackage{etoolbox}
\appto\appendix{\addtocontents{toc}{\protect\setcounter{tocdepth}{1}}}

\appto\listoffigures{\addtocontents{lof}{\protect\setcounter{tocdepth}{1}}}
\appto\listoftables{\addtocontents{lot}{\protect\setcounter{tocdepth}{1}}}

\begin{document} 

\maketitle
\flushbottom



\section{Introduction and summary}

't Hooft anomalies are robust and useful observables in
quantum field theory. 
They are invariant under renormalization group flow
and can be used to constrain the phases of theories
at long distances.
The most familiar type of 't Hooft anomalies
are arguably perturbative anomalies for continuous,
ordinary (0-form) symmetries, which only
occur in even spacetime dimension.
The full set of anomalies, however, is much richer.
This work is concerned with anomalies for 
discrete symmetries,  
generalized or higher-form symmetries \cite{Gaiotto:2014kfa},
and anomalies in the space of
coupling constants of a quantum field theory \cite{Cordova:2019jnf,Cordova:2019uob}.

't Hooft anomalies are particularly helpful in the study of the dynamics
of strongly-coupled field theories in the framework of geometric
engineering. Moreover, anomalies provide an organizing
principle in exploring   the landscape of such theories.
Discrete higher-form symmetries for field theories
engineered by M-theory on a    singular local geometry
have been recently studied in \cite{Morrison:2020ool,Albertini:2020mdx}.

This work focuses on field theories engineered with M5-branes. 
Using M5-branes, one can realize 
6d (2,0) theories of type $A_{N-1}$ \cite{Witten:1995zh,Strominger:1995ac},
as well as 6d (1,0) theories obtained by putting 
the M5-brane stack on top of an orbifold
singularity \cite{Blum:1997mm}. A vast class of 4d theories
is realized by further compactification on a Riemann surface,
possibly with punctures, as first studied for $\cN = 2$ theories
in \cite{Gaiotto:2009we,Gaiotto:2009hg}, and further extended to $\cN = 1$ theories \cite{Maruyoshi:2009uk,Benini:2009mz,Bah:2011je,Bah:2011vv,Bah:2012dg}.
It is beneficial to develop tools to extract
't Hooft anomalies of a field theory engineered using   branes
directly from the topology and geometry of the brane configuration.
Anomaly inflow provides the framework to address this problem.
Building on the results of \cite{Duff:1995wd,Witten:1996hc,Freed:1998tg,Harvey:1998bx} about anomaly inflow onto a stack of M5-branes, systematic tools have been developed 
to compute perturbative 't Hooft anomalies for 0-form symmetries via inflow
for setups engineered with M5-branes \cite{Bah:2018jrv,Bah:2019jts,Bah:2019rgq}
and D3-branes \cite{Bah:2020jas}. 

A more complete understanding 
of the space of quantum field theories
would require 
to extend the scope of  this program 
to include other types of 't Hooft anomalies.
In this paper, we address a class of discrete and higher-form symmetries
for 4d SCFTs engineered with wrapped M5-branes.
In particular, we perform a detailed analysis
for M5-branes probing a $\mathbb Z_2$ singularity, further wrapped
on a Riemann surface. This case study furnishes a controlled
example that exhibits interesting features.
Our strategy and results are summarized below.

\subsubsection*{Summary of results}

For a 4d SCFT  engineered with wrapped M5-branes,
non-trivial information about 't Hooft anomalies
for  discrete symmetries, higher-form symmetries,
and anomalies in the space of coupling constants \cite{Cordova:2019jnf}
can be extracted via anomaly inflow. This is done
by studying
the topological couplings in the 5d 
low-energy effective action originating from reduction of M-theory
on $M_6$, the compact space that encodes the geometry transverse
to the four extended directions of the M5-branes worldvolume.

In our analysis, we include all 5d    $0$-, $1$-, $2$- and $3$-form
gauge fields
associated to expansion of the M-theory 3-form $C_3$
onto  cohomology classes of $M_6$,
as well as 1-form gauge fields associated to isometries of $M_6$.
A crucial role is played by 5d topological mass terms
of BF type between a $1$-form gauge field $\cA_1$ and a $3$-form gauge field
$c_3$,
and between  pairs $(B_{2i}, \widetilde B_2^i)$ of   $2$-form gauge fields,
\beq
S = \int_{\cM_5} \bigg[  - \frac{1}{2\pi} \, k \, c_3 \wedge d\cA_1 
- \frac{1}{2\pi} \,  N \, \widetilde B_2^i \wedge dB_{2i}\bigg]  \ ,
\eeq
where $\cM_5$ is 5d spacetime, and $i$
labels the pairs of 2-form gauge fields.
As we shall see,
for setups with wrapped M5-branes the integers
$k$ and $N$ are determined by the $G_4$-flux quanta
of the system, and  $i = 1,\dots, g$ where $g$ is the genus of the
Riemann surface.
The BF term $\tfrac{1}{2\pi} \, k \, \cA_1 \wedge dc_3$ implies that
the 5d $U(1)$ 0-form gauge symmetry associated to $\cA_1$
is spontaneously broken to a $\mathbb Z_k$ 0-form 
gauge symmetry,
and the 5d $U(1)$ 2-form gauge symmetry associated to $c_3$
is spontaneously broken to a $\mathbb Z_k$ 2-form 
gauge symmetry
(see \emph{e.g.}~\cite{Maldacena:2001ss,Banks:2010zn} for reviews).
In a similar way, for each $i$ the term $\tfrac{1}{2\pi} \, N \, \widetilde B_2^i \wedge dB_{2i}$
signals   the spontaneous breaking of a bulk $U(1)^2$
1-form gauge symmetry to a $(\mathbb Z_N)^2$ 1-form gauge symmetry.
After a choice of topological boundary
conditions for the BF terms is made,
the discrete gauge symmetries in the bulk
are mapped to discrete global symmetries
of the 4d SCFT.
Moreover, the extended operators of the 5d BF theory
are mapped to   defects in the 4d SCFT,
which are charged under the discrete global symmetries.
A similar analysis in the context of $\rm{AdS}_4/{\rm CFT}_3$
has been recently performed for ABJM-type theories \cite{Bergman:2020ifi}.

In order to compute the full set of topological 
terms in five dimensions, including
the contributions of gauge fields associated to isometries of $M_6$
and an arbitrary external spacetime metric,
we use the tools developed in \cite{Bah:2019rgq}.
The 5d topological terms are conveniently 
encoded in a gauge-invariant closed
6-form $I_6^{\rm inflow}$, which is a polynomial in the 5d gauge field strengths. As concrete examples,
we consider 4d SCFTs engineered by M5-branes
wrapped on a Riemann surface \cite{Bah:2011vv,Bah:2012dg},
as well as theories from   M5-branes wrapped on a Riemann surface and
probing a $\mathbb Z_2$ singularity---in this case the gravity dual was identified in \cite{Bah:2019vmq}
to be one of the solutions first discussed in \cite{Gauntlett:2004zh}.
The 6-form $I_6^{\rm inflow}$ for these setups
are given in \eqref{BBBW_result}, \eqref{GMSW_inflow_tot}, respectively.

The 6-form $I_6^{\rm inflow}$ encodes the 't Hooft anomalies
of the 4d SCFT, together with the anomalies of modes that decouple
in the IR. Since the  5d 
bulk theory contains 
 massive 
gauge fields, care has to be taken in reading off
4d 't Hooft anomalies from $I_6^{\rm inflow}$.

If one is interested in perturbative anomalies for
continuous global symmetries of the 4d field theory,
the topologically massive gauge fields in five dimensions
must be integrated out. 
A similar mechanism is at play for 6d (1,0) SCFTs
engineered with M5-branes probing an ALE singularity,
and clarifies how the Green-Schwarz terms in the 8-form
anomaly polynomial \cite{Ohmori:2014kda} are reproduced by inflow.

The  perturbative anomaly polynomial for wrapped
M5-branes probing a $\mathbb Z_2$ singularity,
recorded in \eqref{pert_result},
contains several terms with 0-form gauge fields
(\emph{i.e.}~axions),
with 1-form field strengths.
Following \cite{Cordova:2019jnf}, we interpret such terms as  
anomalies in the space of coupling constants.
The couplings in question are associated to
exactly marginal operators of the 4d SCFT.
We argue that these operators can be thought
of as dimensional reduction on the Riemann surface
of the 6d conserved $U(1)$ currents
associated to the Cartan $U(1)_\mathrm N \times U(1)_\mathrm S$
of the  $SU(2)_\mathrm N \times SU(2)_\mathrm S$
flavor symmetry of the   6d (1,0) theory engineered
by M5-branes on a $\mathbb Z_2$ singularity.

After the continuous
part of a topologically massive gauge field is integrated out,
a discrete gauge field is left over, 
whose precise features depend
on the choice of boundary conditions
for the BF terms.
Hence, 
in order to extract 
4d 't Hooft anomalies
for discrete symmetries
from the 6-form $I_6^{\rm inflow}$,
we need to specify the  boundary conditions.
For definiteness, we focus on the case in which
we assign Dirichlet boundary conditions to the fields
$\cA_1$ and $B_{2i}$ (and free boundary conditions to $c_3$ and $\widetilde B_2^i$), so that the 4d field theory
admits a   $\mathbb Z_k$ global 0-form  symmetry and a  
 $(\mathbb Z_N)^g$ global 1-form  symmetry.
The field $c_3$ acts as a Lagrange multiplier
that imposes a constraint on $\cA_1$.
If we write 
$\cA_1 = \cA_1^{\rm cont} + \boldsymbol{\mathsf A}_1$,
the constraint fixes $\cA_1^{\rm cont}$ in terms of
gauge fields for continuous symmetries,
and forces $\boldsymbol{\mathsf A}_1$ to be a flat
1-form gauge field with holonomies that are $k$-th roots
of unity. Similarly, the Lagrange multiplier $\widetilde B_2^i$
imposes a constraint on $B_{2i} = B_{2i}^{\rm cont} + \boldsymbol{\mathsf B}_{2i}$,
which determines $B_{2i}^{\rm cont}$ in terms of continuous
gauge fields, and forces $\boldsymbol{\mathsf B}_{2i}$
to be a flat 2-form gauge field with holonomies that are $N$-th roots
of unity.

By substituting $\cA_1 = \cA_1^{\rm cont} + \boldsymbol{\mathsf A}_1$,
$B_{2i} = B_{2i}^{\rm cont} + \boldsymbol{\mathsf B}_{2i}$ into the
6-form $I_6^{\rm inflow}$, we obtain a formal expression that
encodes 't Hooft anomalies for both the continuous symmetries
and the $\mathbb Z_k$   0-form  symmetry and   
 $(\mathbb Z_N)^g$   1-form  symmetry.
Discrete anomalies are read off from terms in $I_6^{\rm inflow}$
with $d \boldsymbol{\mathsf A}_1$, $d \boldsymbol{\mathsf B}_{2i}$.
These objects are zero  as differential
forms. To circumvent this difficulty,
we reinterpret the quantity $I_6^{\rm inflow}$ in the framework
of differential cohomology
(see \emph{e.g.}~\cite{Freed:2006yc,Cordova:2019jnf} and appendix \ref{app_diff_cohom} for some
background material).
Differential forms are regarded as a proxy for 
classes in differential cohomology,
and their wedge product is a proxy for the product
in differential cohomology.
A crucial feature of the latter is that
the product of a flat gauge field with other gauge fields
is not necessarily zero.
This approach dates back to  
Dijkgraaf and Witten
\cite{Dijkgraaf:1989pz} and has also been recently used in \cite{Kapustin:2014zva}.

We apply the recipe outlined in the previous paragraphs
to the setup with wrapped M5-branes probing a $\mathbb Z_2$
singularity. The terms  in $I_6^{\rm inflow}$ involving 
$d \boldsymbol{\mathsf A}_1$, $d \boldsymbol{\mathsf B}_{2i}$
are collected schematically in \eqref{schematic_terms},
while the full result is recorded in appendix \ref{sec_app_result}.
We encounter a rich variety of 't Hooft anomalies
involving the discrete symmetries,
including: a cubic   term  in $d \boldsymbol{\mathsf A}_1$;
terms mixing $d \boldsymbol{\mathsf A}_1$ and $d \boldsymbol{\mathsf B}_{2i}$
to the other continuous symmetries, including a
gravitational term $d \boldsymbol{\mathsf A}_1 \, p_1(T)$;
a mixed anomaly between the two discrete symmetries
and a coupling constant.

Finally, we observe that the BF couplings
in the 5d topological bulk theory
can also be used to identify some of the singleton
modes of the 5d supergravity theory.
(By ``singleton modes'' we mean  modes that are pure gauge in the 5d bulk, but propagate 
on the conformal boundary; they are holographically dual
to modes in the 4d field theory that decouple in the IR.)
For setups with wrapped M5-branes with 4d $\cN = 2$ supersymmetry,
the knowledge of singleton modes from BF terms,
combined with supersymmetry, is sufficient to reconstruct
from the gravity side the entire set of modes that decouple on the field theory side. This offers a proof of principle that
one can compute the exact anomaly polynomial
from the gravity dual, including $\cO(1)$ terms
in the number of M5-branes.



\section{Topological mass terms in 5d supergravity}

Let us consider a supersymmetric $AdS_5$
solution of M-theory with internal space $M_6$.
These solutions were classified in \cite{Gauntlett:2004zh}.
We study the 5d supergravity theory 
obtained from reduction of M-theory on  a warped product
of the form $\cM_5 \times_w M_6$,
where external spacetime $\cM_5$ is  negatively curved.
The case $\cM_5 = AdS_5$ is recovered as the
vacuum solution of the 5d supergravity theory.
We restrict our attention to solutions where 
the space $M_6$ is  compact and  smooth, 
and the warp factor   is smooth and non-vanishing.
In this section we focus on the topological couplings
in the low-energy effective action of the 5d supergravity.

In particular, we are interested  in identifying 
the topological mass terms for the $p$-form gauge fields
that arise from Kaluza-Klein expansion of the M-theory 3-form
$C_3$ onto a basis of non-trivial cohomology classes on $M_6$.
If the internal space $M_6$ has isometries,
the 5d supergravity theory contains additional
(possibly non-Abelian) gauge fields.
For the remainder of this section, these gauge fields associated to isometries
of $M_6$ are turned off, since it can be checked
that they do not contribute to 
the topological terms of interest.
 They will be reinstated in section \ref{sec_anomalies}.

\subsection{Ansatz for $G_4$ and dimensional reduction}
\label{sec_ansatz}

The spectrum of the 5d supergravity obtained from reduction of M-theory on 
$M_6$ contains massless Abelian $p$-form gauge fields
coming from the Kaluza-Klein expansion of the M-theory 3-form $C_3$.
These massless $p$-form gauge fields are in 1-to-1 correspondence with
non-trivial cohomology classes of $M_6$.

For each $q=0,\dots,6$ we choose a basis in the 
lattice $H^q(M_6,\mathbb Z)_{\rm free}$,\footnote{This is the finitely
generated free Abelian group defined by the short exact sequence 
\beq
0 \rightarrow {\rm Tor}\, H^q(M_6,\mathbb Z) \rightarrow H^q(M_6,\mathbb Z)
\rightarrow H^q(M_6,\mathbb Z)_{\rm free} \rightarrow 0  \ ,\nn
\eeq
where ${\rm Tor}\, H^q(M_6,\mathbb Z)$ is the torsion
subgroup of $H^q(M_6,\mathbb Z)$.
}
which has rank given by the Betti number $b^q(M_6)$.
The   Betti numbers
of $M_6$ satisfy $b^0(M_6) = b^6(M_6) = 1$,
$b^1(M_6) = b^5(M_6)$, $b^2(M_6) = b^4(M_6)$.
Elements of $H^q(M_6,\mathbb Z)_{\rm free}$ can be identified with
de Rham cohomology classes of closed $q$-forms with integral
periods. As a result, we can represent
a basis of  $H^q(M_6,\mathbb Z)_{\rm free}$ using
a set of closed (but not exact) $q$-forms on $M_6$
with integral periods.
We use the following notation for these forms,
\beq \label{form_table}
\begin{array}{l      r r}
\text{1-forms:} & \qquad \lambda_{1u} \ , & \qquad u = 1,\dots, b^1(M_6) \ , \\
\text{2-forms:} & \qquad \omega_{2\alpha} \ , & \qquad \alpha = 1,\dots, b^2(M_6) \ , \\
\text{3-forms:} & \qquad \Lambda_{3x} \ , & \qquad x = 1,\dots, b^3(M_6) \ , \\
\text{4-forms:} & \qquad \Omega_4^\alpha \ , & \qquad \alpha = 1,\dots, b^2(M_6) \ .
\end{array}
\eeq
The 5d gauge fields originating from $C_3$ 
and their field strengths are denoted as follows,
\beq \label{field_table}
\begin{array}{l   r   c r}
\text{0-form potentials:} & \qquad a_0^x \ , & 
\qquad f_1^x = da_0^x \ , & \qquad x = 1,\dots, b^3(M_6) \ , \\
\text{1-form potentials:} & \qquad A_1^\alpha \ , & 
\qquad F_2^\alpha = dA_1^\alpha \ , & \qquad \alpha = 1,\dots, b^2(M_6) \ , \\
\text{2-form potentials:} & \qquad B_2^u \ , & 
\qquad H_3^u = dB_2^u \ , & \qquad u = 1,\dots, b^1(M_6) \ , \\
\text{3-form potential:} & \qquad c_3 \ , & 
\qquad \gamma_4 = dc_3 \ .  \\
\end{array}
\eeq
Throughout this work, we adopt conventions in which 
the periods of the field strength of an Abelian $p$-form
gauge fields are quantized in units of $2\pi$.
(A 0-form gauge field whose field strength is quantized in units of
$2\pi$ is the same as a compact scalar field with period $2\pi$.)

In string/M-theory compactifications, torsion cycles in the internal
space can be a source of discrete gauge symmetries
\cite{Camara:2011jg, BerasaluceGonzalez:2012vb}. In this work, we do not study the effects of torsion
in the homology of $M_6$. The geometries $M_6$ that
are relevant for the setups with wrapped M5-branes studied in this paper
do not have torsion in homology.

With the notation introduced in \eqref{form_table} and \eqref{field_table},
the M-theory 4-form field strength $G_4 = dC_3$,
including both its background value and fluctuations
associated to cohomology classes   on $M_6$, is given by
\begin{align} \label{G4_ansatz}
\frac{G_4}{2\pi} &=   N_\alpha \, \Omega_4^\alpha
+ \frac{F_2^\alpha}{2\pi} \wedge \omega_{2\alpha}
 + \frac{f_1^x}{2\pi} \wedge \Lambda_{3x}
 + \frac{H_3^u}{2\pi} \wedge \lambda_1{}_u
 + \frac{\gamma_4}{2\pi} \  .
\end{align}
The integers $N_\alpha$  specify
the background flux that threads $M_6$.
The periods of $G_4$ in \eqref{G4_ansatz} are quantized in units of 
$2\pi$.\footnote{The flux quantization condition in M-theory on an orientable spacetime $M_{11}$ can be written as \cite{Witten:1996md}
\beq
\int_{\cC_4} \frac{G_4}{2\pi} = \frac 12 \, \int_{\cC_4} w_4(TM_{11}) \mod 1 \ , \qquad
\text{for any 4-cycle $\cC_4$ in $M_{11}$} \ . \nn
\eeq
where $w_4$ denotes the fourth  Stiefel-Whitney class.
It is known that $w_4$ is zero for a
 spin manifold of dimension $\le 7$
(the argument can be found for instance on page 65 of \cite{Hsieh:2020jpj}).
In our setups the internal space $M_6$ and external spacetime
are   spin manifolds, hence the shift in the quantization
condition of $G_4$ is not important.
This holds true also for the purposes of writing the anomaly polynomial
of a 4d theory using descent: in that case external spacetime is effectively six-dimensional.
}

In our normalization conventions for $G_4$,
the topological terms of the low-energy effective action
of M-theory are\footnote{In these conventions,
the Einstein-Hilbert term and the kinetic term for $G_4$ take the form
\beq
S_{\rm kin} = \int_{M_{11}} \bigg[  2\pi \, (2\pi \, \ell_{\rm P})^{-9} \, R \, *1 - \frac 12 \,
2\pi \, (2\pi  \, \ell_{\rm P})^{-3}
\,  G_4 \wedge * G_4  \bigg] \ , \nn
\eeq
where $\ell_{\rm P}$ is the 11d Planck length. The
action enters the path integral via $e^{iS}$ and is defined mod $2\pi$.}
\beq \label{Mtheory_action_top}
S_{\rm top} = \int_{M_{11}} \bigg[ 
-  \frac{1}{6 \, (2\pi)^2} \, C_3 \wedge G_4 \wedge G_4
- \frac{1}{2\pi} \, C_3 \wedge X_8
\bigg] \ ,
\quad 
X_8 = \frac{p_1^2(TM_{11}) - 4 \, p_2(TM_{11})}{192}   \ .
\eeq
The low-energy effective action for the 5d $p$-form gauge
fields listed in \eqref{field_table} is computed  
via standard Kaluza-Klein reduction.
Recall that external metric fluctuations and gauge fields
associated to isometries of $M_6$ are turned off in this section.
For the purpose of computing the effective action
for the modes   in \eqref{field_table} the term $C_3 X_8$ plays no role.\footnote{Even
after the isometry gauge fields are turned on,
the term $C_3 X_8$ does not yield topological mass terms (\emph{i.e.}~topological
terms quadratic in the external fields) for the cases of interest in this work.
}
The kinetic term for $G_4$ yields standard kinetic
terms for the 5d gauge fields.
The Chern-Simons
coupling $C_3 G_4 G_4$ yields a set of topological
terms in the 5d effective action.
They are most conveniently written
in terms of a gauge-invariant 6-form, 
\beq\label{5dtopAction}
S_\text{top}  = 2\pi \, \int_{\cM_5} I_5^{(0)} \ , \qquad dI_5^{(0)} = I_6 \ ,
\eeq
where the 6-form $I_6$ is given by
\begin{align} \label{5dtop}
I_6   = \frac{1}{(2\pi)^2} \, \bigg[
&- N_\alpha \, \gamma_4 \wedge F_2^\alpha
+ \frac 12 \, N_\alpha \, \cK^\alpha{}_{uv} \, H_3^u \wedge H_3^v
\bigg] \nn \\
  + \frac{1}{(2\pi)^3} \, \bigg[
 & - \frac 16 \, \cK_{\alpha \beta \gamma} \, F_2^\alpha 
 \wedge F_2^\beta \wedge F_2^\gamma 
 + \cK_{xu\alpha} \, f_1^x \wedge F_2^\alpha \wedge H_3^u
 + \frac 12 \, \cK_{xy} \, \gamma_4 \wedge f_1^x \wedge f_1^y
 \bigg]  \ .
\end{align}
The quantities $\cK^{\alpha uv}$, $\cK_{\alpha \beta \gamma}$, 
$\cK_{xu\alpha}$, $\cK_{xy}$ are integer intersection numbers
which can be defined in terms of the closed forms on  $M_6$ 
as
\begin{align} \label{intersection_numbers}
\cK^{\alpha  }{}_{uv} & = \int_{M_6} \Omega_4^\alpha \wedge \lambda_{1u} \wedge \lambda_{1v} \ , &
\cK_{\alpha \beta \gamma} & = \int_{M_6} \omega_{2\alpha} \wedge
\omega_{2\beta} \wedge \omega_{3\gamma} \ , \nn \\
\cK_{xu\alpha} & = \int_{M_6 } \Lambda_{3x} \wedge \lambda_{1u}
\wedge \omega_{2\alpha} \ , &
\cK_{xy} & = \int_{M_6} \Lambda_{3x} \wedge \Lambda_{3y} \ .
\end{align}
These intersection numbers 
  depend only on the cohomology classes
of the internal forms, and not on the specific representatives
used to write down $G_4$ in \eqref{G4_ansatz}.

The first two terms in  \eqref{5dtop} are the sought-for
topological mass terms in the 5d supergravity effective action.
In contrast to the other topological
couplings in \eqref{5dtop}, they are quadratic in the external gauge fields.
We stress that the topological mass terms are due to the background flux quanta
$N_\alpha$.

When $b^2(M_6) \ge 2$ we are free to consider a change of basis
in the lattice $H^2(M_6,\mathbb Z)_{\rm free}$, which
is accompanied by a
change of basis in the external 1-form gauge fields.
A new basis $A_1'^\alpha$
can always be found such that 
$A_1'{}^{\alpha = 1}$ is   the only
1-form gauge field with a topological mass term with  $c_3$.
Since $A_1'{}^{\alpha = 1}$ plays a special role
compared to the vectors $A_1'{}^{\alpha \neq 1}$,
we introduce the notation
\beq \label{the_new_basis}
A_1'^\alpha = ( \cA_1 ,   {\cA}_1^{\widehat \alpha}) \ , \qquad
\hat \alpha = 2, 3, \dots, n \ .
\eeq
With this notation we have
\beq \label{new_basis_rel}
- N_\alpha \, F_2^\alpha \wedge \gamma_4 = - k \, d\cA_1 \wedge \gamma_4 \ ,
\qquad k = {\rm gcd} (N_\alpha) \ ,
\eeq
while the vectors $ {\cA}_1^{\widehat \alpha}$ do not enter the topological
mass terms. 
Further information about the new basis $A_1'^\alpha$
is collected in appendix \ref{sec_goodbasis}.

\subsection{Applications to wrapped M5-branes} \label{sec_wrappedM5s}

In this section we specialize the results of
the previous section to two classes of $AdS_5$ solutions
that are particularly relevant in connection to
4d $\cN = 1$ SCFTs engineered with M5-branes wrapped on a Riemann surface.
More precisely, we consider:
\begin{itemize}
\item M5-branes wrapped on a Riemann surface
without punctures, which correspond to   the solutions of \cite{Bah:2011vv,Bah:2012dg},  referred to as BBBW.
\item M5-branes probing a $\mathbb Z_2$ singularity
and wrapped on a Riemann surface without punctures \cite{Bah:2019vmq},
which correspond to a class of solutions of \cite{Gauntlett:2004zh},
referred to as GMSW.
\end{itemize}

\subsubsection{M5-branes wrapped on a Riemann surface} \label{BBBW_sol_sec}

The BBBW solutions   \cite{Bah:2011vv,Bah:2012dg} 
describe the near-horizon geometry of a stack of M5-branes
wrapped on a genus-$g$ Riemann surface $\Sigma_g$
with a non-trivial topological twist preserving 4d $\cN \ge1 $
superconformal symmetry.
The internal space $M_6$ 
is topologically an $S^4$ bundle over $\Sigma_g$.
Its topology
is encoded in two
integer numbers $p$, $q$ satisfying
\beq \label{BBBW_susy}
p+q = - \chi(\Sigma_g) = 2 \,(g-1) \ .
\eeq
We can regard $S^4 \hookrightarrow M_6 \rightarrow \Sigma_g$
as the unit-sphere bundle associated to a real rank-5
vector bundle $\mathbb R^5 \hookrightarrow \mathscr N \rightarrow \Sigma_g$.
The bundle $\mathscr N$ 
is identified with the normal bundle to the M5-brane stack.
It splits as $\mathscr N = \cL_1 \oplus \cL_2 \oplus \mathscr N_0$,
where $\cL_1$, $\cL_2$ are complex line bundles over $\Sigma_g$,
and $\mathscr N_0$ is a trivial real rank-1 vector bundle.
The integers $p$, $q$ are the Chern numbers of 
the complex line bundles $\cL_1$, $\cL_2$, respectively.

\begin{table}
\begingroup
\renewcommand{\arraystretch}{1.1}
\begin{center}
\begin{tabular}{|c  | c | c| l  |} \hline
 \multirow{3}{*}{$g \ge 2$}  & $p\neq 0$, $q\neq 0$, $p \neq q$ &   $U(1)_1 \times U(1)_2$ & $\cN = 1$ \\
 & $p= 0$ or $q= 0$   & $SU(2)_1 \times U(1)_2$ \;or\; $U(1)_1 \times SU(2)_2$
& $\cN = 2$ MN \\
 & $p= q$    & $SU(2)  \times U(1)$
& $\cN = 1$ MN \\ \hline
$g =0$ & $|p-q|>2$    & $U(1)_1 \times U(1)_2 \times SU(2)_\Sigma$
& $\cN = 1$   \\ \hline
$g =1$ & $p \neq 0$    & $U(1)_1 \times U(1)_2$
& $\cN = 1$   \\ \hline
\end{tabular}
\end{center}
\endgroup
\caption{ 
Summary of the values of $p$, $q$ that yield smooth $AdS_5$ solutions in M-theory. 
Recall $p+q = 2(g-1)$.
In the third column we list the isometries of the internal space $M_6$.
In the last column, MN stands for Maldacena-Nu\~nez and refers to the solutions of~\cite{Maldacena:2000mw}.
} 
\label{BBBW_cases}
\end{table}

In  table \ref{BBBW_cases}  we summarize the 
choices of $g$, $p$, $q$ 
for which a smooth $AdS_5$ M-theory solution exists,
and for each case we list the isometries of the internal space $M_6$.
Some comments are in order. In all cases, $M_6$ admits
at least a $U(1)_1\times U(1)_2$ isometry, which is the subgroup
of the $SO(5)$ isometry of the $S^4$ fiber that is preserved by
the fibration over $\Sigma_g$
for any choice of $p$, $q$. 
When $g\ge 2$, $p=0$, $U(1)_1$ enhances to $SU(2)_1$.
Supersymmetry enhances to $\cN = 2$
and the isometry group $SU(2)_1 \times U(1)_2$ is identified
with the R-symmetry of the SCFT. This setup is
the $\cN = 2$ Maldacena-Nu\~nez
(MN) solution \cite{Maldacena:2000mw}. Similar remarks apply to $g \ge 2$, $q=0$.
In the case $g \ge 2$, $p=q$, the difference of the generators of $U(1)_1$
and $U(1)_2$ enhances to $SU(2)$, which is identified with an enhanced
flavor symmetry of the SCFT side. This is the $\cN = 1$ MN solution \cite{Maldacena:2000mw}.
When $g=0$, the Riemann surface is a round sphere $S^2$.
The space $M_6$ admits an additional $SO(3)_\Sigma$ isometry,
originating from the isometry of $S^2$.
Finally, we would like to emphasize that the case
$g=1$, $p=q=0$,
which corresponds to 4d $\cN = 4$ SYM theory,
is not included in table \ref{BBBW_cases}, because there is no smooth $AdS_5$ M-theory
solution with internal space $S^4 \times T^2$ without any twisting.
The $\cN = 4$ SYM theory is best studied holographically
via the standard $AdS_5 \times S^5$ solution in type IIB string theory.

The number of external $p$-form fields entering the topological terms \eqref{5dtop} in the 5d effective action
is determined by the Betti numbers of $M_6$. The latter do not
depend on the twist parameters $p$, $q$ and are given by
\beq
b^0(M_6) = b^2(M_6) = b^4(M_6) = b^6(M_6) = 1 \ , \quad
b^1(M_6)= b^5(M_6) = 2g \ , \quad
b^3(M_6)  = 0 \ .
\eeq
This claim is verified in appendix \ref{app_BBBW},
where we also construct the associated closed forms
with integral periods.
The fact that $b^4(M_6) = 1$ is consistent with the fact
that BBBW solutions have only one flux parameter,
\beq
N_{\alpha = 1} = N \ ,
\eeq
which is the number of M5-branes in the stack.
We notice  that $b^1(M_6) = 2g$ stems from the fact that
the $2g$ harmonic 1-forms on $\Sigma_g$ can be pulled
back to $M_6$, yielding closed but not exact 1-forms,
whose de Rham classes account for the entire  1-cohomology of $M_6$.
The 5d $p$-form gauge fields originating from the expansion of $C_3$
are
\beq
c_3 \ , \qquad A_1^{\alpha = 1} = A_1 \ , \qquad B_2^u \ , \qquad u = 1,\dots, 2g \ ,
\eeq
while we do not find any 0-form gauge potential.

Making use of the closed forms of appendix \ref{app_BBBW},
we can compute explicitly the intersection numbers \eqref{intersection_numbers}.
The only non-zero intersection pairing is $\cK^{\alpha = 1}{}_{uv}$,
which can be written as
\beq
\cK^{\alpha =1 }{}_{uv}  = \Omega_{uv} = \cC_{1u}^\Sigma \cdot \cC_{1v}^\Sigma \ , \qquad
u,v = 1,\dots, 2g \ .
\eeq
In the previous expression $\cC_{1u}$ denotes a basis of integral
1-homology on $\Sigma_g$,
and $\Omega_{uv}$ is the intersection pairing,
which is antisymmetric and non-degenerate.
The 6-form $I_6$ encoding the topological couplings of the 5d action
as in \eqref{5dtop} is given by
\begin{align} 
I_6 & =\frac{1}{(2\pi)^2} \bigg[
 -  N  \, \gamma_4 \wedge F_2
+ \frac 12 \,  N \, \Omega_{uv}  \, H_3^u \wedge H_3^v
\bigg] \ .
\end{align}

It is useful to choose a basis  $\cC_{1u}^\Sigma$ of integral 1-homology
on $\Sigma_g$ that is based on the standard A and B cycles 
on the Riemann surface.
Correspondingly, we write
\beq \label{nice_basis}
\cC_{1u}^\Sigma = (\cC_i  , \widetilde \cC^i) \ , \qquad 
\begin{array}{l}
\cC_i \cdot \widetilde \cC^j =  - \widetilde  \cC^j \cdot \cC_i = \delta^j_i    \ , \\
\cC_i \cdot \cC_j = 
\widetilde \cC^i \cdot \widetilde \cC^j  = 0 \ ,
\end{array}
\qquad i, j = 1,\dots, g \ ,
\eeq
In other words, 
the intersection pairing $\Omega_{uv}$ in this basis takes the standard form
\beq \label{goodOmega}
\Omega_{uv} = \begin{pmatrix}
0 & \delta_i^j \\
- \delta_i^j & 0
\end{pmatrix} \ .
\eeq
The group of linear transformations of the lattice $H_1(\Sigma_g,\mathbb Z)$
that preserve this form of $\Omega_{uv}$ is $Sp(2g;\mathbb Z)$.
(In our notation, $Sp(2,\mathbb Z) = SL(2,\mathbb Z)$.)
The choice of basis \eqref{nice_basis} implies that
the index $u$ on the 2-form gauge fields $B_2^u$
is split into two sets of $g$ values,
\beq
B_2^u = (B_{2i} , \widetilde B_2^i) \  , \qquad i = 1,\dots,g \ .
\eeq
In this basis, the 6-form $I_6$ reads
\begin{align}  \label{BBBW_top}
I_6 & =\frac{1}{(2\pi)^2} \bigg[  -  N  \, \gamma_4 \wedge F_2
-  N \,  \widetilde H_3^i \wedge H_{3i} \bigg] \ .
\end{align}
 
\subsubsection{M5-branes probing a $\mathbb Z_2$ singularity and
wrapped on a Riemann surface}

Let us now consider a class of solutions first discussed in \cite{Gauntlett:2004zh}.
The space $M_6$ is topologically an $S^2$ bundle
over the   product of two Riemann surfaces. 
If one of the Riemann surfaces is a torus, the setup is best studied
by dualizing the M-theory solution to a type IIB string theory solution.
We thus focus on the case where both Riemann surfaces are non-flat.
There is no smooth solutions if both Riemann surfaces are negatively curved.
We are therefore left with one sphere and one Riemann surface $\Sigma_g$
with $g=0$ or $g\ge 2$.
The line element has the form
\beq \label{schematic_GMSW}
ds^2(M_6) = f_{\varphi}(\mu)\, ds^2(S^2_\varphi) 
+ f_\Sigma(\mu) \, ds^2(\Sigma_g)  
+ f_\mu(\mu) \,  d \mu^2 
+ f_\psi(\mu) \,  D\psi^2 \ .
\eeq
In the previous expression,  $\psi$ is an angular coordinate
with period $2\pi$, while 
$\mu$ is a coordinate on an interval, $\mu \in [\mu_{\rm S}, \mu_{\rm N}]$.
The quantity $ds^2(S^2_\varphi) = d\theta^2 + \sin^2 \theta\, d\varphi^2$
is the standard line element on a unit-radius two-sphere,
while 
$ds^2(\Sigma_g)$ denotes the line element on 
a Riemann surface of constant curvature $\kappa = \pm1$,
respectively.
The functions
$f_{\varphi}$, $f_\Sigma$ are strictly positive on the entire
$\mu$ interval. 
The function $f_\mu$ has poles at $\mu = \mu_{\rm N,S}$,
while $f_\psi$ as zeros at $\mu = \mu_{\rm N,S}$;
as a result, the $\mu$ and $\psi$ coordinate describe
a two-dimensional space $S^2_\psi$ which is topologically a 2-sphere,
with isometry group $U(1)_\psi$.
The circle $S^1_\psi$ shrinks smoothly at $\mu = \mu_{\rm N,S}$.
Finally, the fibration of $S^2_\psi$ over the base $S^2_\varphi \times \Sigma_g$
is encoded in
\beq \label{dDpsi}
dD\psi = - 2 \, V_{\varphi} - \chi \, V_\Sigma  \ , \qquad\chi = 2 - 2g \ , 
\eeq
where $V_{\varphi}$, $V_\Sigma$ are the volume forms
on $S^2_\varphi$, $\Sigma_g$, respectively, normalized according to
\beq
\int_{S^2_\varphi} V_{\varphi}= 2\pi \ , \qquad
\int_{\Sigma_g} V_\Sigma = 2\pi \ .
\eeq
More details on these geometries can be found in appendix \ref{app_GMSW}.

\begin{figure}
\centering
\includegraphics[width = 7.5 cm]{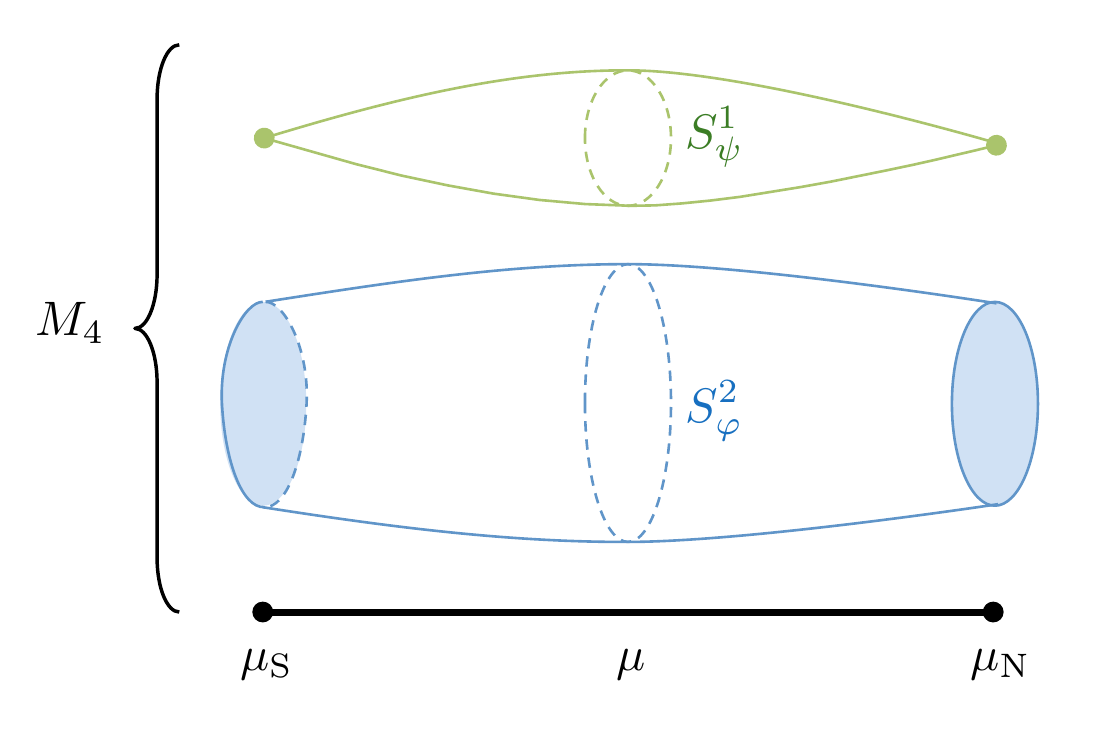}
\caption{Schematic depiction of the space $M_4$ comprised by the 2-sphere
$S^2_\varphi$, the circle $S^1_\psi$, and the $\mu$ interval.
The space $M_4$ is the blow-up resolution of $S^4/\mathbb Z_2$.
The blow-up $\mathbb P^1$'s are identified
with $S^2_\varphi$ at $\mu = \mu_\mathrm N$ and $\mu = \mu_\mathrm S$.
}
\label{GMSW}
\end{figure}

To highlight the interpretation of $M_6$ in terms of wrapped
M5-branes, it is convenient to present $M_6$ as
\beq
M_4 \hookrightarrow M_6 \rightarrow \Sigma_g \ ,
\eeq
where the space $M_4$ consists of the $\mu$, $\psi$ directions and the
2-sphere $S^2_\varphi$. It is depicted schematically in figure \ref{GMSW}.
The space $M_4$ can be identified with the resolution of the
quotient $S^4/\mathbb Z_2$ \cite{Bah:2019vmq}.
Notice that $S^2_\varphi$ does not shrink at $\mu = \mu_{\rm N,S}$
and defines two 2-cycles in $M_4$. The latter
are identified with the resolution cycles originating from
the blow-up of the singularities of $S^4/\mathbb Z_2$.\footnote{The
$\mathbb Z_2$ action is $(y^1, y^2, y^3, y^4, y^5) \mapsto (- y^1, - y^2, - y^3, - y^4, y^5)$
in terms
the  Cartesian   coordinates $y^{1,2,3,4,5}$
of $\mathbb R^5 \supset S^4$.  This action has
two fixed points on $S^4$ at $y^5 =\pm 1$. Near each fixed point
the space looks like $\mathbb R^4/\mathbb Z_2$
and can be resolved by a 2-center ALE Taub-NUT geometry.}
This motivates the interpretation of $M_6$ as near
horizon geometry of a stack of M5-branes
probing a $\mathbb Z_2$ singularity and wrapped on a Riemann surface.

The Betti numbers of $M_6$ are
\begin{align}
b^0(M_6) &= b^6(M_6) = 1 \ , &
b^2(M_6) &= b^4(M_6) = 3 \ , \nn \\
b^1(M_6) &= b^5(M_6) = 2g \ , &
b^3(M_6) &= 4g \ .
\end{align}
This is verified in appendix \ref{app_GMSW},
where we also construct the closed forms needed to represent
all cohomology classes of $M_6$.
In accordance with $b^4(M_6) = 3$,
this class of GMSW solutions has three independent flux parameters.
They can be taken to be 
\beq \label{basis_of_Ns}
N_\alpha = (N, N_+ , N_-) \ , \qquad \alpha = 1,2,3 \ ,
\eeq
where $N$ is the flux through $M_4$ (at a generic point on $\Sigma_g$)
and is identified with the number of M5-branes in the stack,
while $N_\pm = \frac 12 \, (N_\mathrm N \pm N_\mathrm S)$ encode the fluxes through the
2-cycles in $M_4$ at $\mu = \mu_{\rm N,S}$ combined with
$\Sigma_g$. A more detailed discussion
can be found in appendix  \ref{app_GMSW}.
With reference to \eqref{basis_of_Ns}, the three 1-form gauge fields
originating from expansion of $C_3$ onto 2-cohomology classes
are denoted
\beq
A_1^\alpha = (A_1 , A _1^+, A_1 ^-) \ , \qquad
F_2^\alpha = (F_2,  F_2^+, F_2^-) \ .
\eeq

The 1-cohomology classes of $M_6$ are labeled
by the same index $u= 1,\dots, 2g$ that labels the
non-trivial 1-cycles on the Riemann surface.
As in the previous section,
we can choose a canonical basis of 1-cycles on $\Sigma_g$,
and split the index $u$ into two sets of $g$ values.
Accordingly, we have a total of $2g$ 2-form gauge fields,
which we can arrange into two groups of $g$ each,
and similarly for their field strengths,
\beq \label{2forms}
B_2^u = (B_{2i} , \widetilde B_2^i) \ , \qquad
H_3^u = (H_{3i} , \widetilde H_3^i) \ ,  \qquad  i = 1,\dots g \ .
\eeq
By a similar token, 
we organize the $4g$ 0-form gauge fields
associated to 3-cohomology class of $M_6$ into four groups
of $g$ elements. Within each group, we label
0-form fields with the same index $i$ as in \eqref{2forms},
\begin{align}
a_0^{x}   = (a_{0i}^+ , \widetilde a_0^{i+} , a_{0i}^- , \widetilde a_0^{i-}) \ , \qquad
f_1^{x}   = (f_{1i}^+ , \widetilde f_1^{i+} ,  f_{1i}^- , \widetilde f_1^{i-}) \ ,
\qquad i =1, \dots, g \ .
\end{align}

Having introduced our choice of bases in cohomology
and our notation, we can present the expression for $I_6$.
It reads (suppressing wedge products)
\begin{align} \label{GMSW_top}
I_6   = \frac{1}{(2\pi)^2} \, \bigg[
&- (N \, F_2 + N_+ \, F_2^+ + N_- \, F_2^-) \, \gamma_4
- N \, \widetilde H_3^i \, H_{3i} \bigg] \nn \\
 + \frac{1}{(2\pi)^3} \, \bigg[ 
 &- \frac 1 6\, \chi \, (F_2^+)^3
- \frac 12 \, \chi \, F_2^+ \, (F_2^-)^2
+ F_2 \, F_2^+ \, F_2^- 
-   \gamma_4 \, \Big( f_{1i}^+ \, \widetilde f_1^{i-}
- \widetilde f_{1}^{i+} \, f_{1i}^-
 \Big)
 \nn \\
&  + F_2^+ \, \Big( 
 f_{1i}^+ \, \widetilde H_3^i
 - \widetilde f_1^{i+} \, H_{3i}
 \Big)
  + F_2^- \, \Big( 
 f_{1i}^- \, \widetilde H_3^i
 - \widetilde f_1^{i-} \, H_{3i}
 \Big)
 \bigg] \ .
\end{align}
The topological mass terms are collected in the first line.
We summarize the $p$-forms fields and their topological mass terms in
table \ref{GMSW_fields}. The bulk gauge groups
in the last column are explained in greater detail in the next section.

\begin{table}
\begingroup
\renewcommand{\arraystretch}{1.5}
\begin{center}
\begin{tabular}{c | c | c | c  }
multiplicity &   fields & top.~mass terms & 5d bulk gauge symm. \\ \hline \hline
\multirow{2}{*}{  $b^2(M_6)=3$  }  & 
\multirow{2}{*}{    $A_1^\alpha$   }  &  
\multirow{3}{*}{   $\frac{1}{2\pi} \, N_\alpha \, A_1^\alpha \wedge d\gamma_3$   } & 
$U(1)^{2}$ 0-form symm. 
\\
   & 
   & 
    & 
$\mathbb Z_{k}$ 0-form symm. 
\\ \cline{1-2} \cline{4-4}
 $1$  & 
 $\gamma_3$  & 
    & 
$\mathbb Z_{k}$ 2-form symm. \\ \hline
$b^1(M_6) =2g$ & $B_{2i}$, $\widetilde B_2^i$ & $\frac{1}{2\pi} \, N \,  
\widetilde  B_2^i \wedge dB_{2i}$ &
$(\mathbb Z_N \times \mathbb Z_N)^{g}$ 1-form symm.
\\ \hline
$b^3(M_6) = 4g$ & $a_{0i}^\pm$, $\widetilde a_0^{i \pm}$ & --- &
5d axions
\end{tabular}
\end{center}
\endgroup
\caption{ 
Summary of $p$-form
gauge fields in 5d supergravity obtained from expansion of $C_3$ onto cohomology classes
in $M_6$ for 
M5-branes at a $\mathbb Z_2$ singularity
wrapped on a genus-$g$ Riemann surface.
We have defined $k = {\rm gcd} (N_\alpha)$.
} 
\label{GMSW_fields}
\end{table}



\section{BF theory in the bulk and holographic interpretation} \label{sec_holography}

In this section we analyze the 5d dynamics
of the $p$-form gauge fields originating from
the expansion of the M-theory 3-form $C_3$.
We make contact with well-known aspects of BF theories \cite{Maldacena:2001ss, Banks:2010zn}
and   argue that the 5d   theory 
contains gauge fields with discrete   gauge groups.
When the 5d spacetime has a boundary,
the theory has to be supplemented by suitable boundary conditions
and boundary terms, which we partially review. 
Moreover, we describe the
 singleton modes that propagate on the boundary of spacetime.
Finally, we discuss the holographic correspondence between
discrete gauge fields in five dimensions and 
global discrete $p$-form symmetries in the 4d boundary theory,
as well as the holographic interpretation 
of the singleton modes of the 5d bulk theory.
Most of the ideas presented in this section
are modeled on results that have appeared in the literature. Our main goal here is to collect
useful observations to set the stage for the 't Hooft anomaly
discussion of section \ref{sec_anomalies}.

\subsection{Low-energy dynamics in five dimensions}
\label{sec_low_energy}

The relevant couplings in the 
effective action for the $p$-form gauge fields 
coming from the M-theory 3-form
 are
the kinetic terms 
and the topological terms \eqref{5dtopAction}.
At very low-energies, the dynamics is governed
by the topological terms that are quadratic in the $p$-form
gauge fields, as these terms in the 5d action contain only one derivative.
These topological terms are encoded in the
first line of the formal 6-form $I_6$ in \eqref{5dtop}.

We have already argued that the basis of 1-form
gauge fields in \eqref{the_new_basis}
is the best suited for discussing topological mass terms
involving $c_3$. 
We also notice that,
in all setups described in section \ref{sec_wrappedM5s},
the term $\frac 12 \, N_\alpha \, \cK^\alpha{}_{uv} \, H_3^u \wedge H_3^v$
takes the simple form $- N \, \widetilde H_3^i \wedge H_{3i}$,
see \eqref{BBBW_top} and \eqref{GMSW_top}.
For these reasons, 
for the remainder of this section
we consider the 5d topological theory 
defined by the action
\beq \label{simple_5dtopBIS}
S = \int_{\cM_5} \bigg[  - \frac{1}{2\pi} \, k \, c_3 \wedge d\cA_1 
- \frac{1}{2\pi} \,  N \, \widetilde B_2^i \wedge dB_{2i}\bigg] \ .
\eeq
In the previous expression $\cM_5$ denotes external spacetime.
In writing the action \eqref{simple_5dtopBIS} we have chosen
a specific antiderivative $I_5^{(0)}$ of the formal 6-form $I_6$.
If $\cM_5$ has no boundary,
this choice does not matter.
The case $\partial \cM_5 \neq \emptyset$ is discussed below.
Recall that  $i = 1,\dots, g$ and $k = {\rm gcd}(N_\alpha)$
(if $b^2(M_6) = 1$, we define $k = N_{\alpha = 1}$).
The action \eqref{simple_5dtopBIS}
describes a collection of decoupled standard BF theories.
We refer the reader to \emph{e.g.}~\cite{Maldacena:2001ss, Banks:2010zn} 
for   background material on BF theories
and their relation to the St\"uckelberg mechanism.

The fact that the 
1-form gauge fields  $ {\cA}_1^{\widehat \alpha}$
do not enter \eqref{simple_5dtopBIS} means that their dynamics
is governed 
by the kinetic terms  
and the cubic topological couplings in \eqref{5dtopAction}.
As a result, the 1-form gauge fields $ {\cA}_1^{\widehat \alpha}$
are standard $U(1)$ gauge fields.
In contrast, the dynamics of $\cA_1$, $c_3$, $B_{2i}$, and $\widetilde B_2^i$
is governed by
\eqref{simple_5dtopBIS}
and
therefore:
\begin{itemize}
\item $\cA_1$ describes a  1-form gauge field with gauge group $\mathbb Z_k$.
\item $c_3$ describes  a 3-form gauge field with gauge group $\mathbb Z_k$.
\item $B_{2i}$, $\widetilde B_2^i$ describe   2-form gauge fields with gauge group $\mathbb Z_N$.
\end{itemize}
The field $\cA_1$ is a continuum description
of a discrete gauge field because it is a flat connection
and its holonomies are restricted in $\mathbb Z_k \subset U(1)$.
Similar remarks apply to $c_3$, $B_{2i}$, $\widetilde B_2^i$:
for arbitrary cycles $\cC_1$, $\cC_3$, $\cC_2$ in 5d spacetime $\cM_5$,
\begin{align}
\exp \bigg( i \, \int_{\cC_1}  \cA_1 \bigg) & \in \mathbb Z_k \subset U(1) \ , &
\exp  \bigg( i \, \int_{\cC_3}  c_3  \bigg) & \in \mathbb Z_k \subset U(1) \ , \nn \\
\exp  \bigg( i \, \int_{\cC_2}  B_{2i} \bigg) & \in \mathbb Z_N \subset U(1) \ , &
\exp \bigg( i \, \int_{\cC_2}  \widetilde B_2^i  \bigg) & \in \mathbb Z_N \subset U(1) \ .
\end{align}
We stress that, since all these $p$-form gauge fields are flat on-shell,
the holonomies written above only depend on the homology classes
of $\cC_1$, $\cC_3$, $\cC_2$, and not on the specific representatives.
Let us also emphasize that
this description of discrete gauge fields in terms of
local $p$-forms and their holonomies
is convenient for our purposes, but in more general
situations an approach based on cocycles is preferred \cite{Dijkgraaf:1989pz,
 Kapustin:2014zva}.

\subsubsection{Boundary terms and boundary conditions} \label{sec_boundary_cond}

For applications to holography we have to consider 5d spacetimes with a
conformal boundary.
In this case, the bulk action \eqref{simple_5dtopBIS}
has to be supplemented with suitable boundary conditions
and possibly additional boundary terms, in order to ensure a well-defined
variational problem. 
In this section
we describe some sets of boundary conditions that will be relevant
below in the holographic discussion.

\paragraph{Topological boundary conditions.}
Let us first discuss boundary conditions for the $c_3$, $\cA_1$ BF theory.
A simple choice is to assign Dirichlet boundary conditions on 
$\cA_1$, with free boundary conditions for $c_3$. 
The variational problem is well-posed because the
relevant terms in the on-shell variation of the
action   \eqref{simple_5dtopBIS} are
\beq \label{simple_Dirichlet}
\delta S = \int_{\partial \cM_5} 
\frac{1}{2\pi} \, k \, c_3 \wedge \delta \cA_1
+\dots  \ .
\eeq
Let us stress that imposing Dirichlet boundary conditions
for both $\cA_1$ and $c_3$   would   be inconsistent,
since the variational problem defined by the bulk BF action
 \eqref{simple_5dtopBIS} is first-order.
If desired, the roles of $\cA_1$ and $c_3$ can be exchanged.
By adding the boundary term $- \frac{k}{2\pi}\int_{\partial \cM_5} c_3 \wedge \cA_1$
to  \eqref{simple_5dtopBIS}, we can rewrite the relevant terms in the total
action as
\beq
S' = \int_{\cM_5} \bigg[  - \frac{1}{2\pi} \, k \, \cA_1 \wedge dc_3
\bigg]  + \dots \ .
\eeq
In this case we impose Dirichlet boundary conditions on $c_3$,
with free boundary conditions for~$\cA_1$.
The boundary conditions described so far as topological,
since they are invariant under orientation-preserving diffeomorphisms
of $\partial \cM_5$ and   do not require
the choice of a boundary metric.
(See \cite{Kapustin:2010hk} for a classification of topological boundary conditions
in Abelian 3d Chern-Simons theory.)

If the integer $k$ can be factorized as $k = m m'$,
we can also consider a generalization of the above
topological boundary conditions, along the lines of \cite{Gaiotto:2014kfa,Bergman:2020ifi}.
Let us stress that, since $\cA_1$ is a discrete 1-form gauge field,
assigning Dirichlet boundary conditions for $\cA_1$
means specifying its $\mathbb Z_k$ holonomies
around 1-cycles in the boundary $\partial M_5$.
We can partially relax the boundary conditions on $\cA_1$
as follows.
To a given 1-cycle $\cC_1$  in $\partial M_5$
we no longer associate
an element $x\in \mathbb Z_k$,
but rather a coset $[x]_{\mathbb Z_{m}}\in \mathbb Z_k/\mathbb Z_{m}$.
The holonomy $\exp ( i \int_{\cC_1} \cA_1)$ 
is free to take any value $y\in [x]_{\mathbb Z_m}$,
which is the same as $y = x$ mod $m'$.\footnote{For
example,
if $m = 3$, $m' = 4$, $k = 12$, 
the subgroup $\mathbb Z_3 \subset \mathbb Z_{12}$
consists of $\{ 0, 4, 8\}$.
The elements of the quotient $\mathbb Z_{12}/\mathbb Z_3$
are the cosets $[0]_{\mathbb Z_3} = \{0,4,8\} \subset \mathbb Z_{12}$,
$[1]_{\mathbb Z_3} = \{1,5,9\} \subset \mathbb Z_{12}$,
$[2]_{\mathbb Z_3} = \{2,6,10\} \subset \mathbb Z_{12}$,
and $[3]_{\mathbb Z_3} = \{3,7,11\} \subset \mathbb Z_{12}$.
The boundary conditions for $\cA_1$ 
select one coset, for example $[1]_{\mathbb Z_3}$,
leaving the holonomy of $\cA_1$ free to take
any value $y = 1$ mod 4, namely 
$y=1$, $5$, or $9$.
}
Following the terminology of \cite{Bergman:2020ifi},
we say that $\cA_1$ is free in $\mathbb Z_k$
modulo $\mathbb Z_{m'}$.
It is interesting to notice that,
since $\mathbb Z_{k}/\mathbb Z_m \cong
\mathbb Z_{m'}$, the data encoded in the boundary conditions for $\cA_1$
 is the same data that define  a background $\mathbb Z_{m'}$ 1-form gauge
field on the boundary.

In order to have a well-defined variational problem,
we must   partially restrict the field $c_3$.
Its boundary conditions are no longer free.
To a 3-cycle $\cC_3$, 
we assign a coset $[x]_{\mathbb Z_{m'}}$ and
the holonomy $\exp ( i \, \int_{\cC_3} c_3)$
can take any value $y \in  [x]_{\mathbb Z_{m'}}$,
which is the same as   $y = x$ mod $m$.
In short, we say that 
$c_3$ is free
in $\mathbb Z_k$ modulo $\mathbb Z_m$.
Specifying the boundary condition for $c_3$
is the same as choosing a 
background $\mathbb Z_{m}$ 3-form gauge
field on the boundary.

Let us now comment on topological boundary
conditions for the $\widetilde B_2^i$, $B_{2i}$ BF theory.
For a given label $i = 1,\dots, g$,
we may assign Dirichlet boundary conditions for $\widetilde B_2^i$
and free boundary conditions for $B_{2i}$, or \emph{vice versa}.
If $N$ can be factorized
as $N = n n'$, we can also consider boundary conditions in which
$B_{2i}$ is free in $\mathbb Z_N$ modulo
$\mathbb Z_{n'}$,
while $\widetilde B_2^i$ is free in $\mathbb Z_N$ modulo $\mathbb Z_{n}$
(in the same terminology explained above.)

The full set of boundary conditions for the 
$\widetilde B_2^i$, $B_{2i}$ BF theory, however, is richer.
Indeed, we can select suitable linear combinations
$(B_{2i}', \widetilde B_2'^i)$ of the original 2-forms
$(B_{2i}, \widetilde B_2^i)$, and impose 
that $B'_{2i}$ be free in $\mathbb Z_N$ modulo
$\mathbb Z_{n'}$,
and $\widetilde B_2'^i$ be free in $\mathbb Z_N$ modulo $\mathbb Z_{n}$.
Moreover, the duality group $Sp(2g,\mathbb Z)$ acts on the set
of topological boundary conditions.
We leave the problem of  
classifying topological boundary conditions for the 
$\widetilde B_2^i$, $B_{2i}$ BF theory to future work.

\paragraph{The role of kinetic terms.}
Let us close this section by emphasizing that
the discussion of boundary conditions is qualitatively
different if the kinetic terms are included in the analysis.
This is because, if the kinetic terms are retained, the variational
problem is a second-order problem. 
It is therefore possible, for instance, to impose
Dirichlet boundary conditions on all fields.
This point is discussed in 
\cite{Gukov:2004id,Belov:2004ht}   
in the context of 3d and 5d topological theories.
We expect similar features in our 5d BF system.

\subsubsection{Singleton modes propagating on the boundary}
\label{sec_singleton_BF}

When a topological 5d BF theory  with a coupling between a $p$-form
gauge field and a $(4-p)$-form gauge field  is considered in a spacetime with
a boundary, there is a massless $(p-1)$-form gauge field propagating along
the boundary. (Equivalently, the massless mode on the boundary can be thought of
as a $(3-p)$-form gauge field.)
These massless boundary modes are usually referred to as
singletons. 
 For a justification of the previous claims and of the following statements,
see \emph{e.g.}~\cite{Maldacena:2001ss, Banks:2010zn}.
For the case at hand, the singleton modes are:
\begin{itemize}
\item One 0-form gauge field in $\partial \cM_5 $.
\item $2g$ 1-form gauge fields in $\partial \cM_5 $.
\end{itemize}
All these gauge fields are standard $U(1)$ gauge fields,
as opposed to discrete gauge fields.

While the Hilbert space of the singleton fields
is insensitive to the choice of boundary terms and boundary conditions,
its dynamics (\emph{i.e.}~the Hamiltonian on the Hilbert space)
is different for different boundary terms and boundary conditions.
In this work, we refrain from a detailed analysis of the singleton
dynamics. We will be mainly interested in counting 
singletons and discussing their holographic duals.
A thorough analysis of the singleton sector
would require to take into account the kinetic terms,
as in \cite{Moore:2004jv, Belov:2004ht}. We leave such investigation for the future.

\subsection{Holographic interpretation}

Let us now turn to a discussion of the holographic interpretation
of the features of the bulk BF theory listed in the previous section.

\subsubsection{Global discrete symmetries in four dimensions}
\label{sec_global_symm}
We have argued above that the 1-form gauge fields
$  {\cA}_1^{\widehat \alpha}$ are standard $U(1)$
gauge fields in five dimensions. As a result, they are dual to
global $U(1)$ 0-form symmetries in the interacting CFT
living on the boundary.

In contrast, the holographic interpretation of    the discrete gauge fields $\cA_1$, $c_3$,
$B_{2i}$, and $\widetilde B_2^i$ 
is more subtle. We describe it in the purely 
topological BF theory, neglecting kinetic terms.

\paragraph{Holography of the topological BF theory.}
The holographic interpretation of the 5d bulk BF theory
\eqref{simple_5dtopBIS} depends on the choice of boundary conditions.
In other words, different boundary conditions correspond to different
dual CFTs, which may have different global symmetries. This is a standard phenomenon
in the paradigmatic example of $AdS_5 \times S^5$ in type IIB \cite{Witten:1998wy}
and has recently been studied in the context of ABJM theories \cite{Bergman:2020ifi}.

Firstly, let us focus on the $\cA_1$, $c_3$ system.
The holographic interpretation of the   topological
boundary conditions discussed above is as follows.
\begin{enumerate}[(a)]
\item Dirichlet boundary conditions for $\cA_1$ and free boundary
conditions for $c_3$:\\
The dual interacting CFT admits a global
$\mathbb Z_k$ 0-form symmetry.
Specifying the boundary condition for $\cA_1$ is
the same as fixing a configuration for the
4d background 1-form gauge field that couples to this global symmetry.

\item Dirichlet boundary conditions for $c_3$ and free boundary
conditions for $\cA_1$:\\
The dual interacting CFT admits a global
$\mathbb Z_k$ 2-form symmetry.
Specifying the boundary condition for $c_3$ is
the same as fixing a configuration for the
4d background 3-form gauge field that couples to this global symmetry.

\item The case $k = mm'$ with $\cA_1$ free in $\mathbb Z_k$
modulo $\mathbb Z_{m'}$ and $c_3$ free in $\mathbb Z_k$
modulo $\mathbb Z_m$:\\
The dual interacting CFT admits both a global
$\mathbb Z_{m'}$ 0-form symmetry
and a global $\mathbb Z_m$ 2-form symmetry.
Specifying the boundary conditions  for $\cA_1$ and $c_3$ is
the same as fixing a configuration for the
4d background 1-form and 3-form gauge fields
 that couple to these global symmetry.

\end{enumerate}

Case (c) is intermediate between cases (a) and (b).
In case (c), 
there is a mixed 't Hooft anomaly
between the $\mathbb Z_{m'}$ 0-form symmetry and the 
$\mathbb Z_m$ 2-form symmetry.
This 't Hooft anomaly is encoded in 
the 6-form $I_6 = -k\, \frac{dc_3}{2\pi} \wedge \frac{d\cA_1}{2\pi}$,
which is related by descent to the   BF coupling in the 5d bulk action.\footnote{Roughly
speaking, terms in $I_6$ involving two or more Dirichlet fields
are interpreted as   't Hooft anomalies.
If we choose Dirichlet boundary conditions for $\cA_1$,
the field $c_3$ has free boundary conditions.
The 6-form $I_6 = -k\, \frac{dc_3}{2\pi} \wedge \frac{d\cA_1}{2\pi}$ does not encode
a 't Hooft anomaly, and indeed we only have a global
$\mathbb Z_k$ 0-form symmetry.
When we let $\cA_1$ free in $\mathbb Z_k$ modulo
$\mathbb Z_{m'}$
and $c_3$ free in $\mathbb Z_k$ modulo
$\mathbb Z_{m}$, both fields
$\cA_1$ and $c_3$ are ``partially Dirichlet''.
As a result, $I_6 = -k\, \frac{dc_3}{2\pi} \wedge \frac{d\cA_1}{2\pi}$ now encodes
the mixed 't Hooft anomaly between
the 0-form and 2-form global symmetries.
}

Let us also observe that cases (b) and (c) can be obtained
from case (a) via gauging.
More precisely, suppose $k = mm'$. The CFT of case (a)
has a global $\mathbb Z_k$ 0-form symmetry.
We may gauge a subgroup $\mathbb Z_m \subset \mathbb Z_k$
of this global symmetry. 
The gauging is performed by path-integrating over 
the background 1-form gauge field that couples
to the $\mathbb Z_m$ subgroup. This is the same
as modifying the boundary conditions for $\cA_1$:
we go from Dirichlet boundary conditions,
to having $\cA_1$ free in $\mathbb Z_k$ modulo
$\mathbb Z_{m'}$. After gauging, 
the residual global 0-form symmetry is $\mathbb Z_{m'}$.
There is also an emergent global $\mathbb Z_m$ 2-form
symmetry. We recognize the features of the CFT of case (c).
Selecting $m=k$, $m' = 1$ we recover case (b).

The $B_{2i}$, $\widetilde B_2^i$ system can be analyzed in a similar way.
As recalled in section \ref{sec_boundary_cond}, the full set of allowed topological
boundary conditions is rich, 
and their classification is left for future work.
To illustrate the relation between boundary conditions and global
discrete symmetries,
we consider a simple class of boundary conditions, in which we can treat
each label $i = 1,\dots, g$ independently.
One may thus consider the following three scenarios.
\begin{enumerate}[(a${}^\prime$)]
\item Dirichlet boundary conditions for $B_{2i}$ and free boundary
conditions for $\widetilde B_2^i$:\\
The dual interacting CFT admits a global
$(\mathbb Z_N)^g$ 1-form symmetry of ``electric type''.
Specifying the boundary condition for $B_{2i}$ is
the same as fixing a configuration for the
4d background 2-form gauge fields that couple to this global symmetry.

\item  Dirichlet boundary conditions for $\widetilde B_2^i$ and free boundary
conditions for $B_{2i}$:\\
The dual interacting CFT admits a global
$(\mathbb Z_N)^g$ 1-form symmetry of ``magnetic type''.
Specifying the boundary condition for $\widetilde B_2^i$ is
the same as fixing a configuration for the
4d background 2-form gauge fields that couple to this global symmetry.

\item  The case $N = nn'$ with $B_{2i}$ free in $\mathbb Z_N$
modulo $\mathbb Z_{n'}$ and $\widetilde B_2^i$ free in $\mathbb Z_N$
modulo $\mathbb Z_n$:\\
The dual interacting CFT admits both a global
$(\mathbb Z_{n'})^g$ 1-form symmetry of ``electric type''
and a global $(\mathbb Z_n)^g$ 1-form symmetry of ``magnetic type''.
Specifying the boundary conditions  for $B_{2i}$ and $\widetilde B_2^i$ is
the same as fixing  configurations for the
4d background 2-forms 
 that couple to these global symmetries.
\end{enumerate}
As before, the case (c${}^\prime$) is intermediate between (a${}^\prime$) and
 (b$^\prime$), and 
in case (c${}^\prime$)   there is a mixed 't Hooft anomaly
between the $(\mathbb Z_{n'})^g$ and $(\mathbb Z_n)^g$
1-form symmetries. This 't Hooft anomaly
is encoded in the 6-form
$I_6 = N \, \frac{d\widetilde B_2^i}{2\pi  } \wedge \frac{dB_{2i}}{2\pi}$.

\subsubsection{Singleton modes as Goldstone modes}
According to the usual holographic dictionary,
the supergravity theory in the bulk of 5d spacetime $M_5$
is dual to an interacting CFT  living on $\partial M_5$.
The gravity theory in five dimensions
has additional singleton modes, that only propagate on the
conformal boundary of 5d spacetime.
These modes to not gravitate. They are holographically dual
to additional, decoupled free fields in four dimensions.

\paragraph{BF singletons as Goldstone modes.}
In section \ref{sec_singleton_BF} we have identified
a subset of the singleton modes 
for the 5d gravitational theories 
of interest in this work.
More precisely, we have identified the singleton
modes associated to the 5d
 BF theory \eqref{simple_5dtopBIS}. 
For these singleton modes we can 
offer an interpretation in terms of Goldstone's theorem,
as follows.

The singleton mode associated to the 
 BF coupling $ k \, c_3 \wedge d\cA_1$ is a 4d 0-form gauge field,
\emph{i.e.}~an axion.
The BF coupling $ k \, c_3 \wedge d\cA_1$
can be related by dualization to a St\"uckelberg coupling
between the vector $\cA_1$ and the axion dual to $c_3$.
(Useful background material can be found in    \cite{Maldacena:2001ss, Banks:2010zn}.)
The 5d theory describes a $U(1)$ 0-form gauge symmetry
Higgsed down to $\mathbb Z_k$. 
According to the usual holographic dictionary,
the $U(1)$ 0-form gauge symmetry in five dimensions
is dual to a $U(1)$ 0-form global symmetry in four dimensions.
The boundary value of the 5d gauge field $\cA_1$
is identified with the Noether 1-form current $J_1$
for the global $U(1)$ 0-form symmetry in four dimensions.
Since the $U(1)$ 0-form gauge symmetry is
spontaneously broken in five dimensions,
the dual $U(1)$ 0-form global symmetry in four dimensions
is also spontaneouly broken.
As a result, we have a 4d massless Goldstone scalar $\Phi$,
related to the current $J_1$ by the schematic relation
$J_1 \sim d\Phi$.
The Goldstone mode $\Phi$ enjoys a global shift symmetry
$\Phi \rightarrow \Phi + \text{const}$. It is an axion and its interactions
are   derivative interactions.
In the deep IR, $\Phi$ decouples from the rest of the 4d theory.
We identify it with the holographic dual
of the singleton mode from the BF coupling $k \, c_3 \wedge d\cA_1$.

Similar remarks apply to the singleton modes
associated to the BF coupling $N\, \widetilde B_2^i \wedge B_{2i}$.
A general statement is as follows:
\begin{displayquote}
A $D$-dimensional BF coupling 
$B_{D-1-p} \wedge dA_{p}$
between a $p$-form gauge field $A_p$
and a $(D-1-p)$-form gauge field $B_{D-1-p}$
yields a singleton mode which is a massless
$U(1)$ $(p-1)$-form gauge field
in $(D-1)$ dimensions.
It is identified with the Goldstone mode
originating from spontaneous breaking
of a global $(p-1)$-form symmetry in the $(D-1)$-dimensional
dual field theory.
\end{displayquote}
In the case $D = 2p+1$ we can formulate  a similar statement 
regarding Chern-Simons couplings.
\begin{displayquote}
A $(2p+1)$-dimensional Chern-Simons coupling 
$A_p \wedge dA_p$
for a  $p$-form gauge field $A_p$
yields a singleton mode which is a massless \emph{chiral}
$U(1)$ $(p-1)$-form gauge field
in $2p$ dimensions.
It is identified with the Goldstone mode
originating from spontaneous breaking
of a global $(p-1)$-form symmetry in the $2p$-dimensional
dual field theory.
\end{displayquote}
Here a chiral $(p-1)$-form gauge field is
by definition a gauge field whose $p$-form field
strength obeys a self-duality constraint
of the form $*_{2p} F_p = \pm F_p$
or $*_{2p} F_p = \pm i \, F_p$,
depending on the dimension and signature of spacetime.



\section{Extended operators and discrete symmetries}

In this section we review the extended operators 
of the 5d topological BF theory with action
\eqref{simple_5dtopBIS}. We identify the 11d origin of these operators in terms of
wrapped M2-branes. We also consider the interplay between 
these operators and the topological boundary
conditions for the BF system \eqref{simple_5dtopBIS}
and  infer what extended operators are expected in the dual 4d field theories.

\subsection{Extended operators in the BF bulk theory}

A natural set of gauge-invariant
observables in the 5d theory \eqref{simple_5dtopBIS} is given
by the holonomies
of the gauge fields $c_3$, $\cA_1$, $\widetilde B_2^i$,
$B_{2i}$ on cycles in 5d spacetime,
\begin{align} \label{electric_ops}
W_{c}(\cC_3^\mathrm{ext}, n) & = \exp \bigg[ i \,n \, \int_{\cC_3^\mathrm{ext}} c_3\bigg] \ ,  & 
W_{\cA}(\cC_1^\mathrm{ ext}, n)   = \exp \bigg[ i \,n \, \int_{\cC_1^\mathrm{ext} } \cA_1 \bigg] \ , \nn \\
W_{B,\widetilde B}(\cC_2^\mathrm {ext} , n, \widetilde n) & = \exp \bigg[
i  \, \int_{\cC_2^\mathrm {ext} } \Big(  
\widetilde n^i \, B_{2i} - n_i \, \widetilde B_2^i
\Big) \bigg]\ ,
\end{align}
where $n$, $n_i$,  and $\widetilde n^i$ are integers
and $\cC_1^\mathrm {ext} $, $\cC_2^\mathrm {ext} $, $\cC_3^\mathrm {ext} $ and 1-, 2-, 3-cycles in 5d spacetime.
The superscript `ext' stands for external and is inserted to avoid
possible confusions with cycles in the internal geometry $M_6$.
We remind the reader that $p$-form gauge fields
are normalized to have periods quantized in units of $2\pi$.
In the topological BF theory, the operator $W_{c}(\cC_3^\mathrm{ext}, n)$ describes  a 3d defect in 5d spacetime extended along 
 $\cC_3^\mathrm {ext}$ with  electric charge $n$ under $c_3$.
In a similar way, $W_{A}(\cC_1^\mathrm{ ext}, n)$ represents 
a 1d defect extended along $\cC_1^\mathrm{ ext}$ with charge $n$ under $\cA_1$,
while $W_{B,\widetilde B}(\cC_2^\mathrm {ext}, n, \widetilde n)$
describes a 2d defect along $\cC_2^\mathrm {ext}$ with charges $n_i$,
$ \widetilde n^i $ under $B_{2i}$, $\widetilde B_2^i$.
The operators \eqref{electric_ops} will be
referred to as electric operators.

If a defect
charged under $\cA_1$ is transported
around a defect charged under $c_3$,
it acquires a non-trivial $\mathbb Z_k$ Aharonov-Bohm phase.
The latter is encoded in the 
 correlator
\begin{align} \label{correlator_no1}
\langle W_{c}(\cC_3^\mathrm {ext}, n) \, W_{\cA}(\cC_1^\mathrm {ext}, n') \rangle
& \sim \exp \bigg[  i \, \frac{n\, n'}{k} \, L(\cC_3^\mathrm {ext}, \cC_1^\mathrm {ext}) \bigg] \ , 
\end{align}
where $L(\cC_3^\mathrm {ext}, \cC_1^\mathrm {ext})$ is the integer linking number of
$\cC_1^\mathrm {ext}$ and $\cC_3^\mathrm {ext}$ in the ambient 5d spacetime.
By a similar token, the 
Aharonov-Bohm phases of
defects charged under $B_{2i}$, $\widetilde B_{2i}$
are captured by the correlators
\begin{align}
\langle  \label{correlator_no2}
W_{B,\widetilde B}(\cC_2^\mathrm {ext}, n , \widetilde n) \,
W_{B,\widetilde B}(\cC_2^\mathrm {ext}{}', n' , \widetilde n')
\rangle
& \sim \exp \bigg[ i \, \frac{  n_i \, \widetilde n'^i 
- \widetilde n^i \, n'_i }{N} \, L(\cC_2^\mathrm {ext}, \cC_2 ^\mathrm {ext}{}' )  \bigg] \ .
\end{align}
The derivation of \eqref{correlator_no1} and \eqref{correlator_no2}
can be found 
\emph{e.g.} in~\cite{Maldacena:2001ss, Banks:2010zn}.

In addition to the electric
operators in \eqref{electric_ops},
the 5d topological theory also admits ``mixed''
electric-magnetic operators.
If we consider a 2-cycle $\cB_2^\mathrm {ext}$ in external spacetime,
we can define a 't Hooft operator for $\cA_1$ supported on $\cB_2^\mathrm {ext}$.
This is done in the usual way.
We remove a small tubular neighborhood
of $ \cB_2^\mathrm {ext}$ from   5d spacetime. The boundary of the tubular neighborhood
is an $S^2$ bundle over $ \cB_2^\mathrm {ext}$. The 't Hooft operator on 
$  \cB_2^\mathrm {ext}$ is defined by 
performing the path integral over $\cA_1$ with the boundary condition
$\frac{1}{2\pi} \int_{S^2} \cF_2 = 1$.
Because of the $k\, c_3 \wedge d\cA_1$ coupling in the action,
the 't Hooft operator is not gauge invariant.
It must be supplemented
with a charge-$k$ Wilson operator for $c_3$ on a 3-chain $\cC_3^\mathrm {ext}$
such that $\partial \cC_3^\mathrm {ext} =  \cB_2^\mathrm {ext}$ \cite{Gross:1998gk,Witten:1998wy,Hofman:2017vwr}.\footnote{After removing
a small tubular neighborhood $U$ of $\cB_2^\mathrm {ext}$,
the gauge variation of the 5d action 
reads
\beq
\delta S = - \frac{1}{2\pi} \, k \, \int_{\cM_5 \setminus U} \delta c_3 \wedge \cF_2 = 
- \frac{1}{2\pi} \, k \, \int_{\cM_5 \setminus U} d\Lambda_2 \wedge \cF_2
= -k \, \int_{\cB_2^\mathrm {ext}} \Lambda_2 \ , \nn
\eeq
where we have used $\frac{1}{2\pi} \int_{S^2} \cF_2= 1$. The above expression shows that
the gauge variation of the 't Hooft operator for $\cA_1$
can be cancelled by $k$ Wilson operators for $c_3$.}
In a completely analogous fashion,
one can consider a 't Hooft operator for $c_3$
supported on a 0-cycle $\cB_0^\mathrm {ext}$ (a collection of points taken with signs).
To preserve gauge invariance, this must be supplemented with a charge-$k$
Wilson operator for $\cA_1$ supported on a 1-chain $\cC_1^\mathrm {ext}$
with $\partial \cC_1^\mathrm {ext} = \cB_0^\mathrm {ext}$.
Finally, analogous mixed electric-magnetic operators
exist for the $(\widetilde B_2^i, B_{2i})$ system.

\paragraph{M-theory origin of extended operators.}
The purely electric operators $W_c(\cC_3^\mathrm e,n)$ are realized by a stack of $n$
M2-brane probes
sitting at a point in the internal space $M_6$, and extending along
$\cC_3^\mathrm e$ in the external spacetime directions. By a similar token,
the operators $W_\cA(\cC_1^\mathrm e,n)$ are a stack of $n$ probe M2-branes wrapping
a 2-cycle in $M_6$. More precisely, the 2-cycle is $m_\alpha \, \cC_2^\alpha
= \cC_2'{}^{\alpha  = 1}
$, 
where the integers $m_\alpha$ are defined in \eqref{k_and_m_def},
and in the second step we refer to the primed basis of 2-cycles
defined by \eqref{other_rotations}.
Finally, the operators $W_{B,\widetilde B} (\cC_2^\mathrm e, n, \widetilde n)$ originate from
probe M2-branes wrapping a 1-cycle in $M_6$. The charges $\widetilde n^i$,
$n_i$ are identified with the integers that define this   1-cycle,
with respect to a fixed   basis of 1-cycles in $M_6$.
It can also be verified that the 5d Aharonov-Bohm phases
encoded in the correlators \eqref{correlator_no1},
\eqref{correlator_no2} can be reproduced from an 11d perspective,
using the $C_3 G_4 G_4$ coupling in the M-theory low-energy  effective action.

The mixed electric-magnetic operators
are realized using probe configurations with M2-branes ending on M5-branes.
As an example, let us consider
a charge-1 't Hooft operator for $\cA_1$ on $\cB_2^\mathrm {ext}$
together with a charge-$k$ Wilson operator for $c_3$ on $\cC_3^\mathrm{ext}$,
with $\partial \cC_3^\mathrm{ext} = \cB_2^\mathrm {ext}$. This 5d operator 
is realized by one M5-brane wrapping a 4-cycle on $M_6$
and extending along $\cB_2^\mathrm{ext}$.
With reference to the change of basis
discussed in appendix \ref{sec_goodbasis}, we can characterize this 4-cycle as
$\cC_4'^{\alpha = 1}$. Since there are $k$ units of $G_4$-flux
threading the 4-cycle $\cC_4'^{\alpha = 1}$, there is a tadpole in
 the worldvolume theory of the probe M5-brane. This is canceled
by adding $k$ M2-branes ending on the M5-brane. The M2-branes
sit at a point on $\cC_4'^{\alpha = 1} \subset M_6$
and are extended along $\cC_3^\mathrm{ext}$ in the external directions.
In a similar way, one can describe the 11d origin of all
other mixed electric-magnetic operators.

In our discussion so far we have not taken supersymmetry into
consideration. One could determine the BPS conditions for probe
M2-branes and M5-branes by analyzing $\kappa$-symmetry 
on their worldvolumes.

\subsection{Extended operators in the dual field theory}

The   operators of the 5d topological theory
discussed in the previous section can yield   operators
in the dual boundary field theory. 
The choice of boundary
conditions for $\cA_1$, $c_3$, $B_{2i}$, $\widetilde B_2^i$
determines 
whether a 5d operator is allowed to end
on the boundary $\partial \cM_5$  or not.
A similar discussion has recently appeared in 
 \cite{Bergman:2020ifi}  in the context of $AdS_4/{\rm CFT}_3$.

Let us discuss operators constructed with $\cA_1$, $c_3$.
We make contact to the cases (a), (b), (c) discussed in section \ref{sec_global_symm}.
In case (a)  the   operators 
$\cW_\cA(\cC_1^\mathrm e, n)$ 
defined in \eqref{electric_ops}
are allowed to end on $\partial \cM_5$,
while the   operators 
$\cW_c(\cC_3^\mathrm e, n)$
are forbidden from ending on $\partial \cM_5$.\footnote{More precisely,
$\cW_c(\cC_3^\mathrm e, n)$ can end on $\partial \cM_5$
only if $n$ is a multiple of $k$, in which case  the operator $\cW_c(\cC_3^\mathrm e, n)$
is trivial.}
The 5d operator $\cW_\cA(\cC_1^\mathrm e, 1)$ yields a
local operator $\cO$ on $\partial \cM_5$ that has unit charge  under the global
$\mathbb Z_k$ 0-form symmetry of the field theory.
In a similar way, in case (b) it is the operators 
$\cW_c(\cC_3^\mathrm e, n)$ that can end on the boundary,
and $\cW_c(\cC_3^\mathrm e, 1)$ yields  a surface operator on $\partial \cM_5$ 
with unit
  charge under the global
$\mathbb Z_k$ 2-form symmetry of the field theory.
In case (c), with factorization $k = m\, m'$, the operator 
$\cW_\cA(\cC_1^\mathrm e, n)$ is allowed to end on $\partial \cM_5$
if the charge $n$ is a multiple of $m$,
while $\cW_c(\cC_3^\mathrm e, n)$ can end on the boundary
if $n$ is a multiple of $m'$. 
In the boundary field theory we obtain both local operators and
surface operators, compatibly with the global $\mathbb Z_{m'}$ 0-form
 and $\mathbb Z_m$ 2-form   symmetry.

The mixed electric-magnetic operators of the 5d topological bulk theory
act as baryon vertices \cite{Witten:1998xy} from the point of view of the dual field theory. 
For example, in case (a) we can 
consider the operator $\cO^k$ on the boundary and connect it to a point
in the bulk, where a charge-1 't Hooft operator for $c_3$ is supported
(a ``monopole event'').
 The arguments of \cite{Hofman:2017vwr} show that in this case
$\cO^k$ 
acquires a VEV. Notice that this phenomenon does not break the
 global
 $\mathbb Z_k$ 0-form symmetry.
Analogous remarks apply to cases (b) and (c).

Let us now turn to the operators 
$W_{B,\widetilde B}(\cC_2^\mathrm e , n, \widetilde n)$
in \eqref{electric_ops}. 
In section \ref{sec_global_symm} we have defined the 
cases   (a${}^\prime$), (b${}^\prime$), (c${}^\prime$). Notice that we can choose
any of these three options independently for each label $i = 1,\dots,g$.
For the sake of simplicity, let us discuss the situation
in which we choose case (a${}^\prime$) for all $i = 1,\dots,g$.
Other choices of topological boundary conditions for the
$(\widetilde B_2^i, B_{2i})$ system can be discussed in a similar way.

If we select case (a${}^\prime$) for all $i=1,\dots,g$,  
the operator $W_{B,\widetilde B}(\cC_2^\mathrm e , n, \widetilde n)$
is allowed to end on $\partial \cM_5$ if its $\widetilde B_2^i$ charges
$n_i$ are all equal to zero. If this condition is met, we get a line operator 
in the dual field theory. It has   charges $\widetilde n^i$
under the global ``electric'' $(\mathbb Z_N)^g$ 1-form symmetry of the QFT.
If $\widetilde n^i = N \, \widetilde s^i$ for some integers $\widetilde s^i$,
we can connect the line operator on the boundary
to a mixed electric-magnetic operator (baryon vertex) in the bulk.
More precisely, the line operator on $\cM_5$ is connected by a 2d worksheet
to a line $\cB_1^\mathrm {ext}$ in the bulk,
which supports a 't Hooft operator for $\widetilde H_3^i $ with charges $\widetilde s^i$.\footnote{More explicitly, we consider a small tubular neighborhood of 
$\cB_1^\mathrm{ext}$ in $\cM_5$. Its boundary is an $S^3$ bundle over $\cB_1^\mathrm{ext}$. We impose the boundary condition 
$\frac{1}{2\pi}\int_{S^3} \widetilde H_3^i =  \widetilde s^i$.
}
According to the analysis of \cite{Hofman:2017vwr},
the presence of the baryon vertex in the 5d bulk
implies condensation of the line operator with charges 
$\widetilde n^i = N \, \widetilde s^i$.\footnote{A line operator is said to be condensed if  it
obeys a perimeter law \cite{Gaiotto:2014kfa}.}
This condensation does not trigger a spontaneous breaking of
the $(\mathbb Z_N)^g$ 1-form
symmetry.

The analysis of extended operators and boundary conditions
in the $B_i$, $\widetilde B^i$ BF theory
can be used as a tool to access allowed line operators in 4d SCFTs 
from wrapped M5-branes. The goal is a classification 
  that generalizes the results  of
\cite{Aharony:2013hda} beyond Lagrangian gauge theories.
We expect a rich variety of line operators and a non-trivial action 
of the duality group $Sp(2g,\mathbb Z)$ on them.
Notice that this strategy can be applied to both  $\cN = 2$ and
$\cN = 1$ theories. 
We plan to study this problem in greater detail in future work.



\section{'t Hooft anomalies from inflow} 
\label{sec_anomalies}

In this section we compute the inflow anomaly polynomial
for the M-theory setups of interest in this work.
We include all $p$-form gauge fields originating from
expansion of $C_3$ onto cohomology classes of $M_6$,
as well as background fields for isometries of $M_6$,
and an arbitrary background metric.
A systematic method for performing this computation
was developed in \cite{Bah:2019rgq}.
One main novelty here is the interpretation of the terms
involving $\cA_1$, $c_3$, $B_{2i}$, $\widetilde B_2^i$,
which encode 't Hooft anomalies for 
discrete global symmetries. Moreover,
in the case of M5-branes probing a $\mathbb Z_2$ singularity,
we include background 0-form gauge fields and we discuss
their interpretation in terms of anomalies in the space of coupling constants.

\subsection{Inflow anomaly polynomial}

Let us outline the recipe for the computation of the inflow anomaly polynomial.
We refer   to \cite{Bah:2019rgq} for further explanations. 
The input data 
is the internal geometry $M_6$ and the background $G_4$-flux configuration
$\overline G_4$. In the notation of \eqref{G4_ansatz}, the latter is
$\overline G_4/(2\pi) = N_\alpha \, \Omega_4^\alpha$.
Our goal is to compute the 6-form inflow anomaly polynomial
$I_6^{\rm inflow}$. As per usual  descent formalism,
$I_6^{\rm inflow}$ is a closed, gauge-invariant 6-form
that is defined in a fiducial spacetime $\cM_6$,
which is taken to be Euclidean and six-dimensional.

In order to compute 't Hooft anomalies for symmetries associated
to isometries of $M_6$, we have to consider a fibration of $M_6$
over the fiducial spacetime $\cM_6$. The relevant space is therefore
an auxiliary 12-manifold $M_{12}$,
\beq \label{M12aux}
M_6 \hookrightarrow M_{12} \rightarrow \cM_6 \ .
\eeq
The desired 6-form $I_6^{\rm inflow}$ is computed
by fiber integration along $M_6$ of a globally defined 12-form $\cI_{12}$ on $M_{12}$,
\beq \label{I6inflow}
I_6^{\rm inflow}  = \int_{M_6} \cI_{12}  \ .
\eeq
The 12-form $\cI_{12}$ is constructed 
from the class $X_8(TM_{12})$
(see \eqref{Mtheory_action_top} for the definition of $X_8$ in terms
of Pontryagin classes) and from a 4-form $E_4$ on $M_{12}$,
according to
\beq
\cI_{12} = - \frac 16 \, E_4 \wedge E_4 \wedge E_4 - E_4 \wedge X_8 \ .
\eeq
The expression for $E_4$ is discussed below.
It is argued in \cite{Bah:2019rgq} that the 6-form \eqref{I6inflow}
is equal to minus the 't Hooft anomalies of the full
4d theory living on the M5-branes stack.
In the cases of interest in this paper, the 4d theory consists
of an interacting SCFT, together with free decoupled modes.
We may then write
\beq \label{anomaly_balance}
I_6^{\rm inflow} + I_6^{\rm SCFT} + I_6^{\rm decoupl} = 0 \ .
\eeq
We comment further on $I_6^{\rm decoupl}$ in section \ref{sec_singleton_anomalies}.

Let us now turn to a description of the 4-form $E_4$.
It is a globally defined, closed 4-form on $M_{12}$
with integral periods, which can be  written as
\beq \label{E4_expr}
E_4 =   N_\alpha \, (\Omega_4^\alpha)^{\rm eq}
+ \frac{F_2^\alpha}{2\pi} \wedge (\omega_{2\alpha})^{\rm eq}
 + \frac{f_1^x}{2\pi} \wedge (\Lambda_{3x})^{\rm eq}
 + \frac{H_3^u}{2\pi} \wedge (\lambda_1{}_u)^{\rm eq}
 + \frac{\gamma_4}{2\pi} \ .
\eeq
In the previous expression
the forms $F_2^\alpha$, $f_1^x$, $H_3^u$, $\gamma_4$
are closed forms on the base $\cM_6$ of the fibration \eqref{M12aux},
 pulled back to the total space $M_{12}$ (the pullpack 
is implicit in our notation). Exactly as in \eqref{G4_ansatz},
we interpret $F_2^\alpha$, $f_1^x$, $H_3^u$, $\gamma_4$ as the field strengths
of $p$-form background gauge fields on $\cM_6$, so that the periods of 
$F_2^\alpha$, $f_1^x$, $H_3^u$, $\gamma_4$ are quantized in units of $2\pi$.

The forms $(\Omega_4^\alpha)^{\rm eq}$,
$(\omega_{2\alpha})^{\rm eq}$,
$(\Lambda_{3x})^{\rm eq}$,
$(\lambda_1{}_u)^{\rm eq}$ are globally defined,  closed forms on $M_{12}$
with integral periods.\footnote{The label ``eq'' stands for equivariant,
because the forms $(\Omega_4^\alpha)^{\rm eq}$,
$(\omega_{2\alpha})^{\rm eq}$,
$(\Lambda_{3x})^{\rm eq}$,
$(\lambda_1{}_u)^{\rm eq}$
on $M_{12}$ can be regarded as representatives of classes
in the $G$-equivariant cohomology of $M_6$,
where the group $G$ is the isometry group of $M_6$ \cite{Bah:2019rgq}.  }
They can be regarded as a gauge-invariant and closed
extension of the forms 
$\Omega_4^\alpha$,
$\omega_{2\alpha}$,
$\Lambda_{3x}$,
$ \lambda_1{}_u$ on the fiber $M_6$.
Indeed, if the fibration  \eqref{M12aux}  is replaced
by a direct product and all external gauge fields
related to isometries of $M_6$ are turned off,
the forms $(\Omega_4^\alpha)^{\rm eq}$,
$(\omega_{2\alpha})^{\rm eq}$,
$(\Lambda_{3x})^{\rm eq}$,
$(\lambda_1{}_u)^{\rm eq}$ reduce to 
$\Omega_4^\alpha$,
$\omega_{2\alpha}$,
$\Lambda_{3x}$,
$ \lambda_1{}_u$.

We discuss the construction of the
forms  $(\Omega_4^\alpha)^{\rm eq}$,
$(\omega_{2\alpha})^{\rm eq}$,
$(\Lambda_{3x})^{\rm eq}$,
$(\lambda_1{}_u)^{\rm eq}$
in appendix \ref{app_E4}.
We would like to emphasize here, however,
that they are not uniquely determined by  the forms
 $\Omega_4^\alpha$,
$\omega_{2\alpha}$,
$\Lambda_{3x}$,
$\lambda_1{}_u$ on $M_6$.
Different realizations of 
$(\Omega_4^\alpha)^{\rm eq}$ differ by a closed (but not necessarily exact) 4-form
on $M_{12}$, and similarly for the other forms.
We show in appendix \ref{app_E4}
that the ambiguities related to a
specific choice of  $(\Omega_4^\alpha)^{\rm eq}$,
$(\omega_{2\alpha})^{\rm eq}$,
$(\Lambda_{3x})^{\rm eq}$,
$(\lambda_1{}_u)^{\rm eq}$ in $E_4$ can always
be undone by
adding exact pieces to $E_4$
(which do not alter the integral \eqref{I6inflow}) and/or
 performing a field redefinition of the external
field strengths  $F_2^\alpha$, $H_3^u$, $\gamma_4$.
We also verify that the necessary field redefinitions
preserve the lattice of periods of the field strengths.
We conclude that the inflow anomaly polynomial
$I_6^{\rm inflow}$ is unambiguously defined,
up to a choice of basis in the space of external $p$-form
gauge fields on $\cM_6$.

After these preliminary remarks, we can discuss
anomaly inflow for the wrapped M5-brane setups
of interest in this work.

\subsubsection{M5-branes wrapped on a Riemann surface}
In section \ref{BBBW_sol_sec} we have summarized the 
choices of $g$, $p$, $q$ that lead to a smooth supersymmetric $AdS_5$ M-theory
solution, see table \ref{BBBW_cases}. In all cases, $M_6$ admits at least a $U(1)_1 \times U(1)_2$ isometry,
associated to angular directions $\phi_1$, $\phi_2$.
We couple these isometries to external Abelian gauge fields
  $A^{\phi_1}$, $A^{\phi_2}$. We introduce the notation
\beq \label{c1_notation}
c_1^{\phi_1} = \frac{dA^{\phi_1}}{2\pi} = c_1(U(1)_1) \ , \qquad
c_1^{\phi_2} = \frac{dA^{\phi_2}}{2\pi} = c_1(U(1)_2)
\eeq
for the first Chern classes of these background connections.
When $g=0$, the space $M_6$ admits an additional
$SO(3)_\Sigma \cong SU(2)_\Sigma$ isometry. We can couple this isometry to 
a triplet of external gauge fields. We use the notation
$c_2^\Sigma = c_2(SU(2)_\Sigma)$
for the second Chern class of these $SU(2)_\Sigma$ background gauge fields.
Appendix \ref{app_BBBW} contains
 a more detailed discussion of the gauging of isometries of $M_6$.

The derivation of the inflow anomaly polynomial
is reported in appendix \ref{app_BBBW}.
The result reads (suppressing wedge products)
\begin{align}
I_6^{\rm inflow} & = - \frac 23 \, \bigg( N^3 - \frac 14 \, N \bigg) \, 
\bigg[ p \, c_1^{\phi_1} 
\, (c_1^{\phi_2})^2
+ q \, c_1^{\phi_2} \, (c_1^{\phi_1})^2
\bigg]
- \frac 16 \, N \, \bigg[  p \, (c_1^{\phi_1})^3 + q \, (c_1^{\phi_2})^3 \bigg]
\nn \\
&+ \frac{1}{24} \, N \, \bigg[ p \, c_1^{\phi_1}
+ q \, c_1^{\phi_2}  \bigg] \, p_1(T)
+ \frac{1}{6} \, \bigg[
(N^3 \, q^2 - N) \, p \, c_1^{\phi_1}
+ (N^3 \, p^2 - N) \, q \, c_1^{\phi_2}
 \bigg] c_2^\Sigma
\nn \\
& + \frac{1}{(2\pi)^2} \bigg[  -  N  \, \gamma_4 \, F_2
-  N \,  \widetilde H_3^i \, H_{3i} \bigg] \ .   \label{BBBW_result}
\end{align}
This result   includes the terms
originating from $- E_4 \, X_8$ in $\cI_{12}$.
In \eqref{BBBW_result},
 $p_1(T)$ denotes the 
first Pontryagin class of the background metric on spacetime.
Notice that the coefficients of the BF terms are
both equal to $N$ here. 
It is understood that the terms with $c_2^\Sigma$ are only present
if $g=0$. 
We stress that  \eqref{BBBW_result} does not contain
mixed 't Hooft anomalies between the symmetries related to
isometries of $M_6$ and the symmetries
associated to cohomology classes on $M_6$.\footnote{The 
anomaly polynomial   
is sensitive to the choice of 
 the forms $(\Omega_4^\alpha)^{\rm eq}$,
$(\omega_{2\alpha})^{\rm eq}$,
$(\Lambda_{3x})^{\rm eq}$,
$(\lambda_1{}_u)^{\rm eq}$.
There exists a choice such the result takes the form
\eqref{BBBW_result}. As we shall see, in the GMSW setup
such a choice is not possible.
}

\subsubsection*{Interpretation}

If we focus only on continuous symmetries,
the background fields $F_2$, $\gamma_4$, $\widetilde H_3^i$, $H_3$
are set to zero. This follows from the tadpole condition
on $E_4^2 + 2 \, X_8$ discussed in \cite{Bah:2019rgq}. 
Alternatively, we notice
that  $I_6^{\rm inflow}$ in \eqref{BBBW_result} can be regarded 
as collecting all topological terms in the 5d $AdS_5$ effective action.
Enforcing the tadpole condition is equivalent to 
using the 5d EOMs for $A_1$, $c_3$, $\widetilde B_2^i$, $B_{2i}$ that
come from this topological action:
all these fields are flat on-shell.
After integrating out $F_2$, $\gamma_4$, $\widetilde H_3^i$, $H_3$,
the
first two lines of  \eqref{BBBW_result}
reproduce known results \cite{Bah:2012dg,Bah:2019rgq}
and their 
 interpretation 
is standard:
they encode   't Hooft anomalies for the symmetries
$U(1)_1$, $U(1)_2$, $SU(2)_\Sigma$ and Poincar\'e symmetry.

The terms $\gamma_4 \, F_2$ and $\widetilde H_3^i \, H_{3i}$ on the last line
of \eqref{BBBW_result} are a proxy for 't Hooft anomalies involving discrete
global symmetries. More precisely, 
the global symmetry on the boundary SCFT depends on the choice
of boundary conditions for $A_1 \equiv \cA_1$, $c_3$, $B_{2i}$, $\widetilde B_i^2$.
The terms $\gamma_4 \, F_2$ and $\widetilde H_3^i \, H_{3i}$
each encode a mixed 't Hooft anomaly 
if we choose boundary conditions of type (c), (c${}^\prime$) in the terminology
of section \ref{sec_global_symm}.

\subsubsection{M5-branes probing a $\mathbb Z_2$ singularity
and wrapped on a Riemann surface}
If we consider a higher-genus Riemann surface,
the internal space $M_6$ has   isometry group
$U(1)_\psi \times SU(2)_\varphi$.
In the case $g=0$ we have an additional
$SO(3)_\Sigma \cong SU(2)_\Sigma$ isometry.
We introduce the compact notation
\beq
c_1^\psi = c_1(U(1)_\psi) \ , \qquad c_2^\varphi = c_2(SU(2)_\varphi) \ , \qquad
c_2^\Sigma = c_2(SU(2)_\Sigma)
\eeq
for the Chern classes of the background gauge fields
for   $U(1)_\psi$, $SU(2)_\varphi$, $SU(2)_\Sigma$.

The computation of the inflow anomaly polynomial
can be found in appendix \ref{app_GMSW}.
Let us write the result as
\beq \label{GMSW_inflow_tot}
I_6^{\rm inflow} = I_6^{\rm inflow,1} + I_6^{\rm inflow,2} + I_6^{\rm inflow,3} \ ,
\eeq
where $ I_6^{\rm inflow,1}$ encodes the anomalies
involving exclusively the symmetries associated to isometries
 and Poincar\'e symmetry,
 $ I_6^{\rm inflow,2}$  collects all terms that 
only involve $p$-form gauge fields originating from expansion of $C_3$,
and $I_6^{\rm inflow,3}$ contains all other terms.
Explicitly,
\begin{align}
I_6^{\rm inflow,1} & =  \bigg( 
 \frac{1}{3} \, \chi \, N^3 
+  N^2 \, N_- \bigg)\, c_1^\psi  \, c_2^\varphi   
 - \frac 13 \, N_- \, (c_1^\psi)^3
+ \frac {1}{12} \, N_- \, c_1^\psi \, p_1(T)
- \frac{1}{3} \, \chi \, N  \, c_1^\psi \, c_2^\varphi \nn \\
& + \bigg[
\frac 13 \, N_-^3
-  N^2 \, N_-  
- \frac 23  \,  N^3  
+ \frac 23 \, (N+N_-)
 \bigg] \, c_1^\psi \, c_2^\Sigma \ .
\\
I_6^{\rm inflow,2}  & = \frac{1}{(2\pi)^2} \, \bigg[
 - (N \, F_2 + N_+ \, F_2^+ + N_- \, F_2^-) \, \gamma_4
- N \, \widetilde H_3^i \, H_{3i} \bigg]   \\
&  + \frac{1}{(2\pi)^3} \, \bigg[ 
 - \frac 1 6\, \chi \, (F_2^+)^3
- \frac 12 \, \chi \, F_2^+ \, (F_2^-)^2
+ F_2 \, F_2^+ \, F_2^- 
-   \gamma_4 \, \Big( f_{1i}^+ \, \widetilde f_1^{i-}
- \widetilde f_{1}^{i+} \, f_{1i}^-
 \Big)
 \nn \\
&  + F_2^+ \, \Big( 
 f_{1i}^+ \, \widetilde H_3^i
 - \widetilde f_1^{i+} \, H_{3i}
 \Big)
  + F_2^- \, \Big( 
 f_{1i}^- \, \widetilde H_3^i
 - \widetilde f_1^{i-} \, H_{3i}
 \Big)
 \bigg] \ ,
 \nn \\
 I_6^{\rm inflow,3} & =
    - \frac{1}{(2\pi)^2}    \, N_-  \, c_1^\psi \, (F^+_2)^2 
+\frac{1}{2\pi}  \, \bigg(  N \, N_-  
+ \frac 12 \, N^2 \, \chi
- \frac 12 \, \chi 
\bigg)\, \, (c_1^\psi)^2 \, F_2^+ 
  \\
&  - \frac{1}{2\pi} \, N \, \bigg(  N_- \, F^+_2
+   N_+ \, F^- _2 
\bigg) \, c_2^\varphi
+ \frac{1}{24} \, \chi \, \frac{F_2^+}{2\pi} \, p_1(T) 
\nn \\
& + \frac{1}{2\pi} \,  ( N + N_-) \bigg[
N_+ \, F_2^- + N_- \, F_2^+  - N_+ \, F_2
 \bigg] \, c_2^\Sigma 
\nn \\
&+ \frac{1}{(2\pi)^2} \bigg[
 N \, (c_1^\psi)^2
 - N \, c_2^\varphi
  -  2 \,  c_1^\psi \, \frac{ F_2^+ }{2\pi}
  \bigg] \, 
 \Big(
 f_{1i}^+ \, \widetilde f_1^{i+} +  f_{1i}^- \, \widetilde f_1^{i-}
  \Big)
  \nn \\
&   - \frac{1}{(2\pi)^3}  \, c_1^\psi \, F_2^- \,\Big( f_{1i}^+ \, \widetilde f_1^{i-}
- \widetilde f_{1}^{i+} \, f_{1i}^-
 \Big)
 - \frac{1}{(2\pi)^2} \, N  \, c_1^\psi \, \Big( 
 f_{1i}^+ \, \widetilde H_3^i
 - \widetilde f_1^{i+} \, H_{3i}
 \Big) \ . \nn
\end{align}
It should be stressed that, in presenting $I_6^{\rm inflow}$,
we have implicitly chosen a basis of external $p$-form fields
originating from expansion of $C_3$. We are free to consider field
redefinitions that shift these  $p$-form fields
with terms constructed with the background connections for the isometries
of $M_6$. Unlike the BBBW case, however, there is no such redefinition
that can set to zero all mixed terms  
between symmetries originating from isometries,
and symmetries originating from $C_3$.

\subsubsection*{Perturbative anomalies}

Let us extract physical information about perturbative
anomalies for continuous symmetries
from the inflow anomaly polynomial \eqref{GMSW_inflow_tot}.
The gauge fields $B_{2i}$, $\widetilde B_2^i$, $c_3$, together with
  one linear combination of the three vectors $A_1$, $A_1^\pm$,
are topologically massive gauge fields in five dimensions and therefore
cannot be interpreted as background gauge fields for continuous symmetries in the 4d field theory.  If we are only interested in studying
local aspects of 't Hooft anomalies for continuous global symmetries,
we have to eliminate the topologically massive fields from
the anomaly polynomial.
This is done enforcing the tadpole constraints
on $E_4^2 + 2 \, X_8$ discussed in
 \cite{Bah:2019rgq}.
Equivalently, we impose the equations of motion
for $c_3$,  $B_{2i}$, $\widetilde B_2^i$ in the 5d topological
theory defined by $I_6^{\rm inflow}$.
If we do so, we obtain the relations
\begin{align} \label{pert_relations}
0 & = \tfrac{1}{2\pi} \, \Big( N \, F_2 + N_+ \, F_2^+  + N_- \, F_2^- 
\Big) + \tfrac{1}{(2\pi)^2} \,  \Big( f_{1i}^+ \, \widetilde f_1^{i-}
- \widetilde f_{1}^{i+} \, f_{1i}^-
 \Big) \ , \nn \\
 0 & =  \tfrac{1}{2\pi} \,  N \, H_{3i} +
 \tfrac{1}{(2\pi)^2} \, \Big(  F_2^+ \, f_{1i}^+ + F_2^- \, f_{1i}^- \Big)
 - \tfrac{1}{2\pi} \, N \, c_1^\psi \, f_{1i}^+ \ , \nn \\
  0 & =  \tfrac{1}{2\pi} \,  N \, \widetilde H_{3}^i +
 \tfrac{1}{(2\pi)^2} \, \Big(  F_2^+ \, \widetilde f_{1}^{i+} + F_2^- \, 
 \widetilde f_{1}^{i-} \Big)
 - \tfrac{1}{2\pi} \, N \, c_1^\psi \, \widetilde f_{1 }^{i+} \ .
\end{align}
We may solve these relations for $F_2$, $H_{3i}$, $\widetilde H_3^i$.
After plugging the corresponding expressions back into 
$I_6^{\rm inflow}$, the field $\gamma_4$ drops away and we are left
with a polynomial in 
$c_1^\psi$, $c_2^\varphi$, $c_2^\Sigma$, $p_1(T)$,
$F_2^{\pm}$, $f_{1i}^\pm$, $\widetilde f_1^{i\pm}$.
This polynomial encodes the sought-for
perturbative anomalies and reads
\begin{align} \label{pert_result}
I_6^{\rm inflow,pert} & = 
 \bigg( 
 \frac{1}{3} \, \chi \, N^3 
+  N^2 \, N_- \bigg)\, c_1^\psi  \, c_2^\varphi   
 - \frac 13 \, N_- \, (c_1^\psi)^3
+ \frac {1}{12} \, N_- \, c_1^\psi \, p_1(T)
- \frac{1}{3} \, \chi \, N  \, c_1^\psi \, c_2^\varphi \nn \\
& + \bigg[
\frac 13 \, N_-^3
-  N^2 \, N_-  
- \frac 23  \,  N^3  
+ \frac 23 \, (N+N_-)
 \bigg] \, c_1^\psi \, c_2^\Sigma
 \nn \\
& + \frac{1}{(2\pi)^3} \bigg[ - \frac \chi 6 \, (F_2^+)^3
- \bigg( \frac{N_-}{N}  + \frac \chi 2  \bigg) \, F_2^+ \, (F_2^-)^2
- \frac{N_+}{N} \, F_2^- \, (F_2^+)^2
\bigg]
\nn \\ 
& + \frac{1}{(2\pi)^4} \bigg[
- \frac 1N \, (F_2^-)^2 \, f_{1i}^- \, \widetilde f_1^{i-}
- \frac 1N \, (F_2^+) \, f_{1i}^+ \, \widetilde f_1^{i+}
- \frac 2 N \, F_2^+ \, F_2^- \,  \Big( f_{1i}^+ \, \widetilde f_1^{i-}
- \widetilde f_{1}^{i+} \, f_{1i}^-
 \Big)
 \bigg]
 \nn \\
&  - \frac{1}{(2\pi)^2} \, N_- \, c_1^\psi \, (F_2^+)^2
-\frac{1}{(2\pi)^3}  \,  2\, c_1^\psi \, F_2^+ \, f_{1i}^- \, \widetilde f_1^{i-}
+\frac{1}{(2\pi)^2} \,  N \, (c_1^\psi)^2 \, f_{1i}^- \, \widetilde f_1^{i-}
\nn \\
& +\frac{1}{2\pi} \,  \bigg(N \, N_-  + \frac \chi 2 \, N^2 - \frac \chi 2 \bigg) \, (c_1^\psi)^2 \, F_2^+
- \frac{1}{2\pi} \, N \, (N_+ \, F_2^- + N_- \, F_2^+) \, c_2^\varphi
\nn \\
& - \frac{1}{(2\pi)^2} \, N \,\Big( 
  f_{1i}^+ \, \widetilde f_1^{i+}
+ f_{1i}^- \, \widetilde f_1^{i-}
\Big) \, c_2^\varphi
  + \frac{1}{2\pi} \, \bigg(  N \, N_- + N_-^2 + N_+^2 + \frac{N_- \, N_+^2}{N} \bigg) 
\, F_2^+
c_2^\Sigma
\nn \\
&+ \frac{1}{2\pi} \,  \bigg(  N \, N_+  + 2 \, N_+ \, N_-  + \frac{N_+ \, N_-^2}{N}
\bigg) 
\, F_2^-
c_2^\Sigma
 + \frac{1}{2\pi} \, \frac{\chi}{24} \, F_2^+ \, p_1(T)
\nn \\
& +\frac{1}{(2\pi)^2} \bigg( N_+ + \frac{N_- \, N_+}{N} \bigg) \, c_2^\Sigma \, 
 \Big( f_{1i}^+ \, \widetilde f_1^{i-}
- \widetilde f_{1}^{i+} \, f_{1i}^-
 \Big) 
 \ .
\end{align}
The above expression extends the results of \cite{Bah:2019vmq}
with the inclusion of the terms involving
$f_{1i}^\pm$, $\widetilde f_1^{i \pm}$.
Notice the appearance of $1/N$ factors in the 't Hooft anomaly coefficients.
They originate from solving the relations \eqref{pert_relations}.
Physically, they come from integrating out topologically massive modes.
The perturbative anomaly polynomial \eqref{pert_result}
can be used to compute central charges via $a$-maximization;
at leading order in $N$, one finds a perfect match with the dual supergravity
computation based on the GMSW solutions \cite{Bah:2019vmq}.

\subsubsection*{Aside: topologically massive fields and Green-Schwarz terms in six dimensions}
A variant of the mechanism that
generates $1/N$ terms 
in \eqref{pert_result} by integrating out topologically massive fields
is at play in six dimensions.
More precisely, let us 
consider a stack of $N$ M5-branes
probing a $\Gamma_{\rm ADE} \subset SU(2)$ singularity.
The internal geometry is $S^4/\Gamma_{\rm ADE}$.
Upon resolution of the orbifold
singularities at the north and south poles of $S^4$, we get a smooth internal 
internal space $M_4$. At each pole we have a collection
of resolution 2-cycles. Expansion of the M-theory 3-form
$C_3$ in cohomology of $M_4$ yields
an external 3-form gauge field $c_3$ and
a collection of 1-form gauge fields, associated to the resolution
2-cycles at the north and south poles.
In the limit in which the resolution cycles are shrunk
to zero size we have an $G_\Gamma^\mathrm N \times G_\Gamma^{\mathrm S}$ non-Abelian gauge symmetry in the 7d low-energy effective
action, where $G_\Gamma$ is the ADE Lie group
associated to $\Gamma_{\rm ADE}$.
The topological couplings of the 7d effective action
are conveniently encoded in a gauge-invariant 8-form,
which contains the terms
\beq \label{two_terms}
-\frac 12 \, N \, \frac{ \gamma_4^2}{(2\pi)^2} - \frac 14 \, \frac{\gamma_4}{2\pi} \, \bigg[
\frac{{\rm tr} \, (F^\mathrm N)^2}{(2\pi)^2}
- \frac{{\rm tr} \, (F^\mathrm S)^2}{(2\pi)^2}
\bigg] \ .
\eeq
The 4-form $\gamma_4$ is the   field strength of the
3-form gauge field $c_3$, while $F^{\rm N,S}$ is the field
strength of the gauge group $G_\Gamma^{\rm N,S}$. 
The 3-form gauge field $c_3$ is topologically massive
by virtue of the 7d Chern-Simons coupling encoded
in the term $\gamma_4^2$ in \eqref{two_terms}.
In order to study the perturbative anomalies 
for continuous global symmetries of the system, 
we have to integrate out this massive 
field. Eliminating $c_3$ via its classical equation of motion
is the same as enforcing the tadpole constraint of \cite{Bah:2019rgq}.
The terms \eqref{two_terms} are traded for
\beq
\frac{1}{32 \, N} \, \bigg[
\frac{{\rm tr} \, (F^\mathrm N)^2}{(2\pi)^2}
- \frac{{\rm tr} \, (F^\mathrm S)^2}{(2\pi)^2}
\bigg] ^2 \ .
\eeq
In the analysis of \cite{Ohmori:2014kda} this term is interpreted as a
Green-Schwarz term related to the center-of-mass mode of the M5-branes,
see also the recent field-theoretic analysis of \cite{toappear}.
Our analysis of topological mass terms in supergravity
reveals how this term is automatically accounted for
in inflow via integrating out massive modes.

\subsubsection*{Remarks on the background 0-form gauge fields $a_{0i}^{\pm}$, 
$\widetilde a_0^{i\pm}$}

The terms in $I_6^{\rm inflow}$ with $f_{1i}^\pm = d a_{0i}^{\pm}$,
$\widetilde f_1^{i\pm} = d \widetilde a_0^{i\pm}$ should be interpreted along the lines of 
\cite{Cordova:2019jnf} as 't Hooft anomalies in the space of coupling constants.
We can think of $a_{0i}^{\pm}$, 
$\widetilde a_0^{i\pm}$ as background fields for global ``$(-1)$-form
symmetries'' in the 4d field theory.
(We refer the reader to \cite{Cordova:2019jnf} for a careful discussion of the merits
and limitations of the
notion of ``$(-1)$-form
symmetry''.)

Recall that  $a_{0i}^{\pm}$, 
$\widetilde a_0^{i\pm}$ originate form expansion of $C_3$ onto
3-cycles in the internal space $M_6$. 
We can offer an interpretation of $a_{0i}^{\pm}$, 
$\widetilde a_0^{i\pm}$
 in terms of the picture of M5-branes probing a $\mathbb Z_2$
singularity.
The 6d SCFT on the worldvolume of the M5-branes
has an $SU(2)_\mathrm N \times SU(2)_\mathrm S$ global symmetry.
This theory is reduced on $\Sigma_g$ with a non-zero
flavor flux, which breaks $SU(2)_\mathrm N \times SU(2)_\mathrm S$ to the Cartan
subgroup $U(1)_\mathrm N \times U(1)_\mathrm S$.
The 6d background 1-form gauge fields for this 0-form symmetry
can be dimensionally reduced along 1-cycles in  $\Sigma_g$ to yield
0-form gauge fields in four dimensions.
Since $\Sigma_g$ has $2g$ 1-cycles,
this reduction generates a total of $2 \times 2g = 4g$
0-form gauge fields in four dimensions,
which matches the total number of $a_{0i}^{\pm}$, 
$\widetilde a_0^{i\pm}$ fields.

The operators in the 4d SCFT coupled  to $a_{0i}^{\pm}$, 
$\widetilde a_0^{i\pm}$ are exactly marginal operators.
In a schematic semi-Lagrangian language,
the deformation of the SCFT associated to $a_{0i}^{\pm}$, 
$\widetilde a_0^{i\pm}$ takes the form
\beq
\Delta  \cS = \int_{\cM_4} \frac{*_41}{2\pi} \bigg[ a_{0i}^{+} \, \widetilde \cO^{i+}
- \widetilde a_{0}^{i+} \,   \cO_i^+   
+  a_{0i}^{-} \, \widetilde \cO^{i-}
- \widetilde a_{0}^{i-} \,   \cO_i^-   \bigg] \ , \qquad
\Delta(\cO_i^\pm) = \Delta(\widetilde \cO^{i\pm}) = 4   \ .
\eeq
In light of the discussion of the previous paragraph,
we can regard  $\cO_i^\pm$, $\widetilde \cO^{i\pm}$ as coming from the dimensional reduction on $\Sigma_g$
of the 
6d 1-form conserved current operators associated to the 6d
$U(1)_\mathrm N \times U(1)_\mathrm S$ 0-form symmetry.
Schematically,
\beq
*_6 J^\pm_{1, \text{6d}} \sim ( *_4 \cO_i^\pm) \wedge \widetilde \lambda^i
- (*_4  \widetilde \cO^{i\pm}) \wedge \lambda_i \ ,
\eeq
where $\lambda_i$, $\widetilde \lambda^i$ are closed 1-forms on $\Sigma_g$
as in section \ref{sec_ansatz}.
We also notice that $a_{0i}^{\pm}$, 
$\widetilde a_0^{i\pm}$ are compact scalars with period $2\pi$.
This indicates that the   spacetime integrals of the associated
operators $\cO_i^\pm$, $\widetilde \cO^{i\pm}$ 
satisfy a quantization condition of the form
\beq
\int_{\cM_4} \cO_i^\pm  \, *_4 1 \in \mathbb Z \ , \qquad
\int_{\cM_4} \widetilde \cO^{i \pm}  \, *_4 1 \in \mathbb Z \ .
\eeq
Intuitively, the operators 
$\cO_i^\pm$, $\widetilde \cO^{i\pm}$ are analogous to
${\rm tr} \, (FF)$ in gauge theory,
and $a_{0i}^{\pm}$, 
$\widetilde a_0^{i\pm}$ are analogous to $\theta$ angles.

Let us stress that  $a_{0i}^{\pm}$, 
$\widetilde a_0^{i\pm}$ are distinct from the axionic couplings
that originate from the complex structure moduli of the Riemann
surface $\Sigma_g$. The geometric origin of the latter 
resides in a deformation of the metric on $\Sigma_g$.
't Hooft anomalies associated to these coupling constants
have been analyzed in \cite{Tachikawa:2017aux}.

\subsubsection{Anomalies for discrete symmetries}
\label{sec_discrete_anomalies}

The inflow anomaly polynomial balances against
the total 't Hooft anomalies of interacting and decoupled modes in the 4d field theory,
see \eqref{anomaly_balance}.
It should be stressed that the separation into interacting and
decoupling modes does not necessary correspond to a simple
factorization of the partition function in field theory.
From the perspective of the dual gravity theory,
the singleton sector in a string theory/M-theory
compactification  
decouples from the rest of the dynamics of quantum gravity,
but in general
the full quantum gravity partition function does not simply
factorize into a contribution from the singleton sector times
a contribution from interacting modes.
Rather, as argued in \cite{Belov:2004ht},
one expects the total string theory/M-theory partition function to be of the 
schematic form 
\beq \label{Z_factorization}
Z_{\rm tot} \sim \sum_\beta Z^\beta \, Z_\beta^{\rm singleton} \ .
\eeq
In the previous expression, the discrete label $\beta$
enumerates the relevant topological sectors
of string theory/M-theory in the background under consideration.
The quantities $Z_\beta^{\rm singleton}$ encode the contribution
of singleton modes, while $Z^\beta$ encode the contributions of all other interacting
bulk modes.
The holographic duals of  $Z^\beta$ are the conformal blocks
of an interacting 4d CFT, while the holographic duals of
$Z_\beta^{\rm singleton}$ are the conformal blocks of a free 4d theory.

As demonstrated in \cite{Belov:2004ht}, 
the correct strategy to compute $Z_\beta^{\rm singleton}$
on the gravity side is to consider both 
kinetic terms and topological terms in the 5d supergravity effective action.
In this approach the Hamiltonian in the singleton sector
is unambiguously determined. (In contrast, in the purely topological BF theory
without kinetic terms, the Hamiltonian can be modified by adding boundary terms.)
For the setups of interest in this work,
one needs to consider the BF couplings
\eqref{simple_5dtopBIS} supplemented with standard kinetic terms.

In the setups with wrapped M5-branes studied in this work,
the total worldvolume theory has a partition function of the form
\eqref{Z_factorization} with more than one term on the RHS.
Indeed, the different summands labeled by $\beta$ correspond to inequivalent
choices of boundary conditions for the fields
entering the BF couplings \eqref{simple_5dtopBIS}.
In the total worldvolume theory,
the fields $\cA_1$, $c_3$, $B_{2i}$, $\widetilde B_2^i$
are   associated to global continuous $U(1)$ 0-, 2-, and 1-form symmetries.
These $U(1)$ symmetries are spontaneously broken.
The breaking pattern is different 
for the various $\beta$ summands in \eqref{Z_factorization}.
Indeed, we know that interacting  theories associated
to different choices of boundary conditions have different
global discrete symmetries.
For example, with reference to the terminology
of section \ref{sec_global_symm}, the $U(1)$ 0-form symmetry associated to $\cA_1$
is broken to $\mathbb Z_k$ in case (a),
is broken to nothing in case (b), 
and is broken to $Z_{m'}$ in case (c).

The inflow anomaly polynomial $I_6^{\rm inflow}$ in \eqref{BBBW_result}
or \eqref{GMSW_inflow_tot} is interpreted as minus the anomaly polynomial
of the total worldvolume theory \eqref{Z_factorization}. Since this theory
has continuous symmetries, we can describe its anomalies using the  
 language
of differential forms.
If we ignore the specific breaking pattern of the $U(1)$'s to discrete symmetries,
all interacting SCFTs with partition functions $Z^\beta$
have the same perturbative 't Hooft anomalies
for their unbroken continuous symmetries.
For wrapped M5-branes these anomalies are the first two lines of
\eqref{BBBW_result}, while for wrapped M5-branes at a $\mathbb Z_2$
singularity the perturbative anomalies are collected in \eqref{pert_result}.

Extracting the anomalies for discrete symmetries
of a given interacting SCFT $Z^\beta$ is more challenging.
We expect that the language of differential cohomology should give us the proper mathematical
framework to discuss these anomalies.
In appendix \ref{app_diff_cohom}
we provide a brief review of the aspects of differential
cohomology that are relevant for this work.
We use the notation $\check H^\ell(\cM_4)$
to denote the $\ell$-th differential cohomology group 
of external spacetime.
An element of $\check H^\ell(\cM_4)$ models an $(\ell-1)$-form
$U(1)$ gauge field.

As a   first case, let us consider 
the BBBW setup
and  assign Dirichlet
boundary conditions to $\cA_1$ and free boundary conditions 
to $c_3$, case (a) in the terminology of section \ref{sec_global_symm}. 
We can dualize the 3-form gauge field $c_3$ to a 0-form
gauge field $\phi_0$. The effect of the dualization is to convert the
original BF theory \eqref{simple_5dtopBIS} (supplemented by standard kinetic terms)
into a St\"uckelberg
theory written in terms of the combination
$D\phi_0 = d\phi_0 - k \, \cA_1$.
In the deep IR, the 1-form gauge field $\cA_1$
and the 0-form gauge field $d\phi_0$ are subject to the constraint
$D\phi_0 = 0$,~or
\beq \label{A_constraint}
k \, \cA_1 = d\phi_0 \ .
\eeq
The gauge field $k\, \cA_1$
is pure gauge, because it is given in terms a   
\emph{globally defined} closed 1-form $d\phi_0$ with periods that are
quantized in units of $2\pi$. 
Crucially, this does not mean
that $\cA_1$ is   trivial.
Instead, $\cA_1$ is a flat gauge field that
is allowed to have non-trivial holonomies that are 
$k$-th roots of unity.
These features show that the pair $(\cA_1, \phi_0)$
subject to the constraint \eqref{A_constraint} describes a background
1-form $\mathbb Z_k$ gauge field, 
as  in~\cite{Banks:2010zn,Kapustin:2014zva}.\footnote{If $G$ is a finite group, giving a
connection on a   principal $G$-bundle over $\cM_4$ is the same
as specifying an element of
${\rm Hom}(\pi_1(\cM_4), G)$. For the case at hand $G = \mathbb Z_k$ is Abelian, and therefore
(by Hurewicz theorem) we can 
 equivalently consider
${\rm Hom}(H_1(\cM_4), \mathbb Z_k)$. The pair $(\cA_1, \phi_0)$
determines indeed an element of ${\rm Hom}(H_1(\cM_4), \mathbb Z_k)$,
because the holonomies of $\cA_1$ for any 1-cycle
are in $\mathbb Z_k \subset U(1)$, and only depend on the
homology class of the 1-cycle because $\cA_1$ is flat.
If external spacetime $\cM_4$ has no torsion
in homology,  we also have ${\rm Hom}(H_1(\cM_4), \mathbb Z_k) \cong H^1(\cM_4, \mathbb Z_k)$
from the universal coefficient theorem.
}

In the process of dualizing $c_3$ to $\phi_0$,
the BF term $k \, \cA_1 \wedge dc_3$ is removed.
As a result, the anomaly polynomial \eqref{BBBW_result}
does not contain $c_3$ nor $\cA_1$. This is consistent with
the global symmetries of the theory in case (a):
we have a global 0-form $\mathbb Z_k$ symmetry from $\cA_1$,
but no global symmetry from $c_3$,
and thus no mixed anomaly between the two.
Moreover, since \eqref{BBBW_result} lacks mixed terms between
$\cF_2$, $\gamma_4$, and the other field strengths, 
there are no mixed anomalies between the 0-form $\mathbb Z_k$
symmetry and other symmetries.

Similar remarks apply to case (b), in which we assign
Dirichlet boundary conditions to $c_3$.
In this situation we dualize $\cA_1$ to $\phi_2$,
and we impose the constraint
\beq
k \, c_3 = d\phi_2 \ .
\eeq
Thus, the pair $(c_2, \phi_2)$ models a 3-form
gauge field for a $\mathbb Z_k$ symmetry.
The $B_{2i}$, $\widetilde B_2^i$ system is studied in a similar way.

The setup with wrapped M5-branes probing a $\mathbb Z_2$
singularity is considerably richer.
Let us focus on the fields $\cA_1$, $c_3$, and let us impose
Dirichlet boundary conditions on $\cA_1$.
In the total anomaly polynomial 
\eqref{GMSW_inflow_tot} we can collect all terms 
with a $\gamma_4$ factor,
\beq
- \frac{\gamma_4 }{2\pi} \bigg[ 
k \, \frac{\cF_2}{2\pi} + \frac{ f_{1i}^+ \, \widetilde f_1^{i-}
- \widetilde f_1^{i+} \, f_{1i}^-  }{(2\pi)^2}
\bigg] \ ,
\eeq
where we have recalled that $N_\alpha \, F_2^\alpha = k \, \cF_2$.
The dualization of $c_3$ yields a 0-form gauge field $\phi_0$
as before. The analog of the constraint \eqref{A_constraint} reads now
\beq \label{new_constraints}
k \, \cA_1 + A_1^{f \tilde f} = d\phi_0 \ .
\eeq
In the previous expression, $A_1^{f \tilde f}$ denotes
the 1-form gauge field whose field strength satisfies
\beq
\frac{dA_1^{f \tilde f}}{2\pi} =  \frac{ f_{1i}^+ \, \widetilde f_1^{i-}
- \widetilde f_1^{i+} \, f_{1i}^-  }{(2\pi)^2} \ .
\eeq
More precisely, the 0-form gauge fields $a_{0i}^\pm$, $\widetilde a_0^{i \pm}$
can be modeled by elements of the differential cohomology group
$\check H^1(\cM_4)$. In differential cohomology
a well-defined notion of product exists, which 
maps $\check H^1(\cM_4) \times \check H^1(\cM_4)$ to $\check H^2(\cM_4)$.
In other words,   to a pair of 0-form gauge fields one can
associate a 1-form gauge field, see appendix \ref{app_diff_cohom} for
further details.
It is in this sense that $A_1^{f \tilde f}$ is constructed from
$a_{0i}^\pm$, $\widetilde a_0^{i \pm}$.
The relation \eqref{new_constraints} should be interpreted as a relation
between elements of $\check H^2(\cM_4)$. If we take the field strength of
both sides, we get an equation for differential 2-forms,
\beq \label{2form_eq}
k \, \cF_2 + \frac{ f_{1i}^+ \, \widetilde f_1^{i-}
- \widetilde f_1^{i+} \, f_{1i}^-  }{2\pi}= 0 \ .
\eeq
This relation is one of the equations of motions of the topological
theory defined by \eqref{GMSW_inflow_tot}, or equivalently
one of the tadpole constraints on $E_4^2 + 2 \, X_8$.

The 2-form equation \eqref{2form_eq} can be integrated
on any 2-cycle in spacetime. Since $\cF_2$ has periods that are
quantized in units of $2\pi$, we learn that 
$dA_1^{f \tilde f}$ has periods that are quantized in
units of $2\pi\, k$. This indicates that we can introduce
a new 1-form gauge field $\boldsymbol {\mathsf A}_1$ 
defined as
\beq \label{new_gauge_field}
\boldsymbol{\mathsf A}_1 = \cA_1 + \frac 1k \, A_1^{f \tilde f} \ ,
\eeq
and that $d\boldsymbol{\mathsf A}_1$ has periods that are quantized in units of
$2\pi$. It should therefore be possible to model
$\boldsymbol{\mathsf A}_1$ with an element of $\check H^2(\cM_4)$.
The new gauge field $\boldsymbol{\mathsf A}_1$ satisfies
\beq
k \, \boldsymbol{\mathsf A}_1 = d\phi_0 \ .
\eeq
Therefore, the pair $(\boldsymbol{\mathsf A}_1, \phi_0)$ describes a background gauge field
for a $\mathbb Z_k$ 0-form   symmetry.

We can now go back to the anomaly polynomial
\eqref{GMSW_inflow_tot}. Dualization of $\gamma_4$ has removed
all terms with a $\gamma_4$ factor.
There are several other terms, however, that contain $\cA_1$.
We rewrite these terms using \eqref{new_gauge_field} to 
trade $\cA_1$ for $\boldsymbol{\mathsf A}_1$ and $A_1^{f\tilde f}$.
After this rewriting, we find terms with $d\boldsymbol{\mathsf A}_1$, 
for example the term
\beq \label{example_term}
I_6^{\rm inflow} \supset n \, \frac{d\boldsymbol{\mathsf A}_1}{2\pi} \wedge c_2^\varphi \ ,
\eeq
where $n$ is an integer 't Hooft anomaly coefficient.
The 2-form $d\boldsymbol{\mathsf A}_1$ is zero:
how should \eqref{example_term} be interpreted?
We regard the 6-form $I_6^{\rm inflow}$ as the field strength
of a $U(1)$ 5-form gauge field, modeled by an element of
$\check H^6(\cM_6)$. The wedge product in 
\eqref{example_term} is reinterpreted as the product in differential
cohomology. The second Chern class $c_2^\varphi$
admits a natural extension in differential cohomology
and defines an element of $\check H^4(\cM_4)$.
A more detailed discussion of this point
can be found in  appendix \ref{app_diff_cohom}.
The 1-form gauge field $\boldsymbol{\mathsf A}_1$ is thought of as an element of
$\check H^2(\cM_4)$. Their product is thus
an element in $\check H^6(\cM_6)$. Even though
the field strength of this element of $\check H^6(\cM_6)$ is zero
(because $\boldsymbol{\mathsf A}_1$ is flat),
this object is still non-trivial. It encodes a non-zero
't Hooft anomaly between the 
the discrete $\mathbb Z_k$ 0-form symmetry
and the $SU(2)_\varphi$ symmetry.

The ideas outlined in the previous paragraphs
can also be applied to the $B_{2i}$, $\widetilde B_2^i$
system. For example, if we assign Dirichlet boundary conditions
to $B_{2i}$ (for each label $i$), we have to collect all terms
in \eqref{GMSW_inflow_tot} with $\widetilde H_3^i$
an dualize $\widetilde B_2^i$ to a 1-form gauge field
$\phi_{1i}$. We obtain a St\"uckelberg-like system
that enforces a constraint of the form
\beq \label{B2_constraint}
N \, B_{2i} + B_{2i}^{\rm comp} =  d \phi_{1i} \ .
\eeq
The quantity $B_{2i}^{\rm comp}$ is a composite
2-form gauge field, whose field strength satisfies
\beq
dB_{2i}^{\rm comp} = \frac{F_2^+ \, f_{1i}^+ + F_2^- \, f_{1i}^-}{2\pi}
- N \, c_1^\psi \, f_{1i}^+ \ .
\eeq
As in the case of $A_1^{f \tilde f}$, the object
$dB_{2i}^{\rm comp}$ is best thought of as a sum of  products in differential
cohomology of 1-form and 0-form gauge fields.
The relation \eqref{B2_constraint} is interpreted as an equation in 
$\check H^3(\cM_4)$. Taking the field strength of both sides
we get an equation for   3-forms,
which is the second equation of motion in \eqref{pert_relations}.
The periods of $dB_{2i}^{\rm comp}$ are quantized
in units of $2\pi N$, thus it makes sense to consider
$1/N \, B_{2i}^{\rm comp}$.
Reasoning as above,   
a new 2-form gauge field $\boldsymbol{\mathsf B}_{2i}$ can be introduced,
in terms of which \eqref{B2_constraint} takes a simpler form,
\beq
\boldsymbol{\mathsf B}_{2i} = B_{2i} + \tfrac 1N \, B_{2i}^{\rm comp} \  , \qquad
N \, \boldsymbol{\mathsf B}_{2i} = d \phi_{1i} \ .
\eeq
The dualization of $\widetilde B_2^i$ to $\phi_{1i}$
has removed all terms with $\widetilde H_3^i$ from
\eqref{GMSW_inflow_tot}. There are other terms containing $B_{2i}$, however.
We rewrite such terms trading $B_{2i}$ for $\boldsymbol{\mathsf B}_{2i}$.
As before, the terms that contain $d\boldsymbol{\mathsf B}_{2i}$
encode mixed 't Hooft anomalies between the ``electric''
$(\mathbb Z_N)^g$ 1-form symmetry and other symmetries of the field theory.

In appendix \ref{sec_app_result} we present a case study
for a detailed analysis of the anomalies,
along the lines explained in the previous paragraphs.
In particular, we give the full anomaly polynomial
in the case in which we assign Dirichlet boundary
conditions to $\cA_1$ and $B_{2i}$.
We find a rich variety of mixed anomalies involving 
the $\mathbb Z_k$ 0-form symmetry, the $(\mathbb Z_N)^g$ 1-form symmetry,
and the continuous symmetries of the system,
see \eqref{discrete_anomalies_no1}-\eqref{discrete_anomalies_final}.
The terms in the anomaly polynomial
that involve 
$d \boldsymbol{\mathsf A}_1$ and $d \boldsymbol{\mathsf B}_{2i}$
have the following structure,
\begin{align}   \label{schematic_terms}
 I_6^{\rm inflow}& \supset 
a_1 \,  \frac{(d\boldsymbol{\mathsf A}_1)^3  }{(2\pi)^3}
+ a_2 \, c_1^\psi \,  \frac{ (d\boldsymbol{\mathsf A}_1)^2 }{(2\pi)^2}
+ a_{3,\hat \alpha} \,  \frac{\cF_2^{\hat \alpha} \,  (d\boldsymbol{\mathsf A}_1)^2 }{(2\pi)^3}
+ a_4 \,  \frac{ ( f_{1i}^+ \, \widetilde f_1^{i-}
- \widetilde f_1^{i+} \, f_{1i}^- ) \, (d\boldsymbol{\mathsf A}_1)^2  }{(2\pi)^3}
\nn \\
&+ a_5 \,   \frac{  f_{1i}^+ \, \widetilde f_1^{i+}  \, (d\boldsymbol{\mathsf A}_1)^2  }{(2\pi)^3}
 + a_6 \,  c_2^\varphi \,  \frac{d\boldsymbol{\mathsf A}_1  }{2\pi}
 + a_7 \, c_2^\Sigma \,  \frac{d\boldsymbol{\mathsf A}_1  }{2\pi}
   + a_8 \,  (c_1^\psi)^2 \,  \frac{d\boldsymbol{\mathsf A}_1  }{2\pi}
  + a_9 \, p_1(T) \,  \frac{d\boldsymbol{\mathsf A}_1  }{2\pi}
\nn \\
&  + a_{10,\hat \alpha} \, c_1^\psi \,  \frac{\cF_2^{\hat \alpha} \, d\boldsymbol{\mathsf A}_1  }{( 2\pi )^2}
 +   a_{11, \hat \alpha \hat \beta}  \,  \frac{\cF_2^{\hat \alpha} \, \cF_2^{\hat \beta} \, d\boldsymbol{\mathsf A}_1  }{( 2\pi )^3}
 + a_{12} \,  c_1^\psi \, \frac{  f_{1i}^- \, \widetilde f_1^{i-} \, d\boldsymbol{\mathsf A}_1  }{(2\pi)^3}
  +  a_{13, \hat \alpha} \,  \frac{  f_{1i}^+ \, \widetilde f_1^{i+} \, \cF^{\hat \alpha} \, d\boldsymbol{\mathsf A}_1  }{(2\pi)^4}
\nn \\
& + a_{14} \, c_1^\psi \, \frac{ ( f_{1i}^+ \, \widetilde f_1^{i-}
- \widetilde f_1^{i+} \, f_{1i}^- ) \, d\boldsymbol{\mathsf A}_1}{  (2\pi) ^3}
 +  a_{15, \hat \alpha} \,   \frac{ ( f_{1i}^+ \, \widetilde f_1^{i-}
- \widetilde f_1^{i+} \, f_{1i}^- ) \, \cF_2^{\hat \alpha} \, d\boldsymbol{\mathsf A}_1}{  (2\pi) ^3}
\nn \\
& + a_{16} \,     \frac{ ( f_{1i}^+ \, \widetilde f_1^{i-}
- \widetilde f_1^{i+} \, f_{1i}^- )^2 \,  d\boldsymbol{\mathsf A}_1}{  (2\pi) ^5}
+   a_{17} \,   \frac{   f_{1ij}^- \, \widetilde f_1^{j-}  \, ( f_{1i}^+ \, \widetilde f_1^{i-}
- \widetilde f_1^{i+} \, f_{1i}^- ) \,  d\boldsymbol{\mathsf A}_1}{  (2\pi) ^5}
\nn \\
& + a_{18, \hat \alpha} \, \frac{ \cF_2^{\hat \alpha}   \, \widetilde f_1^{i+} \, 
d \boldsymbol{\mathsf B}_{2i}
 }{(2\pi)^3}
  + a_{19, \hat \alpha} \,  \frac{ \cF_2^{\hat \alpha}   \, \widetilde f_1^{i-} \, 
d \boldsymbol{\mathsf B}_{2i}
 }{(2\pi)^3}
 + a_{20} \, c_1^\psi \, \frac{   \widetilde f_1^{i+} \, 
d \boldsymbol{\mathsf B}_{2i}
 }{(2\pi)^2}
\nn \\
&  +  a_{21} \, \frac{ ( f_{1j}^+ \, \widetilde f_1^{j-}
- \widetilde f_1^{j+} \, f_{1j}^-) \,  \widetilde f_1^{i+} \, 
d \boldsymbol{\mathsf B}_{2i}
 }{(2\pi)^4}
 + a_{22} \, \frac{ d\boldsymbol{\mathsf A}_1 \, d\boldsymbol{\mathsf B}_{2i} \, \widetilde f_{1}^{i+}  }{(2\pi)^3}  \ .
\end{align}
In the previous expression
we have made use of the notation introduced in 
\eqref{the_new_basis}, in which the index $\hat \alpha$
refers to the continuous $U(1)^2$ symmetry
associated to two out of the three vectors coming from
expansion of $C_3$ onto cohomology classes.
The explicit expressions of the anomaly coefficients $a_1$, \dots, $a_{22}$
can be read off from \eqref{discrete_anomalies_no1}-\eqref{discrete_anomalies_final}.
Among the various terms in  \eqref{schematic_terms}
we notice in particular: terms that are cubic and quadratic
in $d \boldsymbol{\mathsf A}_1$; the term 
$p_1(T) \, d\boldsymbol{\mathsf A}_1$ describing a mixed discrete-gravitational anomaly; the last term in \eqref{schematic_terms}
 which mixes
the two discrete symmetries with a continuous axionic ``$(-1)$-form symmetry''.

In closing this section, let us comment on
boundary conditions of type (c) or (c${}^\prime$).
Intuitively speaking, in case (c) only a part of the field $c_3$
should be dualized to $\phi_0$, and a part of $\cA_1$ should be dualized
to $\phi_2$. More precisely, we expect a difficulty in using a Lagrangian
formalism to describe this case, analogous for instance 
to the difficulties that one encounters in formulating a 4d $U(1)$ gauge
theory with both the electric and magnetic photon in the Lagrangian.
Even though we are not able to describe the dualization
from BF form to St\"uckelberg form with the same level of detail
as in cases (a) and (b), we can still give an interpretation
of \eqref{GMSW_inflow_tot} in terms of 't Hooft anomalies for discrete symmetries.
As already anticipated in section \ref{sec_global_symm},
the term $k \, \cF_2 \, \gamma_4$ describes a mixed anomaly
between the $\mathbb Z_{m'}$ 0-form symmetry 
and the $\mathbb Z_m$ 2-form symmetry.
By a similar token, all terms involving $\cF_2$ 
signal non-zero 't Hooft anomalies between the $\mathbb Z_{m'}$
0-form symmetry and the other symmetries in the system,
and similarly for terms with $\gamma_4$. Analogous
remarks apply to $B_{2i}$, $\widetilde B_2^i$.
As in the case studied in appendix \ref{sec_app_result}, we find
a rich structure of mixed 't Hooft anomalies.

\subsection{Singletons and 't Hooft anomalies} 
\label{sec_singleton_anomalies}

A better understanding of the decoupled sector
of the 4d field theory is crucial to obtain
a detailed prediction for the anomalies of the interacting CFTs of interest.
The holographic dictionary suggests a general strategy to extract
 information about the decoupling modes
on the field theory side   from the gravity side:
one has to study the singleton modes that propagate on the conformal boundary
of  $\cM_5 \times_w M_6$.

Let us consider the setup with a stack of M5-branes wrapped on a Riemann
surface. In this case we know the decoupling modes on the field theory side. They
are obtained from dimensional reduction on $\Sigma_g$ of a free
6d $\cN = (2,0)$ tensor multiplet, which corresponds to the center-of-mass degrees of freedom
of the M5-brane stack.
Our goal is to compare the set of decoupled fields 
with  singleton modes in 
  $\cM_5 \times_w M_6$, where $M_6$ is the internal space of
BBBW solutions.  

The dimensional reduction of a free 6d $\cN  = (2,0)$ tensor multiplet
on a genus-$g$ Riemann surface with twist parameters $p$, $q$ is discussed
in appendix \ref{app_tensor}. 
Recall that, for any values of $p$, $q$, the internal space has a 
$U(1)_1 \times U(1)_2$ isometry.
We find the following 4d $\cN = 1$ multiplets:
\beq \label{free_fields_from_six}
\begin{array}{rl}
\text{$g$ vector multiplets:} \qquad \qquad &  A_\mu  \,  (0,0)  \ , \  \lambda \,   (1,1) \ 
;  \\[1mm]
\text{one chiral multiplet:}  \qquad \qquad  &  \Phi  \,  (0,0)  \ , \  b_0  \,  (0,0)  \ , \  \psi  \,  (-1,-1)  \ ;     \\[1mm]
\text{$h^0(K^\frac{p}{p+q})$ chiral multiplets:} \qquad \qquad   &  Q  \,  (2,0)  \ , \  \Lambda  \,  (1,-1)   \ ;     \\[1mm]
\text{$h^0(K^\frac{q}{p+q})$ chiral multiplets:}  \qquad \qquad  &  \widehat Q  \,  (0,2)  \ , \ \widehat  \Lambda  \,  (-1,1)       \ . 
\end{array}
\eeq
In the above expressions, $K$ is the canonical bundle of $\Sigma_g$
(we are assuming $g \neq 1$, see appendix \ref{app_tensor} for the case $g=1$).
The scalars $\Phi$, $b_0$ are real, while $Q$, $\widehat Q$ are complex.
The spinors $\lambda$, $\psi$, $\Lambda$, $\widehat \Lambda$
are   Weyl spinors of positive chirality.
For each field, we have included its 
$U(1)_1 \times U(1)_2$ charges.
The combination $U(1)_1 + U(1)_2$
is an R-symmetry, while $U(1)_1 - U(1)_2$
is a flavor symmetry.

The multiplicities and the $U(1)_1 \times U(1)_2$ charges
of the free fields listed in \eqref{free_fields_from_six} are such that their
combined 't Hooft anomalies match exactly with the dimensional
reduction on $\Sigma_g$ of the 8-form anomaly polynomial
of a free 6d $\cN = (2,0)$ tensor multiplet, as expected.
Notice that the anomalies only depend on the difference
$h^0(K^\frac{p}{p+q}) - h^0(K^\frac{q}{p+q})$,
which is fixed by the Riemann-Roch theorem,
see appendix \ref{app_tensor}.

How are the free fields in \eqref{free_fields_from_six} identified with
singleton modes on   $\cM_5 \times_w M_6$?
The $g$ vectors $A_\mu$ (we omit the degeneracy label)
are identified with the $g$ singleton 
1-form gauge field associated to the BF coupling
$N \, \widetilde B_2^i \wedge dB_{2i}$. In a similar way,
the real scalar $b_0$ is identified with the singleton 0-form
gauge field associated to the BF coupling
$k \, \cA_1 \wedge dc_3$.
The origin of the other scalar modes and of the fermions
is different. These fields are identified with suitable
Kaluza-Klein modes of 11d supergravity on  $\cM_5 \times_w M_6$,
whose internal wavefunction is such that they are
pure gauge in the bulk of $\cM_5$, but propagate on the conformal
boundary $\partial \cM_5$.
We may refer to these modes as Kaluza-Klein singletons.
They are well-understood 
 for 
the  $AdS_5 \times S^5$  solution in type IIB supergravity 
\cite{Kim:1985ez}.

As we can see from \eqref{free_fields_from_six},
the Kaluza-Klein singletons   $\Phi$, $\lambda$, $\psi$
sit in supermultiplets that contain the BF singletons.
The existence and charges of these Kaluza-Klein
singletons can be easily determined
by counting BF singletons (which are neutral under 
$U(1)_1 \times U(1)_2)$ and using 4d $\cN = 1$ supersymmetry.
In contrast, the chiral multiplets $(Q,\Lambda)$
and $(\widehat Q, \widehat \Lambda)$ do not contain
BF singletons. It follows that to verify the existence, charges,
and multiplicities of these Kaluza-Klein singletons
we cannot rely on a simple counting of BF terms,
and we rather have to perform a direct analysis of the
Kaluza-Klein spectrum.

The situation is different if we specialize   
 to $g\ge 2$,
$q=0$, \emph{i.e.}~the $\cN = 2$ Maldacena-Nu\~nez solution.
The $U(1)_1$ isometry enhances to $SU(2)_1$ and the internal space $M_6$
contains a round $S^2$.
The free fields in \eqref{free_fields_from_six} can be reorganized 
into 4d $\cN = 2$ multiplets,
\beq \label{free_fields_MN}
\begin{array}{rl}
\text{$g$ vector multiplets:} \qquad \qquad &  A_\mu  \,   [\mathbf 1_0]   \ , \ 
 Q \,   [\mathbf 1_2] \ , \
 \lambda' \, [\mathbf 2_1] \ 
;  \\[1mm]
\text{one hypermultiplet:}  \qquad \qquad  &  b_0  \,  [\mathbf 1_0]  \ , \  \widehat Q\,{}'  \,  [\mathbf 3_0]  \ , \  \psi'  \,  [\mathbf 2_{-1}]  \ .    
\end{array}
\eeq
For each field we have indicated its $SU(2)_1$ representation
and $U(1)_2$ charge. The triplet of real scalars $\widehat Q\,{}'$ 
comes from combining the complex scalar $\widehat Q$
and the real scalar $\Phi$ in \eqref{free_fields_from_six},
while the fermion $\lambda'$ comes from   $\lambda$,
$\Lambda$, and the fermion $\psi'$ comes from   $\psi$,
$\widehat \Lambda$.
It is clear from \eqref{free_fields_MN} that in this 4d $\cN = 2$ setup
all Kaluza-Klein singletons are related to BF singletons
by supersymmetry. 
Thus, the existence, charges, and multiplicities
of the Kaluza-Klein singletons can be easily inferred from 
counting BF singletons and exploiting 4d $\cN = 2$ supersymmetry.

If we consider the setup with wrapped M5-branes probing a
$\mathbb Z_2$ singularity, the task at hand it to identify
singleton modes for GMSW solutions on the gravity side.
A subset of these modes is easily identified:
a real 0-form BF singleton $b_0$ and a set of $g$ real 1-form BF singletons
$A_\mu$.
These fields are neutral under the $SU(2)_\varphi$ flavor symmetry
and the $U(1)_\psi$ isometry, which is an R-symmetry.\footnote{The \emph{superconformal}
R-symmetry is the linear combination of $U(1)_\psi$ and the baryonic
$U(1)^2$ symmetry fixed by $a$-maximization \cite{Intriligator:2003jj}.}
Exploiting 4d $\cN = 1$ supersymmetry,
we predict the following multiplets of singleton modes in GMSW,
\beq \label{free_fields_GMSW_easy}
\begin{array}{rl}
\text{$g$ vector multiplets:} \qquad \qquad &  A_\mu  \,  [\mathbf 1_0]  \ , \  \lambda \,   [\mathbf 1_1] \ 
;  \\[1mm]
\text{one chiral multiplet:}  \qquad \qquad  &  b_0  \,  [\mathbf 1_0]  \ , \  \Phi  \,  [\mathbf 1_0]  \ , \  \psi  \,  [\mathbf 1]_{-1}         \ . 
\end{array}
\eeq
We have indicated the $SU(2)_\varphi$ representation and the
$U(1)_\psi$ charge. (All these fields are neutral
under the baryonic $U(1)^2$ symmetry.)
In analogy with \eqref{free_fields_from_six}, we expect 
additional chiral multiplets of Kaluza-Klein singletons,
whose charges and multiplicities cannot be inferred
from the BF terms alone.
To identify these chiral multiplets, we need to perform
a more detailed study of the Kaluza-Klein spectrum
of GMSW solutions. We plan to address this problem in future work.

It should be stressed, however, that
we do not expect Kaluza-Klein singletons
to exhaust the entire set of singleton modes
for these geometries. This expectation is based
on analogy with D3-brane setups in type II string theory.
As pointed out in \cite{Maldacena:2001ss}, 
if we consider type IIB supergravity on $AdS_5 \times T^{1,1}$,
we only see one vector BF singleton (coming from the 
term $B_2\wedge dC_2$, where $B_2$ in the NSNS 2-form
and $C_2$ is the RR 2-form). On the other hand,
the worldvolume theory on the D3-branes is a $U(N) \times U(N)$
quiver theory (with superpotential)
in which the two $U(1)$'s in $U(N) \times U(N)$ decouple in the IR.
The overall $U(1)$ is identified with the BF vector,
but the relative $U(1)$ does not appear to have an obvious
singleton interpretation within the supergravity approximation.
The geometry $AdS_5 \times T^{1,1}$ can be regarded
as originating from blow-up of a $\mathbb Z_2$ orbifold singularity.
This feature is qualitatively similar to our interpretation of the smooth
GMSW solutions in terms of a blow-up of the $\mathbb C^2/\mathbb Z_2$
singularities at the north and south poles of $S^4$. For this reason, we expect that, in order to capture
all decoupling modes on the worldvolume theory of the M5-brane stack,
one has to analyze singleton modes beyond the supergravity approximation,
including stringy modes. This program could give an exact answer for the decoupled modes,
which, combined with inflow, 
would yield the exact anomaly of the interacting SCFT,
including $\cO(1)$ terms.



\section{Outlook} \label{sec_conclusion}

The problems studied in this work suggest several  
directions for future research. For instance,
a systematic
analysis of topological boundary conditions
for the BF theory $N \, \widetilde B_{2}^i \wedge dB_{2i}$,
including the  action of the duality group
$Sp(2g,\mathbb Z)$,
has not been performed. Such a study
has the potential of furnishing an organizing principle
for the classification of line operators in 4d SCFTs engineered
with M5-branes, with either $\cN = 2$ or $\cN = 1$ 
supersymmetry.

Another problem that deserves further analysis
is the computation of the partition function vector
of the singleton modes in a setup with wrapped M5-branes.
The full partition function from the gravity side
is expected to take the   form
\eqref{Z_factorization},  where the vector
$Z^\beta$ encodes the contribution of interacting bulk modes
(and is dual to the partition function vector of
an interacting SCFT),
while $Z_\beta^{\rm singleton}$ is the partition
function vector of singleton modes.
The latter is computable following
the methods of \cite{Moore:2004jv,Belov:2004ht}.
The action of  $Sp(2g,\mathbb Z)$ 
on the conformal blocks
$Z^\beta$ of the interacting SCFT
can be determined from its action on 
$Z_\beta^{\rm singleton}$.
One may then explore the interplay between
the duality group and 't Hooft anomalies for various global symmetries.

We have observed that BF couplings in the bulk 5d topological
theory account for a set of singleton modes on the gravity side.
On the other hand, additional singleton modes are present, which do
not originate from BF terms. In general $\cN = 1$ setups,
supersymmetry is not sufficient to determine all singleton
modes starting from BF singleton modes.
It would be beneficial to perform a systematic study
of singleton modes in string/M-theory compactifications,
especially in setups with lower amounts of supersymmetry. 
On the basis of the holographic dictionary,
it is expected that singleton modes on the gravity side
should account for all modes that decouple in the IR on the field
theory side.
 A detailed knowledge of 
decoupling modes 
can provide access to precision holography, allowing for
example for a computation of exact anomalies, beyond
the large-$N$ limit, including $\cO(1)$ terms.
The role of
singleton modes in holographic flows is also worth
analyzing further. 

It is natural to wonder how the results of this paper would be 
modified by the inclusion of punctures on the Riemann surface.
In order to address this question
in a more systematic way, a better understanding
of punctures for 4d $\cN = 1$ theories engineered with M5-branes
would be useful.  With regards
to   $\cN = 1$ regular punctures for 6d $(2,0)$ theories of type $A_{N-1}$
on a Riemann surface, our expectation
is that there should be no    mixed anomalies
between the continuous 0-form flavor symmetries at the punctures
and the discrete and higher-form symmetries 
of the system. On the other hand,
we anticipate a much richer structure in setups
with M5-branes probing a $\mathbb Z_2$ singularity
and wrapped on the Riemann surface with punctures.

Finally, it would be useful to extend the discussion
of symmetries and anomalies in geometric engineering, including
other possible sources of internal discrete symmetries 
(such as discrete isometries of the internal space
or torsion cycles \cite{Camara:2011jg,BerasaluceGonzalez:2012vb}),
as well as spacetime discrete symmetries
(such as parity or time reversal).


\section*{Acknowledgments}

We would like to thank
Anindya Dey,
Enoch Leung,
Gregory Moore,
Emily Nardoni,
Sakura Sch\"afer-Nameki,
Thomas Waddleton, and
Peter Weck
for interesting conversations and correspondence. 
The work of IB and FB is supported in part by NSF grant PHY-1820784. RM is supported in part by ERC Grant 787320-QBH Structure
and by ERC Grant 772408-Stringlandscape.


\appendix


\section{A change of basis} \label{sec_goodbasis}

In this appendix we consider an internal space $M_6$ with
$n:= b^2(M_6) \ge 2$. The lattice $H^2(M_6,\mathbb Z)_{\rm free}$
is preserved by the action of $SL(n,\mathbb Z)$.
In terms of the closed 2-forms $\omega_{2\alpha}$, we can 
consider a change of basis of the form
\beq \label{new_omega_basis}
\omega_{2\alpha}' = \omega_{2\beta} \, (M^{-1})^\beta {}_\alpha \ , \qquad
M \in SL(n,\mathbb Z)   \ .
\eeq
This linear transformation is   accompanied by
  transformations on the closed 4-forms
$\Omega_4^\alpha$, as well as on the basis $\cC_2^\alpha$ of 2-cycles in
$M_6$, and the basis $\cC^4_{\alpha}$ of 4-cycles in $M_6$.
In order to  preserve the relations
$\int_{M_6} \omega_{2\alpha} \wedge \Omega_4^\beta = \delta^\beta_\alpha$,
$\int_{\cC_2^\alpha} \omega_{2\beta} = \delta^\alpha_\beta$,
and $\int_{\cC^4_\alpha} \Omega_4^\beta = \delta^\beta_\alpha$,
we must set
\begin{align} \label{other_rotations}
\Omega_4'^\alpha = M^\alpha{}_\beta \, \Omega_4^\beta \ , \qquad
\cC_2' {}^ \alpha= M^\alpha{}_\beta \, \cC_2^\beta \ , \qquad
\cC^4_\alpha{}' = \cC^4_\beta \, (M^{-1})^\beta {}_\alpha \ .
\end{align}
The ansatz \eqref{G4_ansatz} for $G_4$ contains the terms
$A_1^\alpha \wedge \omega_{2\alpha}$ and $N_\alpha \, \Omega_4^\alpha$.
As a result, the linear transformation \eqref{new_omega_basis}
induces a linear transformation 
for the external 1-form gauge fields and flux quanta,
\beq
A_1'{}^\alpha = M^\alpha{}_\beta \, A_1^\beta  \ , \qquad
N'_\alpha = N_\beta \, (M^{-1})^\beta {}_\alpha \ .
\eeq 

To proceed, we define the integers $k$ and $m_\alpha$ via the relations
\beq \label{k_and_m_def}
k = {\rm gcd}(N_1, \dots, N_n) \ , \qquad N_\alpha = k \, m_\alpha \ .
\eeq
It can then be shown that a matrix $M \in SL(n,\mathbb Z)$ exists,
such that
\beq
A_1'{}^{\alpha = 1} =  m_\beta \, A_1^\beta \ .
\eeq
This can be argued as follows. The integers $\{ m_\alpha\}_{\alpha = 1}^n$
are relatively prime.
There must exist labels $\alpha_1, \alpha_2 \in \{1 , \dots, n\}$,
$\alpha_1 \neq \alpha_2$, such that
$m_{\alpha_1}$ and $m_{\alpha_2}$ are relatively prime.
After reordering $\{ m_\alpha\}_{\alpha = 1}^n$ if necessary,
we can take $\alpha_1 = 1$, $\alpha_2 = 2$.
We may then 
consider the following matrix,
\beq{ \small
M^{\alpha}{}_\beta = {  
 \begin{pmatrix}
m_1 & m_2 & m_3 & m_4 & \dots & m_{n-1} & m_n \\
r & s & 0 & 0 & \dots & 0 & 0 \\
0 & 0 & 1 & 0 & \dots & 0 &  0  \\
0 & 0 & 0 & 1 & \dots & 0 & 0  \\
\vdots & \vdots & \vdots & \vdots  && \vdots & \vdots\\
0 & 0 & 0 & 0 & \dots & 1 & 0  \\
0 & 0 & 0 & 0 & \dots & 0 & 1 
\end{pmatrix}
} }\ , \qquad \det M = s \, m_1 - r \, m_2  \ .
\eeq
Since $m_1$ and $m_2$ are relatively prime,
there exist integers $r$ and $s$ satisfying 
$s \, m_1 - r \, m_2 = 1$. 
This follows from B\'ezout's identity in elementary number theory.
Thus $M\in SL(n,\mathbb Z)$
and clearly $A_1'{}^{\alpha = 1} =  m_\beta \, A_1^\beta$.

In the new basis $N'_\alpha$ of flux quanta,
the component $N'_{\alpha = 1} = k$ is the only non-zero component,
and we have
\beq
N_\alpha \, dA_1^\alpha = k \, dA'_1{}^{\alpha = 1}   \ .
\eeq
This is why this basis is best suited to study the topological
terms in the 5d  action.



\section{Cohomology classes and gauging of isometries}

In this appendix we study   non-trivial cohomology classes 
of the internal space $M_6$ for BBBW solutions and GMSW solutions.
To compute the full inflow anomaly polynomial,
we need to activate background gauge fields for the 
isometries of $M_6$. After turning on these gauge fields,
the relevant spacetime is denoted $M_{12}$ and is of the form
\beq \label{myM12}
M_6 \hookrightarrow M_{12} \rightarrow \cM_6 \ ,
\eeq
where $\cM_6$ denotes external spacetime.
For the purposes of computing anomalies in the descent formalism,
we take $X_6$ to be Euclidean and six-dimensional.
We discuss representatives for cohomology classes in $M_6$
and their counterparts in $M_{12}$.  We suppress wedge products throughout this appendix.

\subsection{BBBW solutions} \label{app_BBBW}

\subsubsection{Cohomology classes in $M_6$}

\begin{figure}
\centering
\includegraphics[width = 8 cm]{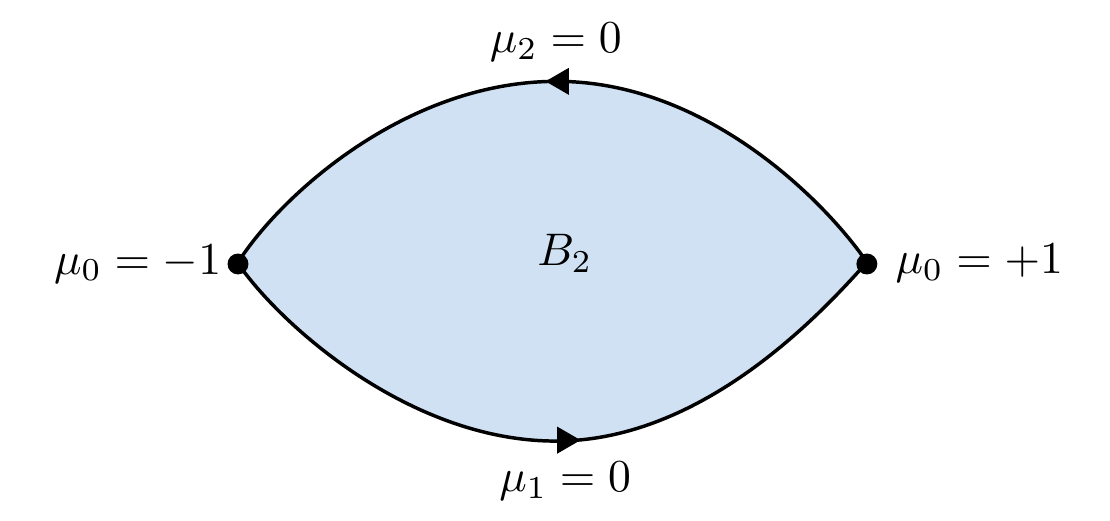}
\caption{Schematic depiction of the space 
$B_2$ described by the constrained coordinates \eqref{mu_coords}
with line element $d\mu_0^2 + d\mu_1^2 + d\mu_2^2$. We have also indicated the
orientation of $\partial B_2$ used throughout this appendix.
}
\label{eye}
\end{figure}

In the BBBW solutions, the internal
space $M_6$ is topologically an $S^4$ fibration over a genus-$g$
Riemann surface $\Sigma_g$.
We refer   to \cite{Bah:2012dg} for the full
expression of the metric on $M_6$.
For the purposes of this work, 
we can use the following simplified line element
on $M_6$, which captures the topology and isometries
of the    metric in \cite{Bah:2012dg},
\begin{align} \label{simple_BBBW}
ds^2(M_6)  &= ds^2(\Sigma_g) 
+ d\mu_0^2 + d\mu_1^2 + d\mu_2^2 
+ \mu_1^2 \, D\phi_1^2 + \mu_2^2 \, D\phi_2^2 \ ,   \\
dD\phi_1 &=  - p  \, V_\Sigma \ ,\qquad
dD\phi_2  =  - q \, V_\Sigma \ , \qquad \int_{\Sigma_g} V_\Sigma = 2\pi \ , \qquad
p+q = - \chi = 2(g-1) \ .   \nn
\end{align}
The angles $\phi_1$, $\phi_2$ have periodicity $2\pi$
and $ds^2(\Sigma_g)$ denotes the 
metric on $\Sigma_g$ with constant curvature $\kappa \in \{0,1,  -1\}$.
We have also 
introduced three constrained coordinates $\mu_0$, $\mu_1$,
$\mu_2$, satisfying
\beq \label{mu_coords}
\mu_0^2 + \mu_1^2 + \mu_2^2 = 1 \ , \qquad
-1 \le \mu_0 \le 1 \ , \qquad 0 \le \mu_1 \le 1
 \ , \qquad 0 \le \mu_2 \le 1 \ .
\eeq
We use $B_2$ to denote the 2d space parametrized
by $\mu_0$, $\mu_1$, $\mu_2$.
The total space of the $S^1_{\phi_1} \times S^1_{\phi_2}$
fibration over $B_2$ is an $S^4$.
We refer to the points   $\mu_0 = \pm1$
as the north and south poles of $S^4$, respectively.
The space $B_2$ is depicted schematically in figure \ref{eye}.

The space $M_6$ admits one   integral 4-homology class.
A representative is obtained by considering the 4-cycle $\cC^4$
that is obtained by taking the $S^4$ fiber
on top of a generic point on the Riemann surface.
Correspondingly, there is one non-trivial 4-cohomology class on $M_6$.
We represent it by a closed but not exact 4-form $\omega_4$,
which integrates to 1 on the $\cC^4$ cycle,
\beq \label{omega4_flux}
\int_{\cC^4} \Omega_4 = 1 \  .
\eeq
The closed 4-form $\omega_4$ is parametrized as follows,
\beq
\Omega_4    = h_2 \,   \frac{D\phi_1}{2\pi} \, \frac{D\phi_2}{2\pi}
+ h_1^1 \, \frac{D\phi_1}{2\pi} \, \frac{dD\phi_2}{2\pi} 
+ h_1^2 \, \frac{D\phi_2}{2\pi} \, \frac{dD\phi_1}{2\pi} \ .
\eeq
In the above expression, $h_2$ is a 2-form on $B_2$,
while $h_1^{1,2}$ are 1-forms on $B_2$. Closure of $\omega_4$ requires
\beq \label{h2_is_exact}
h_2 = dh_1^1 = - dh_1^2 \ .
\eeq
It follows that $d(h_1^1 + h_1^2) = 0$ and therefore 
(since $B_2$ has trivial cohomology) there exists
a 0-form $h_0$ such that
\beq \label{h0_def}
h_1^1 + h_1^2 = - dh_0 \ .
\eeq
Without loss of generality, we can take $h_0$ to satisfy
\beq \label{h0_is_odd}
(h_0)^\mathrm N = - (h_0)^\mathrm S   \ ,
\eeq
where the superscript N, S means evaluation at $\mu_0 = \pm 1$, respectively.
The 1-forms $h_1^1$, $h_1^2$ satisfy additional requirements
that ensure the regularity of $\omega_4$. In particular,
we must demand that $h_1^1$ be zero if restricted to the locus
$\{ \mu_1 = 0\} \subset B_2$, and similarly for $h_1^2$,
\beq \label{h1regularity}
h_1^1  |_{\mu_1 = 0} = 0 \ , \qquad h_1^2 |_{\mu_2 = 0} = 0 \ .
\eeq
This is due to the fact that $S^1_{\phi_1}$ shrinks along $\mu_1 = 0$,
and similarly for $S^1_{\phi_2}$.
We also have to impose \eqref{omega4_flux},
\begin{align} \label{computation}
1 & = \int_{B_2} h_2 = \int_{\partial B_2} h_1^1 = \int_{\{ \mu_1 = 0\}} h_1^1
- \int_{\{\mu_2 = 0\}} h_1^2
=   \int_{\{\mu_2 = 0\}} dh_0  = (h_0)^\mathrm N - (h_0)^\mathrm S \ .
\end{align}
In the first step we have integrated over $\phi_1$, $\phi_2$.
In the second step we used \eqref{h2_is_exact}. The boundary $\partial B_2$
consists of the arcs $\{ \mu_1 = 0\}$ and $\{ \mu_2 = 0\}$,
with a relative minus due to orientation, see figure \ref{eye}.
We conclude recalling \eqref{h1regularity} and \eqref{h0_def}.
We learn that $ (h_0)^\mathrm N - (h_0)^\mathrm S=1$,
which, combined with \eqref{h0_is_odd}, yields
\beq \label{h0_values}
(h_0)^\mathrm N = \frac 12 \ , \qquad (h_0)^\mathrm S = - \frac 12  \ .
\eeq

Poincar\'e duality implies that the space $M_6$ admits 
one non-trivial 2-cohomology class. It can be represented by a closed
but not exact 2-form $\omega_2$, normalized in such a way that  
\beq \label{Poincare_duality}
\int_{M_6} \Omega_4  \, \omega_2 = 1 \ .
\eeq
We parametrize $\omega_2$ as
\beq
\omega_2 = d\bigg[ H_0^1 \, \frac{D\phi_1}{2\pi}
+ H_0^2 \, \frac{D\phi_2}{2\pi} \bigg]  
=  dH_0^1 \, \frac{D\phi_1}{2\pi}
+ dH_0^2 \, \frac{D\phi_2}{2\pi} 
-(  p  \, H_0^1 + q  \, H_0^2  ) \, \frac{V_\Sigma}{2\pi}  \ ,
\eeq
where $H_0^1$, $H_0^2$ are 0-forms on $B_2$.
We might have included an additional term
$\Delta \omega_2 = (\text{const}) \, V_\Sigma$,
but such term can always be reabsorbed by redefining $H_0^1$ or $H_0^2$.
(We do not consider the case $p = 0  =q$ because 
it does not correspond to  
a smooth M-theory solution.)
The 0-forms $H_0^1$, $H_0^2$ satisfy additional
requirements that ensure regularity of $\omega_2$.
In particular, regularity the term $dH_0^1 \wedge D\phi_1$
implies that the function $H_0^1$ restricted to $\{\mu_1 = 0\}$
must be a constant. Similarly, $H_0^2$ restricted to $\{\mu_2 = 0\}$
must be a constant,
\beq \label{H_regularity}
d( H_0^1 |_{\mu_1 = 0}) = 0 \ , \qquad
d( H_0^2 |_{\mu_2 = 0}) = 0  \ .
\eeq
 Since we can connect the north and south
poles with either the arc $\{\mu_1 = 0\}$ or $\{\mu_2 = 0\}$,
we conclude that the values of $H_0^1$, $H_0^2$ at the poles are equal,
\beq \label{north_equals_south}
(H_0^1)^\mathrm N = (H_0^1)^\mathrm S \ , \qquad
(H_0^2)^\mathrm N = (H_0^2)^\mathrm S \ .
\eeq 
Next, let us consider the integral 
 in \eqref{Poincare_duality}. It can be computed with manipulations
similar to those in \eqref{computation}. The result is
\beq
\int_{M_6} \Omega_4 \, \omega_2 = - \Big[ p \, 
(H_0^1)^\mathrm N
+ q \, (H_0^2)^\mathrm N
 \Big] \, \Big[ (h_0)^\mathrm N - (h_0)^{\mathrm S} \Big] \ .
\eeq
Using \eqref{h0_values}, we conclude that, in order to have \eqref{Poincare_duality}, the values
$(H_0^1)^\mathrm N$, $(H_0^2)^\mathrm N$
must satisfy
\beq \label{H_values}
p \, 
(H_0^1)^\mathrm N
+ q \, (H_0^2)^\mathrm N  = -1 \ .
\eeq

Next, let us discuss 1-cohomology classes on $M_6$.
The Riemann surface admits $2g$ independent non-trivial 1-cohomology classes,
which are represented by closed but not exact 1-forms $\lambda_{1u}$, $u = 1,\dots,2g$.
These 1-forms can be pulled back to the total space $M_6$,
yielding 1-forms that we still denote $\lambda_{1u}$ and that are still closed.
It can be checked that they are not exact in $M_6$, and that they furnish representatives
for all 1-cohomology classes of $M_6$.
The associated 1-cycles on $M_6$ are realized by taking a 1-cycle on the Riemann
surface, at the point $\mu_0 = 1$, where both $S^1_{\phi_1}$
and $S^1_{\phi_2}$ shrink. (Choosing $\mu_0 = -1$ yields 1-cycles in $M_6$ that are
homologous to those at $\mu_0 = 1$.)
Finally, the space $M_6$ admits no 3-cohomology class and no 3-cycles.

\subsubsection{Inclusion of background gauge fields for isometries} \label{BBBW_isom_app}
The background gauge fields for the $U(1)_1 \times U(1)_2$ isometry
are denoted $A^{\phi_1}$, $A^{\phi_2}$, with field strenghts
$F^{\phi_1} = dA^{\phi_1}$, $F^{\phi_2} = dA^{\phi_2}$.
After turning on $A^{\phi_1}$, $A^{\phi_2}$
we introduce the 1-forms
\beq \label{gauging_BBBW_angles}
\widetilde D\phi_i = D\phi_i - 2 \, A^{\phi_i} \ , \qquad 
d \widetilde D\phi_1 = - p \, V_\Sigma  - 2 \, F^{\phi_1} \ , \qquad
d\widetilde D\phi_2 = - q \, V_\Sigma  - 2 \, F^{\phi_2} \ .
\eeq
The periods of the field strengths $F^{\phi_1}$, $F^{\phi_2}$
are quantized in units of $2\pi$ and we have\footnote{The
normalization of $A_{\phi_1}$, $A_{\phi_2}$ in 
\eqref{gauging_BBBW_angles} can be checked by matching anomaly inflow
with the know anomaly polynomial of the 4d SCFT.} 
\beq
c_1^{\phi_1} = c_1(U(1)_1) = \frac{F^{\phi_1}}{2\pi} \ , \qquad
c_1^{\phi_2} = c_1(U(1)_2) = \frac{F^{\phi_2}}{2\pi}  \ .
\eeq
In the case $g=0$, the Riemann surface is a round 2-sphere and the space $M_6$ admits an additional $SO(3)_\Sigma \cong SU(2)_\Sigma$
isometry. We find it convenient to describe the 2-sphere as
the locus $y^a \, y_a = 1$ in $\mathbb R^3$, where  $a = 1,2,3$ is a vector index
of $SO(3)_\Sigma$. One can verify that the following 1-forms on $M_6$
are dual to Killing vectors in the metric \eqref{simple_BBBW},
\beq
\epsilon_{abc} \, y^b \, dy^c + \frac 12 \, p \, y_a \, \mu_1^2 \, D\phi_1 
+ \frac 12 \, q \, y_a \, \mu_2^2 \, D\phi_2 \ . 
\eeq
In other words, the $SO(3)_\Sigma$ isometry of the Riemann
surface extends to an isometry of the total space $M_6$ for any
value of $p$, $q$. We couple the $SO(3)_\Sigma$ isometry to a triplet
$A^a$ of external gauge fields. Our conventions are
\beq  \label{Sigma_gauging}
Dy^a = dy^a + \frac 12 \, \epsilon^{abc} \, A_b \, y_c \ , \qquad
F^a = dA^a + \frac 12 \, \epsilon^{abc} \, A_b  \, A_c \ .
\eeq
After turning on $A^a$, the volume form $V_\Sigma$ on the Riemann surface
in \eqref{gauging_BBBW_angles}
must be replaced with the global angular form of $SO(3)_\Sigma$,
\beq
\frac{V_\Sigma}{2\pi} \rightarrow e_2^{\Sigma} =  
\frac{1}{8\pi}  \, \bigg[
 \epsilon_{  a   b   c} \, D   y^{  a} 
\, D  y^{  b} \,   y^{  c} - 2 \,   F_{  a} \,   y^{  a} 
\bigg] 
 \ .
\eeq
The 2-form $e_2^{\Sigma}$ is the closed and $SO(3)_\Sigma$-invariant
completion of $V_\Sigma/(2\pi)$. 
Integrals of powers of $e_2^\Sigma$ on $S^2$ are computed via the Bott-Cattaneo
formula \cite{bott1999integral},
\beq \label{BottCatt}
\int_{S^2} (e_2^\Sigma)^{2s+1} = 2^{-2s} \, \Big[ p_1(SO(3)_\Sigma)\Big]^s \ , \qquad
\int_{S^2} (e_2^\Sigma)^{2s} =0 \ , \qquad s = 0,1,2,\dots
\eeq
and in particular $\int_{S^2} e_2^\Sigma = 1$ in our normalization.
In writing the inflow anomaly polynomial below, we find it convenient to
replace $p_1(SO(3))_\Sigma$ with the second Chern class of $SU(2)_\Sigma$,
according to
\beq \label{p1_to_c2}
p_1(SO(3)_\Sigma) = - 4 \, c_2(SU(2)_\Sigma) \equiv - 4 \, c_2^\Sigma  \ .
\eeq

After activating the gauge fields for isometries of $M_6$
we are effectively considering the auxiliary 12 space $M_{12}$ in \eqref{myM12}.
We have to discuss how the closed forms $\omega_4$, $\omega_2$, $\lambda_{1u}$
extend to closed forms $(\omega_4)^{\rm eq}$, $(\omega_2)^{\rm eq}$, $(\lambda_{1u})^{\rm eq}$
on $M_{12}$.
We start by noting that
the 1-forms $\lambda_{1u}$ are unaffected by 
the gauging of the isometry $U(1)_1 \times U(1)_2$. Since they are only present for
$g\neq0$, the $SO(3)_\Sigma$ isometry plays no role.\footnote{One might
wonder if, in the case $g=1$, the isometries of the $T^2$ base
extend to isometries of $M_6$. We have checked that, contrary to the $g=0$ case,
for $p \neq 0$ one cannot find globally-defined Killing vector fields on $M_6$
that reduce to the Killing vectors on the base $T^2$.} We conclude that
we do not need to modify $\lambda_{1u}$ in any way after 
gauging the isometries of $M_6$,
\beq
(\lambda_{1u})^{\rm eq} = \lambda_{1u} \ .
\eeq

Next, let us consider the 2-form $(\omega_2)^{\rm eq}$. It can be written as
\beq \label{omega2EQ}
(\omega_2)^{\rm eq} = d\bigg[ H_0^1 \, \frac{\widetilde D\phi_1}{2\pi}
+ H_0^2 \, \frac{\widetilde D\phi_2}{2\pi} \bigg] 
+ 2 \, (H_0^1)^\mathrm N \, \frac{F^{\phi_1}}{2\pi} 
+ 2 \, (H_0^2)^\mathrm N \,   \frac{F^{\phi_2}}{2\pi} \ .
\eeq
This is manifestly closed and gauge-invariant. It is also globally defined,
and reduces to $\omega_2$ if all background gauge fields for isometries of $M_6$
are turned off. 
The 2-form $(\omega_2)^{\rm eq}$ in $M_{12}$ should have integral periods.
In particular, we may consider a 2-cycle in external spacetime,
sitting at $\mu_0 = \pm 1$ and a generic point on the Riemann surface in $M_6$. We have defined $(\omega_2)^{\rm eq}$ 
in such a way that its integral
over such cycles is zero. Indeed, the relevant terms are
\beq
(\omega_2)^{\rm eq} = 
2 \,\Big[  (H_0^1)^\mathrm N - H_0^1 \Big] \, \frac{F^{\phi_1}}{2\pi}
+ 2 \,\Big[  (H_0^2)^\mathrm N - H_0^2 \Big] \, \frac{F^{\phi_2}}{2\pi} + \dots
\eeq
and the relation 
\eqref{north_equals_south}
implies that the prefactors of $F^{\phi_i}$ vanish both at $\mu_0 = 1$
and $\mu_0 = -1$.

Finally, let us turn to $(\omega_4)^{\rm eq}$. We parametrize it as
\beq \label{Omega4EQ}
(\Omega_4)^{\rm eq}    = h_2 \,   \frac{\widetilde D\phi_1}{2\pi} \, \frac{\widetilde D\phi_2}{2\pi}
+ h_1^1 \, \frac{\widetilde D\phi_1}{2\pi} \, \frac{d \widetilde D\phi_2}{2\pi} 
+ h_1^2 \, \frac{\widetilde D\phi_2}{2\pi} \, \frac{d\widetilde D\phi_1}{2\pi}
- h_0  \, \frac{d\widetilde D\phi_1}{2\pi} \, \frac{d\widetilde D\phi_2}{2\pi}
 \ .
\eeq
Closure of $(\omega_4)^{\rm eq}$ follows from the relations 
\eqref{h2_is_exact}, \eqref{h0_def}. Moreover
$(\omega_4)^{\rm eq}$ is globally defined and has integral periods in $M_{12}$.

We conclude with two remarks. First, in the case $g=0$ the background
gauge fields for $SO(3)_\Sigma$ are implicitly included in 
\eqref{omega2EQ} and \eqref{Omega4EQ} inside the 1-forms $\widetilde D \phi_{i}$.
Second, the forms $(\omega_2)^{\rm eq}$, $(\omega_4)^{\rm eq}$ 
are not the only possible choices of a closed and gauge-invariant
completion of $\omega_2$, $\omega_4$. As argued in appendix \ref{app_E4},
however, any other choice leads to equivalent results for the
inflow anomaly polynomial.
 
\subsubsection{Computation of the inflow anomaly polynomial}
Our first task is the computation of $- \tfrac 16 \, \int_{M_6} E_4^3$,
where $E_4$ is given by
\beq
E_4 = N \, (\Omega_4)^{\rm eq} + \frac{F_2}{2\pi} \, (\omega_2)^{\rm eq}
+ \frac{H_3^u}{2\pi} \, (\lambda_{1u})^{\rm eq} + \frac{\gamma_4}{2\pi} \ .
\eeq
To compute the integral $\int_{M_6} E_4^3$, we first collect terms with
exactly one $\widetilde D\phi_1$ and one $\widetilde D\phi_2$ factor.
The integral over the Riemann surface for $g \neq 0$ is straightforward;
in the case $g=0$, we perform it with the help of the Bott-Cattaneo formula
\eqref{BottCatt}. We are left with an integral over $B_2$. It is performed
in a similar way as in \eqref{computation}. More precisely, the integrand 2-form is 
cast as a total derivative of a 1-form on $B_2$. Applying Stokes' theorem,
we reduce the problem to an integral over the two arcs 
$\{ \mu_1 = 0$\}, $\{ \mu_2 = 0\}$, with a relative minus sign.
In the computation, we make use of \eqref{h2_is_exact}, 
\eqref{h0_def}, \eqref{h1regularity}, \eqref{h0_values},
\eqref{H_regularity}, \eqref{north_equals_south}, and \eqref{H_values}.
The result reads
\begin{align} \label{noX8}
- \frac 16 \, \int_{M_6} E_4^3 & = 
 - \frac 23 \,  N^3  \, 
\bigg[ p \, c_1^{\phi_1} 
\, (c_1^{\phi_2})^2
+ q \, c_1^{\phi_2} \, (c_1^{\phi_1})^2
\bigg] 
+ \frac{1}{6} \, \bigg[
N^3 \, q^2   \, p \, c_1^{\phi_1}
+ N^3 \, p^2  \, q \, c_1^{\phi_2}
 \bigg] c_2^\Sigma
\nn \\
& + \frac{1}{(2\pi)^2} \bigg[  -  N  \, \gamma_4 \, F_2
-  N \,  \widetilde H_3^i \, H_{3i} \bigg] \ .    
\end{align}
We have used the relation \eqref{p1_to_c2} and we have written
  final expression in the notation   introduced in the main text.
  
To proceed, we need the 8-form class $X_8$, defined in \eqref{Mtheory_action_top}.
The computation of $X_8$ for BBBW setups is reviewed \emph{e.g.}~in \cite{Bah:2019rgq}.
One finds
\begin{align} \label{withX8}
- \int_{M_6} E_4 \, X_8 & = \frac 16 \, N \, 
\bigg[ p \, c_1^{\phi_1} 
\, (c_1^{\phi_2})^2
+ q \, c_1^{\phi_2} \, (c_1^{\phi_1})^2
\bigg]
- \frac 16 \, N \, \bigg[  p \, (c_1^{\phi_1})^3 + q \, (c_1^{\phi_2})^3 \bigg]
\nn \\
&+ \frac{1}{24} \, N \, \bigg[ p \, c_1^{\phi_1}
+ q \, c_1^{\phi_2}  \bigg] \, p_1(T)
- \frac{1}{6} \, N \, \bigg[
 p \, c_1^{\phi_1}
+  q \, c_1^{\phi_2}
 \bigg] c_2^\Sigma
 \ .    
\end{align}
Combining \eqref{noX8} and \eqref{withX8} we get the result
\eqref{BBBW_result} quoted in the main text.

\subsection{GMSW solutions} \label{app_GMSW}

\subsubsection{Cohomology classes in $M_6$}

The exact line element 
on $M_6$ in recorded in 
 \cite{Gauntlett:2004zh}.   For the purposes of this appendix,
we can use the schematic line element in \eqref{schematic_GMSW}
without a detailed knowledge of the $f$ functions.

We can define the following 4-cycles in $M_6$,
\beq
\begin{array}{llll}
\cC^4_{\rm N} & :   &\qquad \qquad  
& \text{$S^2_\varphi \times \Sigma_g$ at $\mu = \mu_\mathrm N$} \ , \\
\cC^4_{\rm S}  & : &\qquad \qquad
& \text{$S^2_\varphi \times \Sigma_g$ at $\mu = \mu_\mathrm S$}  \ ,   \\
\cC^4_{\rm C}  & :&\qquad\qquad  
& \text{$S^2_\varphi \times S^2_\psi$ at a point on $\Sigma_g$} \ , \\
 \cC^4_{\Sigma}  & :&\qquad\qquad  
& \text{$\Sigma_g \times S^2_\psi$ at a point on $S^2_\varphi$} \ . 
\end{array}
\eeq
We recall that $S^2_\psi$ is the two-dimensional space
spanned by the angle $\psi$ and the $\mu$ interval,
with topology of a two-sphere and isometry $U(1)_\psi$.
The 4-cycles $\cC^4_{\mathrm N}$,  $\cC^4_{\mathrm S}$,
$\cC^4_{\mathrm C}$, $\cC^4_{\Sigma}$     
define elements in the integral 4-homology of $M_6$.
They are not all independent, however, since
the following relation holds in homology,
\beq \label{4cycle_combo}
\cC^4_\mathrm N - \cC^4_\mathrm S + \chi \, \cC^4_\mathrm C
+ 2 \, \cC^4_\Sigma = 0 \ .
\eeq
The above can be verified by checking that the linear
combination of 4-cycles on the LHS yields integral zero
when paired with an
arbitrary closed 4-form on $M_6$.
The relation \eqref{4cycle_combo} implies
that $\cC^4_\mathrm N - \cC^4_\mathrm S$
represents an \emph{even} 4-homology class.
(The Euler characteristic $\chi$ is always an even integer.)
The class $\cC^4_\mathrm N + \cC^4_\mathrm S
= (\cC^4_\mathrm N - \cC^4_\mathrm S) + 2\, \cC^4_\mathrm S$
is therefore also an even class.
This observation allows us to choose the following basis of
integral 4-homology,
\beq
\cC^4_{\alpha = 1} = \cC^4_\mathrm C \ , \qquad
\cC^4_{\alpha = 2} = \frac 12 \, (  \cC^4_\mathrm N + \cC^4_\mathrm S ) \ , \qquad
\cC^4_{\alpha = 3} = \frac 12 \, (  \cC^4_\mathrm N - \cC^4_\mathrm S ) \ .
\eeq
We can check that $\cC^4_\mathrm N$, $\cC^4_\mathrm S$,
$\cC^4_\mathrm C$, and $\cC^4_\Sigma$ can all
be written as linear combinations of $\cC^4_\alpha$ with \emph{integer} 
coefficients.
 
To each 4-homology class $\cC^4_\alpha$ there is a corresponding
4-cohomology class, which can be represented by a closed but not exact
4-form $\Omega_4^\alpha$.
The 4-forms $\Omega_4^\alpha$ are parametrized as
\beq
\Omega_4^\alpha = \bigg(dU_{\varphi}^\alpha \, \frac{ V_{\varphi} }{2\pi}
+ dU_\Sigma^\alpha \, \frac{ V_\Sigma }{2\pi} \bigg) \, \frac{D\psi}{2\pi}
+ \bigg( C^\alpha - 2 \, U^\alpha_\Sigma - \chi \, U^\alpha_{\varphi}\bigg) 
\frac{ V_{\varphi} }{2\pi} \,  \frac{ V_\Sigma }{2\pi}   \ ,
\eeq
where $C^\alpha$ is constant and $U_{\varphi}$, $U_\Sigma$ are functions of $\mu$ only, satisfying
\beq
U_{\varphi}^\alpha(\mu_\mathrm N) + U_{\varphi}^\alpha(\mu_\mathrm S) = 0 \ , \qquad
U_{\Sigma}^\alpha(\mu_\mathrm N) + U_{\Sigma}^\alpha(\mu_\mathrm S) = 0 \ .
\eeq
We demand  the standard pairing $\int_{\cC^4_\alpha} \Omega_4^\beta = \delta^\beta_\alpha$
between 4-homology and 4-cohomology classes.
This   fixes the values of $U_{\varphi}(\mu_\mathrm N)$,
$U_{\Sigma}(\mu_\mathrm N)$, $C^\alpha$,
\beq \label{4form_table}
\begin{array}{c  ||  c | c | c }
&  \;\;\;\; \phantom{ \big|} U_{\varphi}^\alpha(\mu_\mathrm N) \phantom{ \big|} \;\;\;\; & 
\;\; \;\;\phantom{ \big|} U_{\Sigma}^\alpha (\mu_\mathrm N)\phantom{ \big|}  \;\;\;\; &
\;\;\;\;  \phantom{ \big|} C^\alpha\phantom{ \big|}   \;\;\;\; \\ \hline
\;\;\;\; \phantom{ \big|} \alpha = 1\phantom{ \big|}  \;\;\;\; & 1/2 & -\chi/4 & 0  \\
\;\;\;\; \phantom{ \big|} \alpha = 2\phantom{ \big|}  \;\;\;\;  & 0 & 0 & 1 \\
\;\;\;\; \phantom{ \big|} \alpha = 3\phantom{ \big|}  \;\;\;\; &  0 & -1/2 & 0 
\end{array} 
\eeq
To each 4-homology class $\cC^4_\alpha$
there   is also  a Poincar\'e dual 2-cohomology class,
which we represent by a closed but not exact 2-form
$\omega_{2\alpha}$.
The parametrization of $\omega_{2\alpha}$ reads
\beq
\omega_{2\alpha} = dH_\alpha \, \frac{D\psi}{2\pi}
+ \bigg( t^{\varphi}_\alpha - 2 \, H_\alpha \bigg) \, \frac{ V_{\varphi} }{2\pi}
+ \bigg( t^{\Sigma}_\alpha - \chi \, H_\alpha \bigg) \, \frac{ V_{\Sigma} }{2\pi} \ ,
\eeq
where $t^{\varphi}_\alpha $, $t^{\Sigma}_\alpha $ are constants
and $H_\alpha$ is a function of $\mu$ only,
satisfying
\beq
H_\alpha(\mu_\mathrm N) + H_\alpha(\mu_\mathrm S) = 0  \ .
\eeq
We impose the relation $\int_{M_6} \Omega_4^\alpha \, \omega_{2\beta}
= \delta^\alpha_\beta$
to fix the values of $H_\alpha(\mu_\mathrm N)$, $t^{\varphi}_\alpha $, 
and $t^{\Sigma}_\alpha $,
\beq \label{2form_table}
\begin{array}{c  ||  c | c | c }
&  \;\;\;\; \phantom{ \big|} H_\alpha(\mu_\mathrm N)\phantom{ \big|}  \;\;\;\; & 
\;\; \;\; \phantom{ \big|} t^{\varphi}_\alpha \phantom{ \big|} \;\;\;\; &
\;\;\;\; \phantom{ \big|}  t^\Sigma_\alpha\phantom{ \big|}   \;\;\;\; \\ \hline
\;\;\;\; \phantom{ \big|} \alpha = 1 \phantom{ \big|} \;\;\;\; & 0 & 0 & 1  \\
\;\;\;\; \phantom{ \big|} \alpha = 2 \phantom{ \big|} \;\;\;\;  & 1/2 & 0 & 0 \\
\;\;\;\; \phantom{ \big|} \alpha = 3 \phantom{ \big|} \;\;\;\; &  0 & -1 & -\chi/2 
\end{array} 
\eeq
We can define the following five 2-cycles inside $M_6$,
\beq \label{all_2cycles}
\begin{array}{llll}
 \cC_2^{\varphi,\mathrm N}  & :&\qquad\qquad  
& \text{$S^2_\varphi$ at $\mu = \mu_\mathrm N$ and at a point on $\Sigma_g$} \ , \\
 \cC_2^{\varphi,\mathrm S}  & :&\qquad\qquad  
& \text{$S^2_\varphi$ at $\mu = \mu_\mathrm S$ and at a point on $\Sigma_g$} \ , \\
 \cC_2^{\Sigma,\mathrm N}  & :&\qquad\qquad  
& \text{$\Sigma_g$ at $\mu = \mu_\mathrm N$ and at a point on $S^2_\varphi$} \ , \\
 \cC_2^{\Sigma,\mathrm S}  & :&\qquad\qquad  
& \text{$\Sigma_g$ at $\mu = \mu_\mathrm S$ and at a point on $S^2_\varphi$} \ , \\
 \cC_2^{\rm fiber}  & :&\qquad\qquad  
& \text{the $S^2_\psi$ fiber at a point on the base $S^2_\varphi \times \Sigma_g$} \ .  
\end{array}
\eeq
They all correspond to  elements in the integral 2-homology of $M_6$,
but they satisfy two linear relations in homology,
\begin{align}
 \cC_2^{\varphi,\mathrm N} -  \cC_2^{\varphi,\mathrm S} + 2 \,  \cC_2^{\rm fiber} & = 0 \ , \nn \\
  \cC_2^{\Sigma,\mathrm N} -  \cC_2^{\Sigma,\mathrm S} + \chi \,  \cC_2^{\rm fiber} & = 0 \ .
\end{align}
The previous relations can be checked by pairing
the combinations of 2-cycles with an arbitrary
closed 2-form on $M_6$.
It is convenient to use a basis of integral 2-homology
$\cC_2^\alpha$ such that
$\int_{\cC_2^\alpha} \omega_{2\beta} = \delta^\alpha_\beta$.
One can check that such a basis is given by
\begin{align}
\cC_2^{\alpha = 1} & =   \cC_2^{\Sigma,\mathrm N} - \frac \chi 2 \,  \cC_2^{\varphi,\mathrm N}  \ , \nn \\
\cC_2^{\alpha = 1} & = \cC_2^{\rm fiber}  \ , \nn \\
\cC_2^{\alpha = 3} & = - \cC_2^{\rm fiber} - \cC_2^{\varphi,\mathrm N}  \,  \ .
\end{align}
Moreover, we verify that all the five 2-cycles
defined in \eqref{all_2cycles} can be written as
\emph{integral} linear combinations
of the basis 2-cycles $\cC_2^\alpha$.

Let us now discuss 3-cycles in $M_6$.
We can define several 3-cycles in terms of the 1-cycles 
on the Riemann surface.
Let $\Omega^{uv}$ denote the inverse of the intersection
pairing $\Omega_{uv}$, with the convention $\Omega_{uv} \, \Omega^{vw} = - \delta_u^w$. Since $\Omega_{uv}$ is integral
and unimodular, so is $\Omega^{uv}$.
From $\int_{\Sigma_g} \lambda_{1u} \, \lambda_{1v} = \Omega_{uv}
= \cC_{1u}^\Sigma \cdot  \cC_{1v}^\Sigma$,
we see that a basis of 1-cycles $\cC_1^{\Sigma u}$ with $\int_{\cC_1^{\Sigma u}} \lambda_{1v} = \delta_v^u$ is given by
$\cC_1^{\Sigma u} = \Omega^{uv} \, \cC_{1v}^\Sigma$.
We use these 1-cycles on the Riemann surface
to construct   
3-cycles in $M_6$,
\beq \label{all_3cycles}
\begin{array}{llll}
\cC_3^{ \mathrm N u}  & :&\qquad\qquad  
& \text{$S^2_\varphi \times \cC_{1}^{\Sigma u}$ at $\mu = \mu_\mathrm N$} \ , \\
\cC_3^{  \mathrm S u}  & :&\qquad\qquad  
& \text{$S^2_\varphi \times \cC_{1 }^{\Sigma u}$ at $\mu = \mu_\mathrm S$} \ , \\
\cC_3^{  \psi u}  & :&\qquad\qquad  
& \text{$S^2_\psi$ fibered over $\cC_{1}^{\Sigma u}$ at a point on $S^2_\varphi$} \ .
\end{array}
\eeq
These 3-cycles represent integral 3-homology classes,
subject to the relations
\beq
\cC_3^{ \mathrm N u}  - \cC_3^{  \mathrm S u}
+ 2 \, \cC_3^{  \psi u} = 0 \ .
\eeq
In total, we have $4g$ independent 3-homology classes.
The above relation implies that
$\cC_3^{ \mathrm N u}  - \cC_3^{  \mathrm S u}$ is an even 3-homology class.
The class $\cC_3^{ \mathrm N u}  + \cC_3^{  \mathrm S u} = (\cC_3^{ \mathrm N u}  - \cC_3^{  \mathrm S u}) + 2 \, \cC_3^{  \mathrm S u}$ is also even.
It follows that the following classes are integral,
\beq
\cC_3^{u \pm} = \frac 12 \, (\cC_3^{\mathrm N u} \pm \cC_3^{\mathrm S u}) \ .
\eeq
We use them as a basis of 3-homology,
\beq
\cC_{3}^x = (  \cC_3^{u +} , \cC_3^{u -}   ) \ .
\eeq
Notice that $\cC_3^{\mathrm N u}$, $\cC_4^{\mathrm S u}$,
and $\cC_3^{\psi u}$ can all be written as linear
combinations of $\cC_{3}^x$ with \emph{integer} coefficients.

The 3-cohomology classes dual to the 3-homology classes
$\cC_3^x$ can be represented by closed but not exact 3-forms
$\Lambda_{3x}$. We parametrize them as follows,
\beq
\Lambda_{3x}  = (\Lambda_{3u+} , \Lambda_{3u-}) \ , \qquad
\Lambda_{3 u \pm } = \bigg( d\cS_\pm \, \frac{D\psi}{2\pi}
- 2 \, \cS_\pm \, \frac{V_{\varphi}}{2\pi} \bigg) \, \lambda_{1u}   \ ,
\eeq
where $\lambda_{1u}$ are the harmonic 1-forms on the Riemann surface,
pulled back to $M_6$,
and $\cS_\pm$ are functions of $\mu$ only.
If we demand $\int_{\cC_{3}^x} \Lambda_{3y} = \delta_y^x$
we derive
\beq \label{3form_table}
\cS_+(\mu_\mathrm N) = - \frac 12 \ , \qquad
\cS_+(\mu_\mathrm S) = - \frac 12 \ , \qquad
\cS_-(\mu_\mathrm N) = - \frac 12 \ , \qquad
\cS_-(\mu_\mathrm S) = + \frac 12 \ .
\eeq
The pairing $\cK_{xy}$ defined in \eqref{intersection_numbers}
takes the form
\beq
\cK_{xy} = \begin{pmatrix}
\cK_{u+, v+} & 
\cK_{u+, v-} \\
\cK_{u-, v+} & 
\cK_{u-, v-} 
\end{pmatrix}
= \begin{pmatrix}
0 & - \Omega_{uv} \\
-\Omega_{uv} & 0
\end{pmatrix} \ ,
\eeq
and is antisymmetric and unimodular.
Finally, we may compute the intersection
numbers $\cK_{xv\alpha}$ in \eqref{intersection_numbers}.
The only non-zero components of 
$\cK_{xv\alpha}$ are
\beq
\cK_{u+ , v , \alpha = 2} = \Omega_{uv} \ , \qquad
\cK_{u-, v, \alpha = 3} = \Omega_{uv} \ .
\eeq

\subsubsection{Inclusion of background gauge fields for isometries}
For $g \ge 2$, the isometry group of $M_6$ is $U(1)_\psi \times SO(3)_\varphi$.
To describe the isometries of $ds^2(S^2_\varphi) = d\theta^2 + \sin ^2 \theta \, d\varphi^2$
it is convenient to introduce three constrained coordinates
\beq
\tilde y^{\tilde a} \, \tilde y_{\tilde a} = 1 \ , \qquad \tilde y^{\tilde a} = (\sin \theta \, \cos \varphi , \sin \theta \, \sin \varphi , \cos \theta) \ , \qquad \tilde a  = 1,2,3 \ .
\eeq
In the previous expression, $\tilde a$ is a vector index of $SO(3)_\varphi$.
The gauging of $U(1)_\psi$ is performed by introducing a background
connection $A^\psi$, while for $SO(3)_\varphi$ we introduce
a triplet $\tilde A^{\tilde a}$ of connections.  
Our conventions for the gauging of the 1-forms $d \tilde y^{\tilde a}$ on $S^2_\varphi$
are
\beq
D\tilde y^{\tilde a} = d\tilde y^{\tilde a} + \frac 12 \, \epsilon^{\tilde a \tilde b \tilde c } \, 
\tilde A_{\tilde b } \, \tilde y_{\tilde c} \ , \qquad
\tilde F^{\tilde a} = d \tilde A^{\tilde a } + \frac 12 \, \epsilon^{\tilde a \tilde b \tilde c } \,
\tilde  A_{\tilde b } \, \tilde A_{\tilde c } \ .
\eeq
It is worth commenting on how the $SO(3)_\varphi$ isometry of
$S^2_\varphi$, considered in isolation, extends to an isometry of the total space $M_6$.
We have verified that the 1-forms
\beq
f_\varphi \, \epsilon_{\tilde a \tilde b \tilde c} \, \tilde y^{\tilde b} \, D\tilde y^{\tilde c}
+ f_\psi   \, \tilde y_{\tilde a} \, D\psi
\eeq
are globally defined on $M_6$ and dual to Killing vectors in the 
line element \eqref{schematic_GMSW}.

In the case $g=0$, the space $M_6$ admits an additional $SO(3)_\Sigma$ isometry,
originating from the isometries of the Riemann surface.
As in section \ref{BBBW_isom_app}, we describe the Riemann surface in terms of
three constrained coordinates $y^a$, where $a = 1,2,3$ is an index of
$SO(3)_\Sigma$ (not to be confused with the $\tilde a$ indices of $SO(3)_\varphi$).
The gauge fields of $SO(3)_\Sigma$ are denoted $A^a$.
The gauging of $dy^a$ is performed as in \eqref{Sigma_gauging}.
The $SO(3)_\Sigma$ isometry of the Riemann surface extends to an isometry of $M_6$
because the following  1-forms are dual to Killing vectors
in the metric \eqref{schematic_GMSW},
\beq
f_\Sigma \, \epsilon_{  a   b   c} \,   y^{  b} \, D  y^{  c}
+ f_\psi   \,   y_{  a} \, D\psi \ .
\eeq

After turning on $A^\psi$, $\tilde A^{\tilde a}$, and (for $g=0$) $A^a$,
the 1-form $D\psi$ is replaced by its gauged version $\widetilde D\psi$, 
which satisfies (cfr.~with the ungauged version \eqref{dDpsi})
\beq \label{dDpsi_gauged}
\frac{d \widetilde D\psi}{2\pi} = - 2 \, e_2^\varphi - \chi \, e_2^\Sigma  
+ 2 \, \frac{F^\psi}{2\pi} \ .
\eeq
In the previous expression $F^\psi = dA^\psi$. The 2-form $e_2^\varphi$
is the global angular form of $SO(3)_\varphi$,
\beq
e_2^{\varphi} =\frac{1}{8\pi}  \, \bigg[
 \epsilon_{\tilde a \tilde b \tilde c} \, D \tilde y^{\tilde a} 
\, D\tilde y^{\tilde b} \, \tilde y^{\tilde c} - 2 \, \tilde F_{\tilde a} \, \tilde y^{\tilde a}
\bigg] \ .
\eeq
It is the closed and gauge-invariant completion of $V_\varphi/(2\pi)$
and satisfies Bott-Cattaneo identities analogous to \eqref{BottCatt}.
The 2-form $e_2^\Sigma$ is understood in different way 
depending on $g \ge 2$ or $g=0$:
for $g\ge 2$ it is simply 
proportional to the volume form $V_\Sigma$,
while for $g=0$ it is the global angular form of $SO(3)_\Sigma$,
\beq
\text{for $g\ge 2$:} \quad e_2^\Sigma = \frac{V_\Sigma}{2\pi} \ , \qquad
\text{for $g = 0$:} \quad e_2^\Sigma =\frac{1}{8\pi}  \, \bigg[
 \epsilon_{  a   b   c} \, D   y^{  a} 
\, D  y^{  b} \,   y^{  c} - 2 \,   F_{  a} \,   y^{  a} 
\bigg]  \ .
\eeq
The   anomaly polynomial will be written in terms of the
second Chern classes of $SU(2)_\Sigma$ and $SU(2)_\varphi$.
They are related to the first Pontryagin
classes of  $SO(3)_\Sigma$ and $SO(3)_\varphi$
by \eqref{p1_to_c2} and the analogous relation for $p_1(SO(3)_\varphi)$.

Let us now turn to a discussion of the extension of the closed 4-forms
$\Omega_4^\alpha$ on $M_6$ to closed 4-forms $(\Omega_4^\alpha)^{\rm eq}$
on $M_{12}$. We define
\beq
(\Omega_4^\alpha)^{\rm eq} = d \bigg[
\Big(U_{\varphi}^\alpha \, e_2^\varphi
+ U_\Sigma^\alpha \, e_2^\Sigma \Big) \, \frac{\widetilde D\psi}{2\pi}
\bigg]
+ C^\alpha  \,
e_2^\varphi \, e_2^\Sigma   \ ,
\eeq 
and we verify that $(\Omega_4^\alpha)^{\rm eq}$ are globally defined, closed 4-forms
on $M_{12}$ with integral periods. 
By a similar token, the   2-forms $\omega_{2\alpha}$ on $M_6$
extend to the   2-forms $(\omega_{2\alpha})^{\rm eq}$, defined as
\beq
(\omega_{2\alpha} )^{\rm eq}= d \bigg[ H_\alpha \, \frac{\widetilde D\psi}{2\pi} \bigg]
+  t^{\varphi}_\alpha  \, e_2^\varphi 
+  t^{\Sigma}_\alpha   \,  e_2^\Sigma \ .
\eeq
The 1-forms $\lambda_{1u}$ are   unaffected by the gauging of the $U(1)_\psi \times SU(2)_\varphi$ isometry,  
\beq
(\lambda_{1u})^{\rm eq} = \lambda_{1u} \ .
\eeq
Finally, the 3-forms $\Lambda_{3 u \pm}$ on $M_6$
extend to the following 3-forms on $M_{12}$,
\beq
(\Lambda_{3u\pm})^{\rm eq} = d\bigg[ \cS_\pm \, \frac{\widetilde D \psi}{2\pi} \bigg] \, 
\lambda_{1u} \ .
\eeq

\subsubsection{Computation of the inflow anomaly polynomial}
Having defined the closed forms $ (\Omega_4^\alpha)^{\rm eq}$,
$(\omega_{2\alpha})^{\rm eq}$, $(\Lambda_{3u\pm})^{\rm eq}$,
and $(\lambda_1{}_u)^{\rm eq}$, 
we have a fully explicit expression for the quantity $E_4$.
The other ingredient for the computation of the inflow anomaly
polynomial is the 8-form $X_8$. Following \cite{Bah:2019vmq},
we can compute it using the following relations among
Pontryagin classes,
\begin{align} \label{Pontr_split}
p_1(TM_{12}) &= p_1(T) + p_1(SO(3)_\varphi) + p_1(SO(3)_\Sigma)
+ \bigg[ \frac{d \widetilde D\psi}{2\pi} \bigg]^2 \ , \nn \\
p_2(TM_{12})  & = \bigg[p_1(T) + p_1(SO(3)_\varphi) + p_1(SO(3)_\Sigma) \bigg] \, 
 \bigg[ \frac{d \widetilde D\psi}{2\pi} \bigg]^2  \ .
\end{align}
In the previous expressions, $p_1(T)$ is the first Pontryagin class of
external spacetime. The terms with $p_1(SO(3)_\Sigma)$ are understood to be 
present only in the case $g=0$.
By combining \eqref{Pontr_split} and \eqref{dDpsi_gauged}
we obtain an explicit expression for the 8-form $X_8$.

The computation of the integrals $\int_{M_6} E_4^3$
and $\int_{M_6} E_4 \, X_8$ is straightforward.
After collecting the terms with exactly one $\widetilde D\psi$ factor,
we can integrate over $S^2_\varphi$ with the help
of the Bott-Cattaneo formula. 
The integral over $\Sigma_g$ receives two contributions:
terms with an odd power of $e_2^\Sigma$ (treated with the Bott-Cattaneo
formula), and terms with exactly two $\lambda_{1u}$ factors (treated
using $\int_{\Sigma_g} \lambda_{1u} \, \lambda_{1v} = \Omega_{uv}$).
We are left with a one-dimensional integral
over the $\mu$ interval, which is evaluated making use of
\eqref{4form_table}, \eqref{2form_table}, and \eqref{3form_table}.
The final result is recorded in the main text.

Let us conclude with a comment on the large $N$ limit.
We define this limit by letting the three flux quanta $N_\alpha$
scale in the same way, $N_\alpha \sim \cO(N)$, with fixed ratios
$N_\alpha/N_\beta$ for $\alpha \neq \beta$.
We assign scaling $N^0$ 
to $p_1(T)$ and to the background fields associated to isometries of $M_6$,
while we assign scaling $N^1$ to all external gauge fields originating from
expansion of $C_3$ onto cohomology classes of $M_6$.
In this way, the $\cO(N^3)$ terms in $I_6^{\rm inflow}$ all originate from
the $E_4^3$ term in $\cI_{12}$, while
the $\cO(N)$ terms originate from $E_4 \, X_8$.



\section{Construction of $E_4$} \label{app_E4}

In this appendix we discuss the construction and properties of the 
forms 
$(\Omega_4^\alpha)^{\rm eq}$,
$(\omega_{2\alpha})^{\rm eq}$,
$(\Lambda_{3x})^{\rm eq}$,
$(\lambda_1{}_u)^{\rm eq}$
that enter the parametrization \eqref{E4_expr} of $E_4$.

Suppose that $M_6$ admits a collection of Killing vectors
$k_I^m$, with $m = 1, \dots, 6$ a curved tangent  index
on $M_6$,
and $I$ labeling a basis of Killing vectors.
The latter obey the Lie algebra
\beq
\pounds_I k_J \equiv \pounds_{k_I} k_J = [k_I, k_J] = f_{IJ}{}^K \, k_K \ ,
\eeq
where $\pounds$ denotes Lie derivative.
The fibration in \eqref{M12aux} includes
arbitrary background gauge fields associated to the
isometries of the $M_6$ fiber. These gauge fields are 1-form gauge fields on the base 
$\cM_6$.
We refer to the operation of turning them on as gauging.
In terms of local coordinates $\xi^m$ on the $M_6$ fiber,
the gauging
is conveniently described by the replacement
\beq \label{eq_replacement}
d\xi^m \;\; \rightarrow \;\;
D\xi^m = d\xi^m + k^m_I \, A^I \ ,
\eeq
where $A^I$ is the external gauge field associated to the Killing
vector $k^m_I$.
In our conventions, the field strength $F^I$ of  
$A^I$ reads (we suppress wedge products throughout this appendix)
\beq \label{non_abelian_F}
F^I = dA^I - \frac 12 \, f_{JK}{}^I \, A^J \, A^K \ .
\eeq
Let $\omega_q$ be a $q$-form on $M_{6}$, 
\beq
\omega_q = \frac{1}{q!} \, \omega_{m_1 \dots m_q}  \, d\xi^{m_1} \dots
d\xi^{m_q } \ ,
\eeq
where the components $\omega_{m_1 \dots m_q}$
depend only on the coordinates $\xi^m$ on $M_{6}$.
We use the symbol $(\omega_q)^{\rm g}$ for the gauged
version of $\omega$, obtained 
by means of the replacement
\eqref{eq_replacement},
\beq \label{def_gauging}
(\omega_q)^{\rm g} = \frac{1}{q!} \, \omega_{m_1 \dots m_q}  \, D\xi^{m_1} \dots
D\xi^{m_q } \ .
\eeq
A useful identity for $(\omega_q)^{\rm g}$ is 
\beq \label{diff_identity}
d (\omega_q)^{\rm g}  + A^I \, (\pounds_I \omega_q)^{\rm g}
= (d\omega_q)^{\rm g} + F^I \, (\iota_I \omega_q)^{\rm g} \ ,
\eeq
where $\iota_I$ denotes the interior product of the vector $k_I^m$ with a 
differential form.

If we choose a metric on $M_6$,
we can select the harmonic
representatives for the de Rham classes defined by
$\Omega_4^\alpha$,
$\omega_{2\alpha}$,
$\Lambda_{3x}$,
$ \lambda_1{}_u$.
A harmonic form is automatically invariant under 
all isometries of $M_6$.\footnote{For example,
if $\omega_2$ is a harmonic 2-form, the
fact that $\pounds_I \omega_2 = 0$ can be seen as follows.
From
$d\omega_2=0$ we derive
$\pounds _I \omega_2 = d(\iota_I \omega_2)$.
Making use of $\nabla_{(m} k_{I|n)}  = 0$ and  $\nabla^m \omega_{  mn} = 0$,
we verify
$(\pounds_I \omega_2)_{mn} = \nabla^p (k_I \wedge \omega_2)_{pmn}$.
We have thus established that the 2-form $\pounds_I \omega_2$
is both exact and co-exact.
It follows that 
 $\int_{M_{10-d}} (\pounds_I \omega_2) *   (\pounds_I \omega_2) = 0$ (no sum over  $I$), which in turn guarantees
$\pounds_I \omega_2 = 0$.
} 
This means that (after adding suitable exact forms, if necessary)
we can take the closed forms $\Omega_4^\alpha$,
$\omega_{2\alpha}$,
$\Lambda_{3x}$,
$ \lambda_1{}_u$ to be invariant 
under all isometries of $M_6$,
\begin{align}
\pounds_I \lambda_{1u}= 0 \ , \qquad
\pounds_I \omega_{2\alpha}= 0 \ , \qquad
\pounds_I \Lambda_{3x} = 0 \ , \qquad
\pounds_I \Omega_4^\alpha = 0 \ .
\end{align}
It follows that the forms
$\iota_I \lambda_{1u}$, $\iota_I \omega_{2\alpha}$,
$\iota_I \Lambda_{3x}$, $\iota_I \Omega_4^\alpha$ are closed.
We may then write
\begin{align} \label{first_layer}
2\pi \, \iota_I \lambda_{1u}   &= c_{uI} \ , &
2\pi \, \iota_I \omega_{2\alpha} + d \omega_{0\alpha I} &= c_\alpha{}^u{}_I \, \lambda_{1u} \ , \nn \\
2\pi \, \iota_I \Lambda_{3x} + d\Lambda_{1xI} & = c_x{}^\alpha{}_I \, \omega_{2\alpha}
\ , &
2\pi \, \iota_I \Omega_4^\alpha + d\Omega_{2 I}^\alpha & = 
c^{\alpha x}{}_I \, \Lambda_{3x} \ ,
\end{align}
where $c_{uI}$, $c_\alpha{}^u{}_I$,
$c_x{}^\alpha{}_I$, $c^{\alpha x}{}_I$ are constants.
For example, we have observed that $2\pi \, \iota_I \omega_{2\alpha}$
is a closed 1-form, and that the de Rham classes of $\lambda_{1u}$
furnish a basis of $H^1(M_6,\mathbb R)$. It follows there are
suitable constants $c_\alpha{}^u{}_I$ such that
that the difference $2\pi \, \iota_I \omega_{2\alpha}  - c_\alpha{}^u{}_I \, \lambda_{1u}$
is exact. Similar remarks apply to the other expressions in \eqref{first_layer}.
The forms  $\omega_{0\alpha I}$,
$\Lambda_{1xI}$, $\Omega_{2I}^\alpha$ are only defined modulo addition of a closed
form. Without loss of generality, they can be taken to satisfy
\begin{align} \label{nice_Lie_der}
\pounds_I \omega_{0 \alpha J} = f_{IJ}{}^K \, \omega_{0 \alpha K} \ , \qquad
\pounds_I \Lambda_{1xJ} = f_{IJ}{}^K \, \Lambda_{1xI} \ , \qquad
\pounds_I \Omega_{2J}^\alpha = f_{IJ}{}^K \, \Omega_{2K}^\alpha \ .
\end{align}
Symmetrizing in $IJ$ and using \eqref{first_layer} we derive  
\begin{align} \label{second_layer}
0 &= 2\pi \,   \pounds_{(I} \omega_{0 \alpha |J) }  
=  c_\alpha{}^u{}_{(I} \, c_{u |J)} \ , \nn \\
0 & = 2\pi \, \pounds_{(I} \Lambda_{1x |J)} 
= c_x{}^{\alpha}{}_{(I} \, \Big[ c_\alpha{}^u{}_{J)} \, \lambda_{1u}
- d\omega_{0\alpha |J)} \Big]
+ 2\pi \, d \iota_{(I} \Lambda_{1x |J)}  \ , \nn \\
0 & =2\pi\, \pounds_{(I} \Omega_{2 |J) }^\alpha =
c^{\alpha x}{}_{(I} \, \Big[  c_x{}^\beta{}_{J)} \, \omega_{2\beta}
- d\Lambda_{1x |J)} \Big] 
+ 2\pi \, d \iota_{(I} \Omega_{2 |J) }^\alpha \ .
\end{align}
If we integrate the second relation on a non-trivial 1-cycle in $M_6$,
only the term with $\lambda_{1u}$ contributes. It follows that its coefficient must be zero.
Similar remarks apply to the third line. We conclude that 
the constants $c$   satisfy
\beq
  c_\alpha{}^u{}_{(I} \, c_{u |J)} = 0 \ , \qquad
  c_x{}^{\alpha}{}_{(I} \,     c_\alpha{}^u{}_{J)} = 0 \ , \qquad
  c^{\alpha x}{}_{(I} \,  c_x{}^\beta{}_{J)}  = 0 \ .
\eeq
It follows from \eqref{second_layer}
that the forms $2\pi \, \iota_{(I} \Lambda_{1x |J)} - c_x{}^\alpha{}_{(I} \, \omega_{0 \alpha |J)}$
and $2\pi \, \iota_{(I} \Omega_{2|J)}^\alpha - c^{\alpha x}{}_{(I} \, \Lambda_{1x|J)}$ 
are closed.
We can therefore write
\beq \label{final_layer}
2\pi \, \iota_{(I} \Lambda_{1x |J)} = c_x{}^\alpha{}_{(I} \, \omega_{0 \alpha |J)}
+ b_{xIJ} \ , \qquad
2\pi \, \iota_{(I} \Omega_{2|J)}^\alpha + d \Omega_{0IJ}^\alpha = c^{\alpha x}{}_{(I} \, \Lambda_{1x|J)}
+ b^{\alpha u}{}_{IJ} \, \lambda_{1u} \ ,
\eeq
where $b_{xIJ}$ and $b^{\alpha u}{}_{IJ}$ are constants
and $\Omega_{0IJ}^\alpha$ are 0-forms,
defined up to a constant.

We can now  write the forms
$(\Omega_4^\alpha)^{\rm eq}$,
$(\omega_{2\alpha})^{\rm eq}$,
$(\Lambda_{3x})^{\rm eq}$,
$(\lambda_1{}_u)^{\rm eq}$.
They are given by
\begin{align}
(\Omega_4^\alpha)^{\rm eq} & = (\Omega_4^\alpha)^{\rm g}
+ \frac{F^I}{2\pi} \, (\Omega_{2I}^\alpha)^{\rm g}
+ \frac{F^I}{2\pi} \, \frac{F^J}{2\pi} \, \Omega_{0IJ}^\alpha \ , \nn \\
(\Lambda_{3x})^{\rm eq} & = (\Lambda_{3x})^{\rm g}
+ \frac{F^I}{2\pi} \, (\Lambda_{1xI})^{\rm g}
 \ , \nn \\
(\omega_{2\alpha})^{\rm eq}  & = (\omega_{2\alpha})^{\rm g}
+ \frac{F^I}{2\pi} \, \omega_{0\alpha I} \ , \nn \\
(\lambda_1{}_u)^{\rm eq}  & =(\lambda_1{}_u)^{\rm g} \ .
\end{align}
Making use of the identity
\eqref{diff_identity}, the Bianchi identity for $F^I$,
 and the relations \eqref{first_layer}, \eqref{final_layer}
 we compute
\begin{align}
d(\lambda_1{}_u)^{\rm eq}  & = \frac{F^I}{2\pi}  \, c_{uI} \ , \nn \\
d(\omega_{2\alpha})^{\rm eq}  & = \frac{F^I}{2\pi}  \, c_\alpha{}^u{}_I \, (\lambda_1{}_u)^{\rm eq} \ , \nn  \\
 d(\Lambda_{3x})^{\rm eq}  & =\frac{F^I}{2 \pi} \, 
 c_x{}^{\alpha}{}_I \, ( \omega_{2\alpha})^{\rm eq}
  +  \frac{F^I}{2\pi} \,  \frac{F^J}{2\pi} \,  b_{xIJ} \ , \nn \\
 d(\Omega_4^\alpha)^{\rm eq}  &  =  \frac{F^I}{2\pi} \, c^{\alpha x}{}_I \, (\Lambda_{3x})^{\rm eq}
 +  \frac{F^I}{2\pi} \, \frac{F^J}{2\pi} \, b^{\alpha u}{}_{IJ} \, (\lambda_{1u})^{\rm eq} \ .
\end{align}
For the spaces $M_6$ of interest in this work,
all $c$ and $b$ constant vanish, and we verify closure
of  $(\Omega_4^\alpha)^{\rm eq}$,
$(\omega_{2\alpha})^{\rm eq}$,
$(\Lambda_{3x})^{\rm eq}$,
$(\lambda_1{}_u)^{\rm eq}$,
as anticipated in the main text.
If we were to study a setup with non-zero $c$ or $b$
constants, we could still make use of \eqref{E4_expr},
but we would have to modify the Bianchi identities
for the external field strengths,
\begin{align}
df_1^x &= - N_\alpha\, c^{\alpha x}{}_I \, F^I  \ , &
dF_2^\alpha   & = c_x{}^\alpha{}_I \, f_1^x \, F^I \ , \nn \\
dH_3^u
  &= 
- c_\alpha{}^u{}_I \, F^\alpha_2 \, F^I  
 - N_\alpha \, b^{\alpha u}{}_{IJ} \, F^I \, F^J \ , &
 d\gamma_4 
 & = 
  c_{uI} \, H_3^u \, F^I
+ b_{xIJ} \, f_1^x \, F^I \, F^J
  \ .
\end{align}
We leave further investigation of this case to future work.

We noticed above that the forms
$\Omega_{2I}^\alpha$, 
$\Lambda_{1xI}$, $\omega_{0\alpha I}$
are only defined up to addition of a closed form.
We can parametrize this ambiguity by writing
\beq
\Omega_{2I}^\alpha{}' = \Omega_{2I}^\alpha 
+ d \cY_{1I}^\alpha + \nu^{\alpha \beta}{}_I \, \omega_{2\beta} \ , \qquad
\Lambda_{1xI}{}' = \Lambda_{1xI} 
+ d \cY_{0xI}
+ \nu_x{}^u{}_I \, \lambda_{1u} \ , \qquad
\omega_{0\alpha I} {}' = \omega_{0\alpha I} + \nu_{\alpha I}
\eeq
where the $\nu$ parameters are constant,
and the $\cY$ forms can be taken to satisfy relations
analogous to \eqref{nice_Lie_der}. 
Since we have a new $\Omega_{2I}^\alpha$,
we have to determine a new $\Omega_{0IJ}^\alpha$,
by solving the second relation in \eqref{final_layer}.
For simplicity, we only consider the situation in which the $c$ and $b$
constants are zero. We can then write
\beq
\Omega_{0IJ}^\alpha{}' = \Omega_{0IJ}^\alpha
+ 2\pi \, \iota_{(I} \cY_{1 |J)}^\alpha
+  \nu^{\alpha \beta}{}_{(I}  \, \omega_{0\alpha |J)}
+ \tau_{IJ}^\alpha  \ ,
\eeq
where $\tau_{IJ}^\alpha$ are arbitrary constants.
If we insert the primed objects into the expression of $E_4$,
we obtain a new realization of $E_4$, denoted $E_4'$,
\begin{align}
E_4' & = 
  N_\alpha \, (\Omega_4^\alpha)^{\rm eq}
+ \bigg[
\frac{F_2^\alpha}{2\pi}
+ N_\beta \, \nu^{\beta \alpha}{}_I\, \frac{F^I}{2\pi} 
\bigg] \, (\omega_{2\alpha})^{\rm eq}
 + \frac{f_1^x}{2\pi} \, (\Lambda_{3x})^{\rm eq}
 \nn \\
& + \bigg[
 \frac{H_3^u}{2\pi}
 +  \nu_x{}^u{}_I \, \frac{f_1^x}{2\pi} \, \frac{F^I}{2\pi}
 \bigg] \, (\lambda_1{}_u)^{\rm eq}
 + \bigg[
 \frac{\gamma_4}{2\pi} 
 + N_\alpha \, \tau^\alpha_{IJ} \, 
\frac{F^I}{2\pi} \, \frac{F^J}{2\pi}
+   \nu_{\alpha I} \, \frac{F_2^\alpha}{2\pi} \, \frac{F^I}{2\pi}
\bigg]
\nn \\
& + d\bigg[ \frac{F^I}{2\pi} \, N_\alpha \, (\cY_{1I}^\alpha)^{\rm g}
- \frac{F^I}{2\pi} \, \frac{f_1^x}{2\pi} \, \cY_{0xI} 
\bigg]
 \ .
\end{align}
The last line collects the total derivative of a
globally defined 3-form on $M_{12}$.
Adding an exact piece to $E_4$ has no effect on the computation of $I_6^{\rm inflow}$.
We see that, up to this immaterial total derivative,
$E_4'$ has the same form as $E_4$,
if we perform a redefinition of the external gauge fields,
\begin{align} \label{external_redef}
\frac{F_2^\alpha{}'}{2\pi} &= \frac{F_2^\alpha}{2\pi}
+ N_\beta \, \nu^{\beta \alpha}{}_I\, \frac{F^I}{2\pi}  \ , &
 \frac{H_3^u{}'}{2\pi} &=  \frac{H_3^u}{2\pi}
 +  \nu_x{}^u{}_I \, \frac{f_1^x}{2\pi} \, \frac{F^I}{2\pi} \ , \nn \\
 \frac{\gamma_4'}{2\pi}  & =  \frac{\gamma_4}{2\pi} 
 + N_\alpha \, \tau^\alpha_{IJ} \, 
\frac{F^I}{2\pi} \, \frac{F^J}{2\pi}
+   \nu_{\alpha I} \, \frac{F_2^\alpha}{2\pi} \, \frac{F^I}{2\pi}
\ .
\end{align}
Let us stress that the constants 
$\nu^{\beta \alpha}{}_I$, $\nu_x{}^u{}_I$,
$\nu_{\alpha I}$, $\tau_{IJ}^\alpha$ are not completely arbitrary: they must be chosen in such a way that
$E_4'$ has integral periods.
Let us assume that the normalization of the
Killing vectors in \eqref{eq_replacement} has been chosen
in such a way that $F^I$ has periods that are quantized
in units of $2\pi$ (here we are assuming an Abelian
isometry group for simplicity).
The $\nu$ and $\tau$ constants have to be chosen in such a way that
\beq
N_\beta \, \nu^{\beta \alpha}{}_I \in \mathbb Z \ , \qquad
\nu_x{}^u{}_I  \in \mathbb Z \ , \qquad
N_\alpha \, \tau^\alpha_{IJ}  \in \mathbb Z \ , \qquad
 \nu_{\alpha I} \in \mathbb Z \ .
\eeq
It then follows that the redefinition \eqref{external_redef}
preserves the lattice of periods
of the external gauge fields.\footnote{We notice that
the field redefinitions for $H_3^u$
and $\gamma_4$ are non-linear.
For example, the quantity $\frac{f_1^x}{2\pi} \, \frac{F^I}{2\pi}$
can be regarded as the 3-form field strength
of a ``composite'' 2-forms gauge field
constructed from $a_0^x$ and $A^I$.
This notion of product of a $p$-form gauge field
and a $q$-form gauge field to yield a $(p+q+1)$-gauge field
can be made mathematically precise 
 in the framework of differential
cohomology, see appendix \ref{app_diff_cohom}.
}



\section{Aspects of differential cohomology} \label{app_diff_cohom}

In this appendix we give a brief review of some basic aspects of
differential cohomology.  We follow a presentation based on   Cheeger-Simons
differential characters \cite{10.1007/BFb0075216}. Introductions aimed at physicists
can be found \emph{e.g.}~in \cite{Freed:2006yc,Cordova:2019jnf}.

\subsubsection*{Cheeger-Simons differential characters}

A degree-$\ell$ Cheeger-Simons differential character $\chi$
on a manifold $\cM$
is a group homomorphism
$\chi \in {\rm Hom}(Z_{\ell-1}(\cM), U(1))$
with the following property: there exists a globally defined $\ell$-form
$F_\chi$ such that 
\beq \label{chi_def}
\chi(\partial B_\ell) = \exp \bigg[2\pi i \, \int_{B_\ell} F_\chi \bigg] \ , \qquad
\text{for all $B_\ell \in C_\ell(\cM)$}  \ .
\eeq
The notation $C_\ell(\cM)$ stands for the group of
$\ell$-chains in $\cM$, while $Z_{\ell-1}(\cM)$ denotes the
group of $(\ell-1)$-cycles. (Chains and cycles are understood
in the context of  smooth singular homology.)
One can verify from the definition of $\chi$
that the $\ell$-form $F_\chi$ is uniquely determined,
is closed, and has integral periods.
The set of degree-$\ell$ Cheeger-Simons
differential characters has a natural Abelian group structure.
We find it convenient to adopt
an additive notation, and write
\beq \label{additive_notation}
(\chi_1 + \chi_2)(\Sigma_{\ell-1}) := \chi_1(\Sigma_{\ell-1})  \,
 \chi_2(\Sigma_{\ell-1})  \ , \qquad
\Sigma_{\ell-1}\in Z_{\ell-1}(\cM) \ .
\eeq
In this notation, $\chi= 0$ means that
$\chi$ associates $1 \in U(1)$ to every
$\Sigma_{\ell-1} \in Z_{\ell-1}(\cM)$.
The group of 
degree-$\ell$ Cheeger-Simons
differential characters 
is denoted $\check H^\ell(\cM)$.
(Contrary to ordinary cohomology groups,
 $\check H^\ell(\cM)$ is usually infinite-dimensional.)

The mathematical object $\chi$ models an $(\ell-1)$-form
$U(1)$ gauge field, or more precisely,
the equivalence class of an $(\ell-1)$-form
$U(1)$ gauge field up to gauge transformations.
To see this, we interpret the map 
$\chi : Z_{\ell-1}(\cM) \rightarrow U(1)$
as the map that to each $(\ell-1)$-cycle in spacetime $\cM$
assigns the    holonomy of the gauge field
on that cycle. The globally-defined, closed $\ell$-form
$F_\chi$ with integral periods is identified with the
field strength of the $(\ell-1)$-form gauge field.
(Notice that, in the main text, field strengths are normalized
to have periods that are quantized in units of $2\pi$.)
The equation \eqref{chi_def}
encodes the expected physical 
relation between the holonomy of a gauge field 
along a boundary of a chain, and the flux of its 
field strength through that chain.

A differential character $\chi \in \check H^\ell(\cM)$
determines uniquely an element $a_\chi \in H^\ell(\cM,\mathbb Z)$,
called the characteristic class of $\chi$.\footnote{This can be seen as follows.
Every group homomorphism $\chi : Z_{\ell -1}(\cM) \rightarrow U(1)$
admits a (non-unique) lift, \emph{i.e.}~a group
homomorphism $T: C_{\ell-1}(\cM) \rightarrow \mathbb R$,
such that $\chi = \exp (2\pi i T)$. From \eqref{chi_def} one shows that
$\delta T = F_\chi - c$ for some $c\in Z^\ell(\cM, \mathbb Z)$
(the group of integer cocycles on $\cM$).
While $T$ and $c$ are not uniquely determined,
the cohomology class $a_\chi := [c]\in H^\ell(\cM,\mathbb Z)$
is uniquely fixed by $\chi$.
} 
The characteristic class $a_\chi$ and the field strength
$F_\chi$ satisfy the following compatibility condition,
\beq \label{compatibility}
[F_\chi]_{\rm dR} = \varrho (a_\chi) \ .
\eeq
The notation $[F_\chi]_{\rm dR}  \in H^\ell(\cM, \mathbb R)$
stands for the de Rham class of the closed form $F_\chi$,
while $\varrho$ is the natural map
\beq \label{natural_rho}
\varrho : H^\ell(\cM, \mathbb Z) \rightarrow H^\ell(\cM, \mathbb R) \ .
\eeq
The relation \eqref{compatibility} might erroneously suggest
that all interesting information about $a_\chi$ is already
contained in the field strength $F_\chi$.
Crucially, however, the map $\varrho$ 
forgets torsion: $a_\chi$ is determined
by $F_\chi$ only up to torsion
elements in integer cohomology,
\emph{i.e.}~up to an element of ${\rm Tor} \, H^\ell(\cM, \mathbb Z)$.
This additional data encoded in $a_\chi$
(and missed by $F_\chi$) is particularly
important if the spacetime manifold $\cM$ has
torsion in homology,
as already emphasized for instance in \cite{Dijkgraaf:1989pz}.

A differential character $\chi$ is called topologically trivial
if $a_\chi = 0$. 
It can be proven that $a_\chi = 0$ if and only $\chi$ can be written in terms of a globally
defined $(\ell-1)$-form $A$ as 
\beq \label{top_trivial}
\chi(\Sigma_{\ell-1}) = \exp \bigg[ 2\pi i \, \int_{\Sigma_{\ell-1}  }A \bigg] \ ,
\qquad \Sigma_{\ell-1} \in Z_{\ell-1}(\cM) \ .
\eeq
In this case, $F_\chi = dA$ and  \eqref{chi_def} 
follows from Stokes' theorem. Moreover,
 \eqref{compatibility} is   
satisfied because $[F_\chi]_{\rm dR} = 0$ (since
$F_\chi = dA$ and $A$ is globally defined).

A differential character $\chi$ is called flat if $F_\chi = 0$.
It can be proven that flat characters are
identified with elements of the (ordinary)
cohomology group $H^{\ell-1}(\cM, U(1))$.
Interestingly, there exist flat but topologically non-trivial
characters. Indeed,  
$H^{\ell-1}(\cM, U(1))$ 
is a compact Abelian group
that generically has more than one connected component.
The connected component of the identity
consists of characters that are both flat and topologically trivial
(we may refer to them as Wilson lines).
The connected components of 
$H^{\ell-1}(\cM, U(1))$  are labeled by ${\rm Tor} \, H^\ell(\cM,\mathbb Z)$.
This fits with the fact that $a_\chi$ for a flat character
$\chi$ is an element of ${\rm Tor} \, H^\ell(\cM,\mathbb Z)$
(this follows from \eqref{compatibility} and $F_\chi = 0$).

Let us emphasize that $\chi \in \check H^\ell(\cM)$
contains more information than 
its field strength $F_\chi$ and its characteristic class
$a_\chi$. 
In fact, $F_\chi$ and $a_\chi$ are unaffected
if we shift $\chi$ by a Wilson line.

The language of differential characters offers a uniform
way to describe $U(1)$ $p$-form gauge fields,
including 0-form fields. In fact, one can prove
that $\check H^1(\cM)$ is the same as the group of
smooth maps from $\cM$ to $S^1$. 
This mathematical
fact fits with the physics picture of a 0-form gauge field
as a circle-valued  scalar field.

\subsubsection*{Product in differential cohomology}

There is a notion of product in differential cohomology
compatible with the grading by the degree $\ell$,
\beq
\star : \check H^{\ell_1}(\cM) \times \check H^{\ell_2} (\cM) \rightarrow
H^{\ell_1 + \ell_2} (\cM) \ .
\eeq
With reference to the additive notation 
of \eqref{additive_notation}, the product $\star$ is distributive,
\beq
(\chi_1 + \chi_2) \star \chi_3 = \chi_1 \star \chi_3 + \chi_2 \star \chi_3 \ , \qquad
\chi_1, \chi_2 \in \check H^{\ell_1}(\cM) \ , \qquad
\chi_3 \in \check H^{\ell_2}(\cM) \ .
\eeq
The product is graded commutative, like the wedge product
of differential forms,
\beq
\chi_1 \star \chi_2 = (-)^{\ell_1 \, \ell_2} \, \chi_2 \star \chi_1 \ , \qquad
\chi_1 \in \check H^{\ell_1}(\cM) \ , \qquad
\chi_2 \in \check H^{\ell_2}(\cM)  \ .
\eeq
The field strength and characteristic class of 
the character $\chi_1 \star \chi_2$
are determined by those of $\chi_1$, $\chi_2$ via
\beq
\chi_3 := \chi_1 \star \chi_2 \ , \qquad 
F_{\chi_3} = F_{\chi_1} \wedge F_{\chi_2} \ , \qquad
a_{\chi_3} = a_{\chi_1} \smile a_{\chi_2} \ ,
\eeq
where in the last relation $\smile$ denotes the cup product
in integer cohomology.
If $\chi_1$ is topologically trivial, 
then $\chi_3$ is topologically trivial, for any $\chi_2$. Indeed, 
if $\chi_1$ is determined by the $(\ell_1-1)$-form
$A_1$, then $\chi_3$ is determined by the $(\ell_1 + \ell_2-1)$-form
$A_1 \wedge F_{\chi_2}$.
By a similar token, if $\chi_1$ is flat, so is $\chi_3$,
for any $\chi_2$. If we regard the flat character 
$\chi_1$ as an element of $H^{\ell_1-1}(\cM,U(1))$,
and the flat character $\chi_3$
as an element of $H^{\ell_1 + \ell_2-1}(\cM,U(1))$,
then we can write
$\chi_3 = \chi_1 \smile a_{\chi_2}$,
where $\smile: H^{\ell_1-1}(\cM,U(1)) \times H^{\ell_2}(\cM,\mathbb Z)
\rightarrow H^{\ell_1 + \ell_2-1}(\cM,U(1))$
is a well-defined cup product in cohomology.

\subsubsection*{Cheeger-Simons characters and $\mathbb Z_k$ gauge fields}

In section \ref{sec_discrete_anomalies} we have encountered 
a constrained 1-form gauge field $\cA_1$, subject to
\eqref{A_constraint}. 
If we describe the (gauge-equivalence class of the)
1-form gauge field $\cA_1$
with a differential character $\chi \in \check H^2(\cM)$,
we  have the correspondence
\beq
k \, \cA_1 = d\phi_0  \quad \leftrightarrow \quad 
k \, \chi := \underbrace{ \chi + \dots + \chi}_{\text{$k$ times}} = 0 \in \check H^2(\cM) \ .
\eeq
Indeed, we have argued that
$k \, \cA_1 = d\phi_0$ means that $k\, \cA_1$ is pure gauge.
Since differential characters are gauge-equivalence
classes of gauge fields, $k \, \cA_1$ is described by the zero
character $0 \in \check H^\ell(\cM)$.
The equation $k \, \chi = 0$ implies
\beq
F_\chi = 0 \ , \qquad k \, a_\chi = 0  \ , \qquad
\chi(\Sigma_{1}) \in \mathbb Z_k \subset U(1) \quad \text{for all $\Sigma_{1}
\in Z_1(\cM)$} \ .
\eeq
Crucially, $k \, a_\chi = 0$ does not imply 
 $a_\chi = 0$,
but merely that $a_\chi$ is a $k$-torsion element in
integer cohomology. 
Even if $a_\chi = 0$
(which is the case if $\cM$ has no torsion in homology),
the character $\chi$ can be non-zero:
it is a Wilson line with holonomies in $\mathbb Z_k \subset U(1)$
determined by a globally-defined closed 1-form.
In physics terms, we may simply write
$\cA_1 = \frac 1k \, d\phi_0$ \cite{Banks:2010zn}.

There exist other realizations of the differential
cohomology groups $\check H^\ell(\cM)$,
for instance in terms of Hopkins-Singer cocycles
\cite{Hopkins:2002rd} or Deligne-Beilinson cocycles,
see \emph{e.g.}~\cite{Bauer:2004nh,Kapustin:2014zva} for a review.
Loosely speaking, in these formalisms  one can model
not only the gauge-equivalence
class of a gauge field, but the gauge field itself.
In these mathematical frameworks
we can give a precise definition to $\cA_1$ and $d\phi_0$
separately, and impose the relation $k \, \cA_1 = d\phi_0$.
This approach is taken in \cite{Kapustin:2014zva}
using Deligne-Beilinson cocycles.

\subsubsection*{Cheeger-Simons characters and 
characteristic classes}

The notions of Chern classes, Pontryagin classes, Euler classes
admit a natural generalization in the framework of differential cohomology.
For definiteness, let us focus on Chern classes; analogous remarks
hold for other characteristic classes.
Our exposition follows \cite{myBunke}.

Let $\cV$ be a complex rank-$n$ vector bundle over $\cM$,
with structure group $U(n)$, equipped with a hermitian
fiber metric and a connection $\nabla$
compatible with the fiber metric.
The curvature of $\nabla$ is the 2-form $F_\Delta$ on $\cM$.
In our conventions, $F_\Delta$ is antihermitian.
The Chern forms $c_{k}(\nabla)$ are defined via
\beq
\det \bigg( \mathbb I  + \frac{i \, F_\nabla}{2\pi} \bigg) = 1 + c_1(\nabla)
+ c_2(\nabla) +  \dots \ , \qquad
c_k(\nabla) \in \Omega^{2k}_\mathbb Z (\cM) \ .
\eeq
The $2k$-form $c_k(\nabla)$ is closed and has integral periods.
If we choose a different connection $\nabla'$ on the same vector bundle,
the form $c_k(\nabla')$ is generically different from $c_k(\nabla)$,
but they differ by an exact piece. 
Their de Rham classes are the same, allowing us to define
\beq
c_k^\mathbb R(\cV) = [c_k(\nabla)]_{\rm dR} \in H^{2k}(\cM, \mathbb R) \ .
\eeq
The superscript $\mathbb R$ on 
$c_k^\mathbb R(\cV)$ is inserted to emphasize that
it is an object in the \emph{real} cohomology of $\cM$.
It is known, however, that $c_k^\mathbb R(\cV)$ admits 
an integral refinement: an \emph{integer} cohomology class
$c_k(\cV)$ can be defined, such that
\beq
c_k^\mathbb R(\cV) = \varrho\big( c_k(\cV) \big) \ , \qquad
 c_k(\cV)  \in H^{2k}(\cM, \mathbb Z) \ ,
\eeq
where $\varrho$ is the map \eqref{natural_rho}.
The integer class $c_k(\cV)$ contains more information
than the real class $c_k^{\mathbb R}(\cV)$.
For example, if the bundle $\cV$ can be equipped with a flat connection,
$c_k^{\mathbb R}(\cV) = 0$ but $c_k(\cV)$ can be a non-trivial
element in ${\rm Tor} \, H^{2k}(\cM,\mathbb Z)$.

By definition, a differential refinement of the $k$-th Chern class
is a map that sends a pair $(\cV, \nabla)$ to an element
$\check c_k(\nabla) \in \check H^{2k}(\cM)$, satisfying the following properties:
\begin{enumerate}[(i)]
\item The field strength of the differential character $\check c_k(\nabla)$
is the Chern form $c_k(\nabla) \in \Omega^{2k}_{\mathbb Z}(\cM)$.
\item The characteristic class of the differential character $\check c_k(\nabla)$
is the integral Chern class $c_k(\cV) \in H^{2k}(\cM,\mathbb Z)$.
\item For every smooth map $f: \cM' \rightarrow \cM$, one has 
$f^* \check c_k(\nabla) = \check c_k (f^* \nabla)$.
\end{enumerate}
In the last point, $f^* \check c_k(\nabla) \in \check H^{2k}(\cM')$
is the pullback from $\cM$ to $\cM'$ of the differential character
$c_k(\nabla)$, while $\check c_k (f^* \nabla)$ denotes the element
of $\check H^{2k}(\cM')$ that is associated to the pullback vector
bundle $f^* \cV$ equipped with the pullback connection
$f^*\nabla$.
It can be proven that Chern classes admit a unique
differential refinement.
Similar theorems hold for Pontryagin classes and Euler classes.

Notice that the differential refinement $\check c_k(\nabla)$
retains information about the specific choice of connection
$\nabla$. In more physical terms,
$\check c_k(\nabla)$ has information about the specific
$U(n)$
background gauge field configuration,
whereas the integral Chern class $c_k(\cV)$
only depends on the topology of the bundle $\cV$.



\section{Case study: wrapped M5-branes at a $\mathbb Z_2$ singularity} \label{sec_app_result}

In this appendix we consider the total anomaly polynomial
\eqref{GMSW_inflow_tot} for wrapped M5-branes at a $\mathbb Z_2$ singularity
and we extract physical information about 't Hooft anomalies for discrete symmetries. More precisely, we consider the case in which we assign
Dirichlet boundary conditions to $\cA_1$ and $B_{2i}$
(for each label $i = 1,\dots,g$). The interacting SCFT
has therefore a global $\mathbb Z_k$ 0-form symmetry
and an ``electric'' global $(\mathbb Z_N)^g$ 1-form symmetry.

Our strategy is as follows:
\begin{enumerate}
\item Perform an $SL(3,\mathbb Z)$ transformation on the 1-form 
gauge fields $A_1^\alpha = (A_1, A_1^+, A_1^-)$
to single out the linear combination that enters the BF coupling with
$\gamma_4$, as described in appendix \ref{sec_goodbasis}.
The new basis is denoted $(\cA_1, \cA_1^+, \cA_1^-)$.

\item Collect all terms with $\gamma_4$ and $\widetilde H_3^i$
and define new gauge fields $\boldsymbol{\mathsf A}_1$, $\boldsymbol{\mathsf B}_{2i}$
in such a way that these terms take the form
\beq
I_6^{\rm inflow} \supset - k \, \frac{\gamma_4}{2\pi} \, \frac{d \boldsymbol{\mathsf A}_1}{2\pi}    - N \, \frac{\widetilde H_3^i}{2\pi} \, \frac{d \boldsymbol{\mathsf B}_{2i}}{2\pi} \ .
\eeq
As explained in section \ref{sec_discrete_anomalies}, we can then
dualize $c_3$ and $\widetilde B_2^i$ to $\phi_0$, $\phi_{1i}$, respectively.
We get a 
St\"uckelberg-type theory for the pairs $(\boldsymbol{\mathsf A}_1, \phi_0)$,
$(\boldsymbol{\mathsf B}_{2i}, \phi_{1i})$ with constraints
\beq
k \, \boldsymbol{\mathsf A}_1 = d \phi_0 \ , \qquad N \, \boldsymbol{\mathsf B}_{2i} = d\phi_{1i} \ .
\eeq
Thus $\boldsymbol{\mathsf A}_1$ is the background gauge field for the global
$\mathbb Z_k$ 0-form symmetry
and $\boldsymbol{\mathsf B}_{2i}$ is the background gauge field for the global
$(\mathbb Z_N)^g$ 1-form symmetry.

\item Remove all terms with $\gamma_4$ and $\widetilde H_3^i$
from $I_6^{\rm inflow}$, and write the rest of  $I_6^{\rm inflow}$
in terms of $\cA_1^\pm$, $\boldsymbol{\mathsf A}_1$, $\boldsymbol{\mathsf B}_{2i}$,
  $a_{0i}^\pm$, $\widetilde a_0^{i\pm}$, $p_1(T)$,
and the background gauge fields for isometries of $M_6$.

\end{enumerate}
Let us address each step in turn.

Given the flux quanta $N_\alpha = (N, N_+, N_-)$ we define the 
integers $k$, $m$, $m_\pm$ via
\beq
k = {\rm gcd}(N, N_+, N_-) \ , \qquad N = k \, m \ , \qquad
N_\pm = k \, m_{\pm} \ .
\eeq
As a simplifying technical assumption, we suppose that $m$ and $m_+$
are relatively prime. (Other cases are studied in a similar way.)
It follows that integers $r$, $s$ exist such that
\beq
m \, s - m_+ \, r  = 1 \ .
\eeq
The integers $r$, $s$ are not uniquely determined
by this equation. We suppose that a choice for $r$, $s$ is made
and kept fixed throughout. 
The change of basis of the 1-form gauge fields
can be written as
\beq
\begin{pmatrix}
A_1 \\ A_1^+ \\ A_1^-
\end{pmatrix}
= M^{-1} \, 
\begin{pmatrix}
\cA_1 \\   \cA^{+}_1   \\   \cA^{-}_1
\end{pmatrix} \ , \qquad
M = \begin{pmatrix}
m & m_+ & m_- \\
r & s & 0 \\
0 & 0 & 1
\end{pmatrix} \ .
\eeq
The field strengths of $\cA_1$, $  \cA^{\pm}_1$
are denoted $\cF_2$, $  \cF_2^{\pm}$.

Next, we examine the terms in \eqref{GMSW_inflow_tot} with $\gamma_4$
and $\widetilde H_3^i$. We find
\begin{align}
I_6^{\rm inflow} &  \supset -   \frac{\gamma_4}{2\pi} \, \bigg[
k \, \frac{\cF_2}{2\pi} + \frac{ f_{1i}^+ \, \widetilde f_1^{i-}
- \widetilde f_1^{i+} \, f_{1i}^-  }{(2\pi)^2}
\bigg] \nn \\
& -   \frac{\widetilde H_3^i}{2\pi} \, \bigg[
N \, \frac{H_{3i}}{2\pi}
+ \left( \frac{  m\, \cF_2^+ + m_- \, r \, \cF_2^- - r \, \cF_2    }{2\pi}
- N \, c_1^\psi
  \right) \,   \frac{ f_{1i}^+ }{2\pi}
+ \frac{\cF_2^-}{2\pi}  \,  \frac{ f_{1i}^- }{2\pi}
\bigg] \ .
\end{align}
This means that the new gauge fields $\boldsymbol{\mathsf A}_1$
are defined by
$\boldsymbol{\mathsf B}_{2i}$ satisfy
\begin{align} \label{new_fields}
k \, \frac{d \boldsymbol{\mathsf A}_1}{2\pi}  & = k \, \frac{\cF_2}{2\pi} + \frac{ f_{1i}^+ \, \widetilde f_1^{i-}
- \widetilde f_1^{i+} \, f_{1i}^-  }{(2\pi)^2} \ , \nn \\
N \, \frac{d \boldsymbol{\mathsf B}_{2i}}{2\pi}  & = 
N \, \frac{H_{3i}}{2\pi}
+ \left( \frac{  m\, \cF_2^+ + m_- \, r \, \cF_2^- - r \, \cF_2    }{2\pi}
- N \, c_1^\psi
  \right) \,   \frac{ f_{1i}^+ }{2\pi}
+ \frac{\cF_2^-}{2\pi}  \,  \frac{ f_{1i}^- }{2\pi}
 \ .
\end{align}

The final step is to remove the terms in
the anomaly polynomial with $\gamma_4$, $\widetilde H_3^i$,
and write the rest using \eqref{new_fields} to trade
 $\cF_2$, $H_{3i}$ for
$d\boldsymbol{\mathsf A}_1$, $d \boldsymbol{\mathsf B}_{2i}$. The   result is quite lengthy:
we present it as the sum of several contributions, listed as follows.
\begin{itemize}
\item {\bf Terms containing only   fields for isometries of $M_6$ and Poincar\'e symmetry:}
\begin{align}
I_6^{\rm inflow}& \supset - \frac 13 \, N_- \, (c_1^\psi)^3
+ \frac{1}{12} \, N_- \, c_1^\psi \, p_1(T)
+ \bigg(
N^2 \, N_-  + \frac{\chi }{3} \, N^3 
  - \frac{\chi }{3} \, N 
 \bigg) c_1^\psi \, c_2^\varphi
 \nn \\
 &+ \bigg(
 - \frac 23 \, N^3 
 - N^2 \, N_- + \frac 13 \, N_-^3
 + \frac 23 \, N 
 + \frac 23 \, N_-
\bigg) \, c_1^\psi \, c_2^\Sigma  \ .
\end{align}
\item 
{\bf Terms with three factors $  \cF_2^{\pm}$:}
\begin{align}
 I_6^{\rm inflow}& \supset  \frac{1}{(2\pi)^3} \, \bigg[
- \frac \chi 6 \, m^3  \, (\cF_2^+)^3 
- \frac 12 \, m \, (2 \, m_+ + m \, m_- \, r \,\chi) \, \cF_2^- \, (\cF_2^+)^2
\nn \\
& - \frac 16 \, m_- \, r \, ( 6 \, m_- \, s + 3 \, \chi + m_-^2 \, r^2 \, \chi ) \,  (\cF_2^-)^3
\nn \\
& - \frac 12 \, (
2 \, m_- \, (m_+ \, r + m \, s) + m \, \chi + m \, m_-^2 \, r^2 \, \chi
) \, \cF_2^+ \, (\cF_2^-)^2
 \bigg] \ .
\end{align}
\item 
{\bf Terms with two factors $  \cF_2^{\pm}$  
and two factors $f_{1i}^\pm$, $\widetilde f_1^{i\pm}$:}
\begin{align}
 I_6^{\rm inflow}& \supset  \frac{1}{(2\pi)^4} \, \bigg[
 - \frac{1}{k \, m} \, (\cF_2^-)^2 \, f_{1i}^- \, \widetilde f_1^{i-}
 - \frac{1}{k \, m} \, (m\, \cF_2^+ + m_- \, r \, \cF_2^-)^2 \, f_{1i}^+ \, \widetilde f_1^{i+}
 \nn \\
& - \frac{1}{2 \, k \,m} \, \Big\{ m^3 \, r \, \chi \, (\cF_2^+)^2  
+ 2 \, m \,  (1 + m_+ \, r + m \, s + m \, m_- \, r^2 \, \chi) \, \cF_2^+ \, \cF_2^- 
\nn \\
& + r \, (m_- \, (2 + 4 \, m \, s) + m\, \chi + m \, m_-^2 \, r^2 \, \chi) \,
(\cF_2^-)^2 
\Big\} \, \Big(  f_{1i}^+ \, \widetilde f_1^{i-}
- \widetilde f_1^{i+} \, f_{1i}^-  \Big)
 \bigg]  \ . 
\end{align}
\item 
{\bf Terms with one factor $  \cF_2^{\pm}$  
and four factors $f_{1i}^\pm$, $\widetilde f_1^{i\pm}$:}
\begin{align}
 I_6^{\rm inflow}& \supset    \frac{1}{(2\pi)^5} \, \bigg[
- \frac{2 \, r}{k^2 \, m} \, (m \, \cF_2^+ + m_- \, r \, \cF_2^-) \,  
f_{1j}^+ \, \widetilde f_1^{j+} \,  \Big(  f_{1i}^+ \, \widetilde f_1^{i-}
- \widetilde f_1^{i+} \, f_{1i}^-  \Big)
  \\
& - \frac{r}{2 \, k^2 \, m} \, \Big( 
m^2 \, r \, \chi \, \cF_2^+ 
+ (2 + 2 \, m \, s + m \, m_- \, r^2 \, \chi) \, \cF_2^-
\Big) \, 
 \Big(  f_{1i}^+ \, \widetilde f_1^{i-}
- \widetilde f_1^{i+} \, f_{1i}^-  \Big)^2
 \bigg] \nn \ .
\end{align}
\item 
{\bf Terms with six factors $f_{1i}^\pm$, $\widetilde f_1^{i\pm}$:}
\begin{align}
 I_6^{\rm inflow}& \supset  \frac{1}{(2\pi)^6} \,\bigg[
- \frac{r^2}{k^3 \, m} \, f_{1j}^+ \, \widetilde f_1^{j+} \,  \Big(  f_{1i}^+ \, \widetilde f_1^{i-}
- \widetilde f_1^{i+} \, f_{1i}^-  \Big)^2
- \frac{r^3 \, \chi}{6 \, k^3} \, 
\Big(  f_{1i}^+ \, \widetilde f_1^{i-}
- \widetilde f_1^{i+} \, f_{1i}^-  \Big)^3
 \bigg] \ .
\end{align}
\item
{\bf Terms with one factor $c_1^\psi$ and $  \cF_2^{\pm}$ and/or
$f_{1i}^\pm$, $\widetilde f_1^{i\pm}$:}
\begin{align}
 I_6^{\rm inflow}& \supset 
- \frac{1}{(2\pi)^2} \, k \, m_- \, (m\, \cF_2^+ + m_- \, r \, \cF_2^- )^2  \, c_1^\psi
- \frac{1}{(2\pi)^3} \, 2 \,  (m\, \cF_2^+ + m_- \, r \, \cF_2^- ) \, f_{1j}^- \, \widetilde f_1^{j-} \, c_1^\psi
\nn  \\
& - \frac{1}{(2\pi)^3} \, 2 \, m_- \, r \, (m\, \cF_2^+ + m_- \, r \, \cF_2^- )  \, \Big(  f_{1i}^+ \, \widetilde f_1^{i-}
- \widetilde f_1^{i+} \, f_{1i}^-  \Big) \, c_1^\psi
  \\
&+ \frac{1}{(2\pi)^4} \bigg[ - \frac{2 \, r}{k} \, f_{1j}^- \, \widetilde f_1^{j-}  \, \Big(  f_{1i}^+ \, \widetilde f_1^{i-}
- \widetilde f_1^{i+} \, f_{1i}^-  \Big) \, c_1^\psi
- \frac{m_- \, r^2}{k} \, \Big(  f_{1i}^+ \, \widetilde f_1^{i-}
- \widetilde f_1^{i+} \, f_{1i}^-  \Big)^2  \bigg] \, c_1^\psi \ . \nn
 \end{align}
\item
{\bf Terms with two factors $c_1^\psi$ and $  \cF_2^{\pm}$ and/or
$f_{1i}^\pm$, $\widetilde f_1^{i\pm}$:}
\begin{align}
 I_6^{\rm inflow}& \supset  
 \frac{1}{2\pi} \,  \frac 12 \, (m\, \cF_2^+ + m_- \, r \, \cF_2^- ) \, \Big[ 
 k^2 \, m \, (2 \, m_- + m \, \chi) - \chi
 \Big] \, (c_1^\psi)^2
 \nn \\
& + \frac{1}{(2\pi)^2} \,  \frac{r}{2 \, k} \,  \Big[ 
 k^2 \, m \, (2 \, m_- + m \, \chi) - \chi
 \Big] \, \Big(  f_{1i}^+ \, \widetilde f_1^{i-}
- \widetilde f_1^{i+} \, f_{1i}^-  \Big) \, (c_1^\psi)^2
\nn \\
& + \frac{1}{(2\pi)^2} \, k \, m \, f_{1j}^- \, \widetilde f_1^{j-}  \, (c_1^\psi)^2 \ .
\end{align}
\item
{\bf Terms with one factor $c_2^\varphi$ and $  \cF_2^{\pm}$ and/or
$f_{1i}^\pm$, $\widetilde f_1^{i\pm}$:}
\begin{align}
 I_6^{\rm inflow}& \supset   
\frac{1}{2\pi} \,  \bigg[ - k^2 \, m^2 \, m_- \, \cF_2^+ - k^2 \, m \, (m_+ + m_-^2 \, r) \, \cF_2^- \bigg] \, c_2^\varphi
\nn \\
& +\frac{1}{(2\pi)^2} \, \bigg[ - k \, m \, m_- \, r \,  \Big(  f_{1i}^+ \, \widetilde f_1^{i-}
- \widetilde f_1^{i+} \, f_{1i}^-  \Big)
- k \, m \, (
f_{1i}^+ \, \widetilde f_1^{i+} + f_{1i}^- \, \widetilde f_1^{i-}
) \bigg] \, c_2^\varphi
\end{align}
\item
{\bf Terms with one factor $c_2^\Sigma$ and $  \cF_2^{\pm}$ and/or
$f_{1i}^\pm$, $\widetilde f_1^{i\pm}$:}
\begin{align}
 I_6^{\rm inflow}& \supset 
 \frac{1}{2\pi} \,  k^2 \, (m + m_-) \, \bigg[ 
 ( m \, m_- + m_+^2 ) \, \cF_2^+ + (m_+ + m_-^2 \, r + m_- \, m_+ \, s) \, \cF_2^-
 \bigg] \, c_2^\Sigma
 \nn \\
& +\frac{1}{(2\pi)^2} \,  k \, (m + m_-) \, (m_- \, r + m_+ \, s) \, \Big(  f_{1i}^+ \, \widetilde f_1^{i-}
- \widetilde f_1^{i+} \, f_{1i}^-  \Big) \, c_2^\Sigma \ .
 \end{align}
\item
{\bf Terms with one factor $p_1(T)$ and $  \cF_2^{\pm}$ and/or
$f_{1i}^\pm$, $\widetilde f_1^{i\pm}$:}
\begin{align}
 I_6^{\rm inflow}& \supset 
 \frac{1}{2\pi} \,  \frac{\chi}{24} \,  (m \, \cF_2^+ + m_- \, r \, \cF_2^-)\, p_1(T) 
+ \frac{1}{(2\pi)^2} \, \frac{\chi \, r}{24 \, k} \,  \Big(  f_{1i}^+ \, \widetilde f_1^{i-}
- \widetilde f_1^{i+} \, f_{1i}^-  \Big) \, p_1(T)
 \end{align}
\item {\bf Terms cubic and quadratic in $d\boldsymbol{\mathsf A}_1$:}
\begin{align} \label{discrete_anomalies_no1}
 I_6^{\rm inflow}& \supset 
 \frac{1}{(2\pi)^3} \, \frac{r^3 \, \chi}{6} \, (d\boldsymbol{\mathsf A}_1)^3 
 - \frac{1}{(2\pi)^2} \, k \, m_- \, r^2 \, c_1^\psi \,  (d\boldsymbol{\mathsf A}_1)^2
 \nn \\
 & + \frac{1}{(2\pi)^3} \bigg[ 
 - \frac \chi 2 \, m \,r^2 \, \cF_2^+
 - \frac r2 \, ( 2 \, s + m_- \, r^2 \, \chi ) \, \cF_2^-
 \bigg] \,  (d\boldsymbol{\mathsf A}_1)^2
 \nn \\
& + \frac{1}{(2\pi)^4} \, \bigg[ - \frac{r^3 \, \chi}{2\,k} \,   \Big(  f_{1i}^+ \, \widetilde f_1^{i-}
- \widetilde f_1^{i+} \, f_{1i}^-  \Big)  
- \frac{r^2}{k \, m} \, f_{1i}^+ \, \widetilde f_1^{i+} 
\bigg] \, (d\boldsymbol{\mathsf A}_1)^2 \ .
\end{align}
\item 
{\bf Terms  linear in $d \boldsymbol{\mathsf A}_1$, without $f_{1i}^\pm$, $\widetilde f_1^{i\pm}$ or $d\boldsymbol{\mathsf B}_{2i}$:}
\begin{align}
 I_6^{\rm inflow}& \supset 
  + \frac{1}{2\pi} \, \bigg[
k^2 \, m \, m_- \, r \, c_2^\varphi
- k^2 \, (m + m_-) \, (m_- \, r + m_+ \, s) \, c_2^\Sigma
- \frac{r \, \chi}{24} \, p_1(T)
\nn \\
& + \frac r2 \, (\chi - k^2 \, m \, (2 \, m_- + m \, \chi)) \, (c_1^\psi)^2
\bigg] \, d \boldsymbol{\mathsf A}_1
\nn \\
& + \frac{1}{(2\pi)^3} \, \bigg[
 \frac{m^2 \, r \, \chi}{2} \, (\cF_2^+)^2
 + \frac r2 \, (4 \, m_- \, s + \chi + m_-^2 \, r^2 \, \chi) \, (\cF_2^-)^2 
\nn \\
& + (m_+ \, r + m \, (s + m_- \, r^2 \, \chi)) \, \cF_2^+ \, \cF_2^-
\bigg] \, d \boldsymbol{\mathsf A}_1
\nn \\
& +\frac{1}{(2\pi)^2} \,  2 \, k \, m_- \, r \, c_1^\psi \, (m \, \cF_2^+ + m_- \, r \, \cF_2^-) \, d\boldsymbol{\mathsf A}_1 \ .
\end{align}
\item 
{\bf Terms linear in $d\boldsymbol{\mathsf A}_1$ with two or four factors $f_{1i}^\pm$, $\widetilde f_1^{i\pm}$:}
\begin{align}
 I_6^{\rm inflow}& 
\supset \frac{1}{(2\pi)^3} \,  2 \, r \, c_1^\psi \, f_{1i}^- \, \widetilde f_1^{i-} \, d\boldsymbol{\mathsf A}_1
 + \frac{1}{(2\pi)^3} \, 2 \, m_- \, r^2 \, c_1^\psi \,   \Big(  f_{1i}^+ \, \widetilde f_1^{i-}
- \widetilde f_1^{i+} \, f_{1i}^-  \Big)  
\, d\boldsymbol{\mathsf A}_1
  \\
& +\frac{1}{(2\pi)^4} \,  \frac{2 \, r}{k \, m} \,  f_{1i}^+ \, \widetilde f_1^{i+}\, (m \, \cF_2^+ + m_- \, r \, \cF_2^-) \, d\boldsymbol{\mathsf A}_1
 \nn \\
& +\frac{1}{(2\pi)^4} \,  \frac{r}{k \, m} \, \Big[ 
m^2 \, r \, \chi \, \cF_2^+ + (1 + 2 \, m \, s + m \, m_- \, r^2 \, \chi) \, \cF_2^-
\Big] \, \Big(  f_{1i}^+ \, \widetilde f_1^{i-}
- \widetilde f_1^{i+} \, f_{1i}^-  \Big)   \, d\boldsymbol{\mathsf A}_1
\nn \\
& + \frac{1}{(2\pi)^5} \,  \frac{r^3 \, \chi}{2 \, k^2} \,   \Big(  f_{1i}^+ \, \widetilde f_1^{i-}
- \widetilde f_1^{i+} \, f_{1i}^-  \Big) ^2 \, d\boldsymbol{\mathsf A}_1
+\frac{1}{(2\pi)^5} \,  \frac{2 \, r^2}{k^2 \, m} \,  f_{1i}^+ \, \widetilde f_1^{i+}\,  \Big(  f_{1i}^+ \, \widetilde f_1^{i-}
- \widetilde f_1^{i+} \, f_{1i}^-  \Big) \, d\boldsymbol{\mathsf A}_1
\ . \nn
\end{align}
\item {\bf Terms with $d \boldsymbol{\mathsf B}_{2i}$:}
\begin{align}  \label{discrete_anomalies_final}
 I_6^{\rm inflow}& \supset 
  \frac{1}{(2\pi)^3} \,  \cF_2^- \, d\boldsymbol{\mathsf B}_{2i} \, \widetilde f_{1}^{i-}
 +  \frac{1}{(2\pi)^3} \,  (m \, \cF_2^+ + m_- \, r \, \cF_2^-) \, d\boldsymbol{\mathsf B}_{2i} \, \widetilde f_{1}^{i+}
-  \frac{1}{(2\pi)^2} \,  k \, m \, c_1^\psi \, d\boldsymbol{\mathsf B}_{2i} \, \widetilde f_{1}^{i+}
\nn \\
&+ \frac{1}{(2\pi)^4} \, \frac r k \, \Big(  f_{1j}^+ \, \widetilde f_1^{j-}
- \widetilde f_1^{j+} \, f_{1j}^-  \Big) \, d\boldsymbol{\mathsf B}_{2i} \, \widetilde f_{1}^{i+}
- \frac{1}{(2\pi)^3} \, r \, d\boldsymbol{\mathsf A}_1 \, d\boldsymbol{\mathsf B}_{2i} \, \widetilde f_{1}^{i+} \ .
\end{align}

\end{itemize}
We have a rich variety of 't Hooft anomalies
involving the $\mathbb Z_k$ 0-form symmetry and the
$(\mathbb Z_N)^g$ 1-form symmetry.



\section{Free tensor multiplet reduction on $\Sigma_g$ with topological twist}
\label{app_tensor}
A free 6d $\cN = (2,0)$ tensor multiplet consists of: a chiral 2-form
$b_{\mu\nu}$
which is a singlet of $SO(5)_R$; a symplectic Majorana-Weyl fermion $\chi$
in the representation $\mathbf 4$ of $USp(4)_R \cong SO(5)_R$; 
five real scalar fields $\phi^1$, \dots, $\phi^5$ in the vector representation   of $SO(5)_R$.
In this appendix we study the reduction of this multiplet on a 
genus-$g$ Riemann surface $\Sigma_g$ with a non-zero $SO(5)_R$
background connection. The latter is encoded in the twist parameters
$p$, $q$ defined in section \ref{BBBW_sol_sec} and satisfying $p+q = -\chi$.

Since the chiral 2-form $b_{\mu\nu}$ is a singlet of $SO(5)_R$, it is unaffected
by the topological twist. Its reduction on a genus-$g$ Riemann surface yields the following massless fields:
$g$ real 4d vectors and one real 4d scalar $b_0$.  
The reduction of the 6d fermion $\chi$ and the 6d scalars $\phi^1$, \dots, $\phi^5$, on the other hand,
is sensitive to the twist parameters. We collect 
all  massless 4d fields, their origins, and their multiplicities 
 in   table \ref{free_tensor}.

\begin{table}
\begingroup
\renewcommand{\arraystretch}{1.3}
\begin{center}
\begin{tabular}{ c | ccc | c | c|c  }
6d origin  & $U(1)_1$ &  $U(1)_2$ & $U(1)_\Sigma$ & $U(1)_\Sigma'$  & 4d field & multiplicity \\ \hline\hline
\multirow{4}{*}{$\chi$} & $+1$  & $+ 1$  & $+\frac 12$ & $1$ & $\lambda_\alpha$ & $g$     \\
  & $-1$ & $-1$ & $+\frac 12$  & $0$  & $\psi_\alpha$ & $1$     \\
  & $+1$  & $- 1$ & $+\frac 12$ &  $\frac{p}{p+q}$ & $\Lambda_\alpha$ & $h^0(K^{\frac{p}{p+q} })$
 \\
  & $-1$  & $+1$  & $+\frac 12$ &  $\frac{q}{p+q}$ &   $\widehat \Lambda_{   \alpha}$ & 
$h^0(K^{\frac{q}{p+q}})$ \\ \hline
$\phi^1 + i \, \phi^2$ & $+2$ & $0$ & $0$ & $\frac{p}{p+q}$ & $q$ & $h^0(K^{\frac{p}{p+q}})$ \\
$\phi^3 + i \, \phi^4$ & $0$ & $+2$ & $0$ & $\frac{q}{p+q}$ & $\widehat q$ & $h^0(K^{\frac{q}{p+q}})$ \\
$\phi^5$ & $0$ & $0$ & $0$ & $0$ & $\Phi$ & $1$ \\ \hline
\multirow{2}{*}{$b_{\mu\nu}$} & $0$ & $0$ & $\pm1$ & $\pm1$ & $A_\mu$ & $g$ \\
 & $0$ & $0$ & $0$ & $0$ & $b_0$ & $1$
\end{tabular}
\end{center}
\endgroup
\caption{ 
Massless 4d fields originating from dimensional reduction of a 6d
$\cN = (2,0)$ free tensor multiplet on a genus-$g$ Riemann surface ($g\neq 1$)
with twist parameters $p$, $q$ satisfying $p+q = 2(g-1)$.
} 
\label{free_tensor}
\end{table}

As we can see from the charges of $\phi^1 +  i\, \phi^2$
and $\phi^3 + i \, \phi^4$,
the subgroup $U(1)_1 \subset SO(5)_R$ is identified with rotations in the 12 plane,
and $U(1)_2 \subset SO(5)_R$ is identified with rotations in the 34 plane.
They are both normalized in such a way that their minimal charge is $\pm 1$.
The notation $U(1)_\Sigma$ refers to  local frame rotations on the Riemann surface.
Its normalization is such that a chiral spinor on $\Sigma_g$ has $U(1)_\Sigma$ charge $\pm \frac 12$.
The symbol $U(1)'_\Sigma$ stands for the twisted
local frame rotations on the Riemann surface.
More precisely,
\beq
t_\Sigma' = t_\Sigma  + \frac{p}{2(p+q)} \, t_1 + \frac{q}{2(p+q)} \, t_2 \ ,
\eeq
where $t'_\Sigma$, $t_\Sigma$, $t_1$, $t_2$
are the generators of $U(1)'_\Sigma$, $U(1)_\Sigma$, $U(1)_1$, $U(2)_2$, respectively.
Our discussion applies to $g \neq 1$. The case $g=1$ is discussed at the end of this appendix.

In table \ref{free_tensor}, the 4d fields
$\lambda_\alpha$, $\psi_\alpha$, $\Lambda_\alpha$, $\widehat \Lambda_\alpha$
are Weyl spinors of positive chirality,
$q$ and $\widehat q$ are complex scalars, $\Phi$ and $b_0$ are real scalars,
and $A_\mu$ are real vectors.
For each 4d field, the $U(1)_\Sigma'$ charge determines the bundle of which the
corresponding internal
wavefunctions must be a section. Massless fields originate from covariantly constant sections,
or equivalently holomorphic sections. The symbol $K$ stands for the canonical
bundle on $\Sigma_g$.

The fields listed in table \ref{free_tensor} are organized into the following multiplets
of 4d $\cN = 1$ supersymmetry:
\begin{itemize}
\item $(A_\mu, \lambda_\alpha)$: a collection of $g$ vector multiplets;
\item $(\Phi, b_0, \psi_\alpha)$: one chiral multiplet with $U(1)_1 \times U(1)_2$ charges $(0,0)$;
\item $(q, \Lambda_\alpha)$: a collection of $h^0(K^{\frac{p}{p+q}})$ 
chiral multiplets with $U(1)_1 \times U(1)_2$ charges $(2,0)$;
\item $(\widehat q, \widehat \Lambda_\alpha)$: a collection of $h^0(K^{\frac{q}{p+q}})$ 
chiral multiplets with $U(1)_1 \times U(1)_2$ charges $(0,2)$.
\end{itemize}

If an integer $m$ divides $2(g-1)$, it is possible to define an $m$-th root $K^{\frac 1m}$ of the 
canonical bundle, but the root is not unique.
Since $p+q = 2(g-1)$, the bundles $K^{\frac{p}{p+q}}$, $K^{\frac{q}{p+q}}$
can be defined, but we would require more data to fully specify them.
(For example, for $p=q$ the additional data is a choice of spin structure on $\Sigma_g$.)
Even though we are not able to determine the
multiplicities $h^0(K^{\frac{p}{p+q}})$ and $h^0(K^{\frac{q}{p+q}})$ 
without further input,
 the Riemann-Roch theorem implies the   relation
\beq \label{difference_rel}
h^0(K^{\frac{p}{p+q}}) - h^0(K^{\frac{q}{p+q}}) = \frac 12 \, (p-q) \ .
\eeq
Notice that, since $p+q$ is an even integer, so is $p-q$, so the RHS is an integer.
To justify \eqref{difference_rel}, we notice that the Riemann-Roch theorem
can be stated as
\beq \label{RiemannRochBIS}
h^0(\cL) - h^0(\cL^{-1} \otimes K) = \deg (\cL) + 1 - g \ ,
\eeq
where $\cL$ is a line bundle on $\Sigma_g$. If we set 
$\cL = K^{\frac{p}{p+q}}$, then we have 
$\cL^{-1} \otimes K = K^{\frac{q}{p+q}}$. Moreover,
$\deg (K) = 2(g-1)$ gives $\deg (\cL) = 2(g-1) \, \frac{p}{p+q} = p$,
and \eqref{RiemannRochBIS} implies \eqref{difference_rel}.

Interestingly, the 't Hooft anomaly polynomial of the 4d fields listed in table
\ref{free_tensor} (with those  $U(1)_1$, $U(1)_2$ charge assignments) only depends on the difference $h^0(K^{\frac{p}{p+q}}) - h^0(K^{\frac{q}{p+q}})$.
We can thus make use of \eqref{difference_rel} and verify that
the anomaly polynomial computed from table \ref{free_tensor}
matches exactly with the integration over $\Sigma_g$
of the 8-form anomaly polynomial of a free 6d $\cN = (2,0)$ tensor multiplet.

We also notice that if we set $q=0$, $p = - \chi$, we get
a number 
$h^0(K^{\frac{p}{p+q}}) = h^0(K) = g$ of chiral multplets with $U(1)_1 \times U(1)_2$
charges $(2,0)$, and a number  
$h^0(K^{\frac{q}{p+q}}) = h^0(K^0) = 1$ of chiral multplets with $U(1)_1 \times U(1)_2$
charges $(0,2)$. 
The 4d multiplets $(A_\mu, \lambda_\alpha)$
and $(q,\Lambda_\alpha)$ fit into $g$ $\cN = 2$ vector multiplets,
while $(\Phi, b_0, \psi_\alpha)$ and $(\widehat q, \widehat \Lambda_\alpha)$
fit into one $\cN = 2$ hypermultiplet.
It should be stressed that, because of its charge assignments,
the contribution of this  hypermultiplet   to the 4d 't Hooft anomalies 
is equal to $-1$ times the contribution of a vector multiplet.

We may also consider the case $p=q=- \chi/2$.
The chiral multiplets $(q,\Lambda_\alpha)$, $(\widehat q, \widehat \Lambda_{\alpha})$
have the same multiplicity and fit into a doublet of the enhanced flavor
symmetry $SU(2)_F$. In contrast, the chiral
multiplet $(\Phi, b_0, \psi_\alpha)$ is a singlet of $SU(2)_F$.
(The Cartan of $SU(2)_F$ is proportional
to the difference $t_1 - t_2$.)

Finally, let us comment on the case $g=1$, $p \neq 0$. The Riemann surface is flat and
its canonical bundle is trivial. The total covariant derivative on $T^2$
(in local flat coordinates) has no spin connection term but includes
 the terms originating from the background $U(1)_1 \times U(1)_2 \subset SO(5)_R$ fields. It 
takes the schematic form $D_m = \partial_m + p \, A_m \, (t_1 - t_2)$,
where $A_m$ is a local antiderivative of the volume form on $T^2$.
With reference to table \ref{free_tensor},
the modes of $\chi$ with  $t_1 = t_2 = \pm 1$
are unaffected by the $U(1)_1 \times U(1)_2$ background.
To get massless modes in four dimensions, we take
their internal  wavefunction to be a covariantly constant spinor on $T^2$.
Since $T^2$ is flat, a covariantly constant spinor is constant,
yielding a multiplicity 1 for both $\lambda_\alpha$ and $\psi_\alpha$.
(We select periodic boundary conditions on both 1-cycles of $T^2$.)
As before, the fermion $\lambda_\alpha$ combines with $A_\mu$ in one vector multiplet,
and $\psi_\alpha$ combines with $b_0$ and $\Phi$ in one chiral multiplet
with $U(1)_1 \times U(1)_2$ charges $(0,0)$.
The mode of $\chi$ with $(t_1, t_2) = (+1,-1)$, denoted $\Lambda_\alpha$ in table \ref{free_tensor},
is affected by the topological twist, and behaves as a section of
$\cL^p$, where $\cL$ is a degree-one line bundle on $T^2$.
The same holds true for the scalar $\phi^1 + i \, \phi^2$.
The fields $\Lambda_\alpha$, $\phi^1 + i \, \phi_2$ fit into chiral multiplets
with $U(1)_1 \times U(1)_2$ charges $(2,0)$ and multiplicity
$h^0(\cL^p)$. In a similar way,
$\widehat \Lambda_\alpha$ and $\phi^1 - i \, \phi^2$ fit into 
 chiral multiplets
with $U(1)_1 \times U(1)_2$ charges $(0,2)$ and multiplicity
$h^0(\cL^{-p})$. 
The difference between $h^0(\cL^p)$ and 
$h^0(\cL^{-p})$ can be computed using 
\eqref{RiemannRochBIS}, with the replacement $\cL \rightarrow \cL^p$,
to give
\beq  
h^0(\cL^p) - h^0(\cL^{-p} ) = p  \ .
\eeq
As in the $g \neq 1$ case, this relation can be used to verify that the
't Hooft anomalies of the 4d fields match with the result obtained
by integration over the Riemann surface of the 8-form anomaly
polynomial of a free 6d $\cN = (2,0)$ tensor multiplet.




\bibliographystyle{./ytphys}
\bibliography{./refs}

\end{document}